\documentstyle[prl,aps,epsf]{revtex}

\newcommand{\be}{\begin{equation}}
\newcommand{\bea}{\begin{eqnarray}}
\newcommand{\ba}{\begin{array}}
\newcommand{\bean}{\begin{eqnarray*}}
\newcommand{\ee}{\end{equation}}
\newcommand{\eea}{\end{eqnarray}}
\newcommand{\ea}{\end{array}}
\newcommand{\eean}{\end{eqnarray*}}

\def \dsl {\partial \kern-.55em{/}}
\def \Dsl {D \kern-.65em{/}}
\def \qsl {q \kern-.45em{/}}
\def \slp {p \kern-.45em{/}}
\def \psl {p \kern-.45em{/}}

\begin{document}

 \begin{center}
{\Large {\bf The 
pinch technique at two-loops: \\[0.2cm]
The case of mass-less Yang-Mills theories}}\\[0.4cm]
{\large Joannis Papavassiliou} \\[0.3cm]
{\em Departamento de F\'{\i}sica Te\'orica, Univ. Valencia\\
E-46100 Burjassot (Valencia), Spain}\\[0.3cm]

\end{center}

\vskip0.5cm     \centerline{\bf   ABSTRACT}  \noindent
The  generalization  of the    pinch  technique  beyond one   loop  is
presented. It   is  shown that   the crucial  physical  principles  of
gauge-invariance,  unitarity,  and gauge-fixing-parameter independence
are instrumental for accomplishing  this   task, and is explained   
how the aforementioned requirements single out at two loops
exactly  the same algorithm which has   been used to  define the pinch
technique  at   one loop, without  any   additional   assumptions. The
two-loop construction of the pinch technique  gluon self-energy,
and quark-gluon vertex are  carried out in detail
for  the case of  mass-less Yang-Mills  theories,  such as perturbative
QCD.   We present two  different but complementary derivations.  First
we  carry out the construction   by  directly rearranging two-loop
diagrams.    The  analysis reveals    that,  quite  interestingly, the
well-known one-loop correspondence between the pinch technique and the
background field method  in the Feynman gauge  {\it persists} also  at
two-loops.  Since we     use dimensional  regularization    the entire
construction does not depend on the value  of the space-time dimension
$d$.  The renormalization (when $d=4$) is discussed  in detail, and is
shown to   respect  the  aforementioned correspondence.    Second,  we
present an absorptive   derivation,  exploiting the unitarity of    the
$S$-matrix and the underlying BRS symmetry; at this stage we deal only
with  tree-level and one-loop physical  
amplitudes.  
The gauge-invariant sub-amplitudes defined  by
means of  this  absorptive construction  correspond  precisely to  the
imaginary parts  of the $n$-point functions
defined  in the    full two-loop
derivation, thus furnishing 
a  highly non-trivial self-consistency check
for the entire method. Various future applications are
briefly discussed.

\vskip1.0cm
e-mail: Joannis.Papavassiliou@cern.ch;@uv.es
\vskip1.0cm
FTUV-99-12-15
\newpage

\setcounter{equation}{0}
\section{Introduction}
\indent

The pinch technique (PT) was introduced by Cornwall
\cite{PTCPT},  as   the first   step  in the
development    of   a   self-consistent   truncation   scheme  for the
Schwinger-Dyson equations of QCD \cite{PT}.   It was further developed
in   \cite{PT2},  and was generalized    to theories  with spontaneous
symmetry breaking \cite{pap1}, and  subsequently to the full  Standard
Model \cite{DS,papsm}.  Various phenomenological applications
were presented \cite{DKS}, 
and  the one-loop correspondence  between
the  PT and the  background field method (BFM) \cite{BFM1,Abb,AGS} was
established   \cite{BFMPT}.  The important   r\^ole  of unitarity   and
analyticity was  appreciated  and exploited   in a series   of  papers
\cite{PP1,PP2,PRW,PP3},  which allowed   the   generalization of   the
Breit-Wigner formalism   for  unstable  particles  to  the non-abelian
sector of the Standard Model.  Parallel developments took place in the
field of  finite temperature QCD \cite{Tqcd},  the concept  of the QCD
effective  charge  \cite{PT}  was examined in detail \cite{NJW2}, and
subsequently applied to the field of renormalon calculus
\cite{SPEDR}, and the   generalized PT  was  presented \cite{GPT}.   
In addition, a
formalism  for studying   resonant   CP violation \cite{CP}   has been
developed \cite{CPPT}, and various issues  related to the
high-energy behaviour of the W-fusion process
\cite{RKWJS} have been resolved \cite{Wfus}.

The  PT reorganises   systematically a given   physical amplitude into
sub-amplitudes,   which  have  the     same  kinematic  properties  as
conventional $n$-point functions, (propagators, vertices, boxes)
\cite{KL}, and are in addition endowed with 
important physical properties.
The essential ingredient one uses in this rearrangement 
is the full exploitation of the elementary 
Ward identities (WI)
of  the  theory,  in  order to enforce crucial cancellations. 
The various types of
physical problems for which the  PT can serve as a useful
tool have been discussed  frequently in the literature, most recently
in \cite{papCor} and \cite{pap2L1}.  

So far the PT program has been carried out only at one-loop order, and
its generalization to higher orders has been a long-standing question
\cite{NJWMontp}. The general methodology  of how the extension   
of the PT proceeds  at two-loops has  been presented in a recent brief
communication \cite{pap2L1}.  There it was explained how the 
one-loop formalism,
when properly  interpreted  and  suitably  adopted  to   the  two-loop
context,  leads naturally to the PT  extension.  In this paper we will
address the technical aspects of  this  procedure in detail, and  will
discuss extensively the plethora of physical issues involved.

We will   present  two independent   constructions,  which are however
inextricably connected. The first one deals directly with the two-loop
$S$-matrix  element  for the    scattering process  $q  \bar{q} \to  q
\bar{q}$,   exactly   as was the   case   in  the early   one-loop  PT
applications.  The first   crucial ingredient is  that of gauge-fixing
parameter (GFP) independence. $S$-matrix elements are guaranteed to be
GFP  independent; in fact,   as  happens in  the  one-loop  case,  the
cancellation    of all  GFP  dependent   terms  proceeds by exploiting
tree-level WI only, without having to carry out sub-integrations. This
property    is important,   because it   preserves   the  diagrammatic
representation of  the S-matrix, as well  as the kinematic identity of
the sub-amplitudes  appearing in it  (propagators,vertices,boxes). One
can therefore, without loss of generality,  choose a convenient gauge,
such as  the  renormalizable Feynman  gauge  (RFG),  provided that one
considers  the entire set of  two-loop  Feynman graphs contributing to
the given $S$-matrix element

In the  one-loop  case the  next step   would  be to split  the   bare
three-gluon vertex appearing inside   the one-loop quark-gluon  vertex
into a pinching and  non-pinching contribution. This splitting is very
special because it guarantees  that  the resulting  effective  Green's
functions satisfy  (at  one-loop) naive, QED-like  WI
\cite{PT}, instead of  the usual  Slavnov-Taylor  identities \cite{ST}.    
As a consequence the effective one-loop gluon self-energy captures the
running of the QCD coupling, exactly as happens with the photon vacuum
polarization in  QED.    In the two-loop   case  the  construction  is
lengthier but   the  crucial operation is precisely     the same.  One
carries out the aforementioned splitting to  all vertices whose one of
the  incoming    momenta  is the    physical   momentum  transfer  (or
center-of-mass energy) of  the process. As we will  see in detail this
splitting is sufficient to give rise  to two-loop PT effective Green's
functions which  have  the exact  same  properties as  their  one-loop
counterparts.

The reason why all other  three-gluon vertices inside the loops should
remain unchanged (no splitting) can  be understood by resorting to the
special unitarity properties that the PT sub-amplitude must satisfy, a
point  which brings us to the  second derivation.  For this derivation
we employ the unitarity and analyticity  properties of the $S$-matrix,
very   much in the spirit  presented  in the  more  recent one-loop PT
literature \cite{PP2,PRW}.  
There, the  precise diagrammatic  correspondence  between
the  one-loop  (forward)  process $q\bar{q}   \to  q\bar{q}$  and  the
tree-level process $q\bar{q} \to  gg$ provided very  useful, stringent
constraints on the entire  construction, rendering the method  all the
more powerful.   These  constraints are  automatically encoded in  the
two-loop  PT construction presented here;  in fact  they are even more
constraining than  in  the previous order,  for  reasons that we  will
describe  qualitatively now, and  in great detail in  the main body of
the paper.  The  imaginary parts of  the two-loop PT Green's functions
are related by the optical  theorem to precisely identifiable and very
special  parts   of two  different   on-shell processes,  the one-loop
process  $q\bar{q} \to gg$  and the  tree-level  process $q\bar{q} \to
ggg$.   In particular, the  two-particle
 Cutkosky cuts  of the two-loop PT
self-energy are related to the {\it PT rearranged} $s$-channel part of
the one-loop   $q\bar{q}   \to gg$,   while,  at the  same  time,  the
three-particle
Cutkosky cuts of the same quantity  are related to the {\it
PT rearranged} $s$-channel  part of the tree-level 
$q\bar{q} \to  ggg$; the latter is
the  {\it  exact} analogue of the  PT  rearranged $s$-channel  part of
the tree-level $q\bar{q} \to gg$, already studied in the
literature cited above.

The paper is  organized as follows: In section  II we present a  brief
overview  of  the one-loop  PT    algorithm,   and discuss  the   most
characteristic  properties  of  the   one-loop PT  effective   Green's
functions.  Here we follow the  original PT formulation \cite{PT,PT2},
and postpone the  overview of the  one-loop absorptive PT construction
\cite{PP2,PRW}  until section V.  In  section III  we present the full
two-loop  construction.    This section  contains  three sub-sections,
where we deal separately  with the one-particle reducible  graphs, the
two-loop quark-gluon  vertex, and the two-loop  gluon self-energy.  In
the  second sub-section we use  at intermediate  steps various results
for the one-loop three-gluon vertex which  are derived in section VI,
and  are  more naturally  integrated  in the analysis presented there.
Section III contains the main results of this paper; a major highlight
is the proof that the correspondence between the PT and the Background
Field Method Feynman gauge (BFMFG)  persists at two-loops.  In section
IV we carry out the  renormalization program in  detail.  We show that
the two-loop PT Green's function   can be renormalized by  judiciously
rearranging   the   existing counterterms,  and  that   this procedure
respects  the correspondence between PT  and  BFMFG established in the
previous section.  In section  V we present  the general formalism and
methodology of the absorptive PT  construction; this second derivation
is completely independent of the first, but at the same time is deeply
connected to  it.  In this   section we first  derive various formulas
which will be used    in the next    section, and present   a thorough
overview of the one-loop case.  In section VI  we present the two-loop
absorptive derivation.   There are  two   sub-sections: the  first one
contains a detailed  discussion of the  one-loop process $q\bar{q} \to
gg$, and  its  r\^ole in  enforcing  the unitarity  of the  individual
two-loop  PT  Green's functions; the   second sub-section contains the
study of the PT rearranged tree-level process $q\bar{q}
\to  ggg$, which demonstrates the crucial function of this process
in realizing the aforementioned unitarity properties.
Finally, in section  VII we present our conclusions, and discuss
possible connections with other work as well as future applications.

\setcounter{equation}{0}
\section{Overview of the one-loop case}

In this section we briefly review the one-loop PT construction
and establish some useful notation. 

The fundamental tree-level three-gluon vertex 
$\Gamma_{\alpha\mu\nu}^{(0)}(q,p_1,p_2)$ is given by the
following manifestly Bose-symmetric expression
(all momenta are incoming, i.e. $q+p_1+p_2 = 0$)
\be
\Gamma_{\alpha\mu\nu}^{(0)}(q,p_1,p_2)
= (q-p_1)_{\nu}g_{\alpha\mu} + (p_1-p_2)_{\alpha}g_{\mu\nu}
 + (p_2-q)_{\mu}g_{\alpha\nu} \, .
\ee
$\Gamma_{\alpha\mu\nu}^{(0)}(q,p_1,p_2)$
may be split into two parts \cite{TH}
\be
\Gamma_{\alpha\mu\nu}^{(0)}(q,p_1,p_2) 
= 
\Gamma_{F\alpha\mu\nu}^{(0)}(q,p_1,p_2) + 
\Gamma_{P\alpha\mu\nu}^{(0)}(q,p_1,p_2) \, ,
\label{decomp}
\ee
with 
\bea
\Gamma_{F\alpha\mu\nu}^{(0)}(q,p_1,p_2) &=& 
(p_1-p_2)_{\alpha} g_{\mu\nu} + 2q_{\nu}g_{\alpha\mu} 
- 2q_{\mu}g_{\alpha\nu} \, , \nonumber\\
\Gamma_{P\alpha\mu\nu}^{(0)}(q,p_1,p_2) &=&
 p_{2\nu} g_{\alpha\mu} - p_{1\mu}g_{\alpha\nu} \, .  
\label{GFGP}
\eea
The vertex $\Gamma_{F\alpha\mu\nu}^{(0)}(q,p_1,p_2)$ 
is Bose-symmetric only with respect to the
$\mu$ and $\nu$ legs.
The
first term in $\Gamma_{F\alpha\mu\nu}^{(0)}$ is a convective
vertex describing the coupling of a gluon to a scalar field, 
whereas the
other two terms originate from gluon spin or magnetic moment. 
$\Gamma_{F\alpha\mu\nu}^{(0)}(q,p_1,p_2)$
coincides with the BFMFG bare vertex involving one
background ($q$) and two quantum ($p_1$,$p_2$) gluons
\cite{BFMPT}. 
Evidently the above decomposition assigns a special r\^ole 
to the $q$-leg,
and allows $\Gamma_{F\alpha\mu\nu}^{(0)}$ 
to satisfy the Ward identity
\be 
q^{\alpha} \Gamma_{F\alpha\mu\nu}^{(0)}(q,p_1,p_2) = 
(p_2^2 - p_1^2)g_{\mu\nu} \, ,
\label{WI2B}
\ee
where the right hand-side (RHS) is the difference of two-inverse
propagators in the FG, and vanishes ``on shell'',
i.e. when $p_1^2=p_2^2=0$.
As has been explained in detail in \cite{PP2,PRW}, 
and as we will discuss extensively later on,
the
splitting of   
$\Gamma_{\alpha\mu\nu}^{(0)}(q,p_1,p_2)$ into 
$\Gamma_{F\alpha\mu\nu}^{(0)}(q,p_1,p_2)$
and $\Gamma_{P\alpha\mu\nu}^{(0)}(q,p_1,p_2)$
given in Eq. (\ref{decomp}) has a natural 
interpretation in the context of the
 {\it tree-level} process $q(P)\bar{q}(P') \to g(p_1) + g(p_2)$
(annihilation channel),
leading to an interesting connection with the
optical theorem. 

Consider next the $S$-matrix element for the  
quark ($q$)-antiquark ($\bar{q}$) elastic scattering process 
$q(P)\bar{q}(P')\to q(Q)\bar{q}(Q')$ in QCD;
we set $q= P'-P= Q'-Q$, and $s=q^2$ is  
the square of the momentum transfer. One could equally well
study the annihilation channel, in which case $s$ would be
the centre-of-mass energy. 
We will work in the RFG; this constitutes
no loss of generality, as long as one studies the 
entire gauge-independent process. It is a straightforward but tedious 
exercise to convince one-self that 
through pinching, i.e.
by simply exploiting the fundamental WI of Eq.\ (\ref{PTWI}), 
one can arrive at the 
set of diagrams of the RFG starting from 
the set of diagrams at any other value of 
$\xi$ (see for example \cite{pap2}), or even from 
the diagrams corresponding to non-covariant
gauge-fixing schemes \cite{masken}.

Let us define 
\bea
t_{\mu\nu}(q) &=& q^2 g_{\mu\nu} - q_{\mu}q_{\nu} \, ,\\
d(q) &=& \frac{-i}{q^2} \, , \\
S_0(p) &=&  \frac{i}{ \not\! p -m }\, ,\\
\rho_{\mu\nu}(p) &=&
\gamma_{\mu} S_0(p)\gamma_{\nu}\, .
\label{somedef}
\eea
$t_{\mu\nu}(q)$ is the {\it dimensionful} transverse tensor,
and $S_0(p)$ is
the tree-level quark propagator.
In what follows we will use the short-hand notation
$ [dk] = \mu^{2\epsilon}\frac{d^d k}{(2\pi)^d}$ \, 
with $d=4-2\epsilon$ the dimension of space-time
and $\mu$ the 't Hooft mass. Furthermore we define
the scalar quantities
\bea
J_1(q,k) &=&  g^2 C_A [k^2 (k+q)^{2}]^{-1} \, , \nonumber\\
J_2(q,k) &=&  g^2 C_A [k^4 (k+q)^{2}]^{-1} \, ,\nonumber\\
J_3(q,\ell,k)  &=&
\frac{i}{2} g^2
C_A [k^2 (k+\ell)^2 (k+\ell-q)^2] ^{-1} \, ,
\label{somedef1}
\eea
where $C_A$ is the Casimir eigenvalue of the adjoint
representation ( $C_A=N$ for $SU(N)$). The quantities
$J_2(q,k)$ and $J_3(q,\ell,k)$  will be used in later sections.

We then implement the vertex decomposition of  Eq.(\ref{decomp}) 
inside the non-Abelian
one-loop quark-gluon vertex 
graph of Fig.1a, 
to be denoted by 
$\Gamma_{\alpha}^{(1),nab}(Q,Q')$, 
where now 
 $p_{1\mu}=-k_{\mu}$, $p_{2\nu}=(k-q)_{\nu}$.
The $\Gamma_{P\alpha\mu\nu}^{(0)}(q,p_1,p_2)$ term
triggers the elementary Ward identity 
\be
\not\!  k = (\not\! k + \not\! Q -m) - (\not\! Q -m) \, ;
\label{PTWI}
\ee
thus, a self-energy like piece is generated  (Fig.1c), 
which is to be alloted
to the conventional self-energy. 
In particular,
\bea
\Gamma_{\alpha}^{(1),nab}(Q,Q') &=& 
\widehat{\Gamma}_{\alpha}^{(1),nab}(Q,Q') +
\frac{1}{2}\, V^{(1)}_{P\alpha\sigma}(q) \gamma^{\sigma}
\nonumber\\
&& 
+X_{1\alpha}^{(1)}(Q,Q')(\not\! Q' -m)
+   (\not\! Q -m) X_{2\alpha}^{(1)}(Q,Q')\, ,
\label{PTact}
\eea
where
\bea
\widehat{\Gamma}_{\alpha}^{(1),nab}(Q,Q')
&=& \int [dk] J_1(q,k) \Gamma_{F\alpha\mu\nu}^{(0)}(q,-k,k-q)
\rho^{\mu\nu}(Q') \, ,\nonumber\\
V^{(1)}_{P\alpha\sigma}(q) &=& 2 \int [dk] J_1(q,k)\, 
g_{\alpha\sigma}  \, , \nonumber\\
X_{1\alpha}^{(1)}(Q,Q') &=& \int [dk] J_1(q,k)\gamma_{\alpha} S_0(Q')
 \, , \nonumber\\
X_{2\alpha}^{(1)}(Q,Q') &=& \int [dk] J_1(q,k) S_0(Q')\gamma_{\alpha} 
\label{somedef2} \, .
\eea

The terms in the second line on the RHS of 
Eq.(\ref{PTact})
vanish for on-shell external fermions.
The 
(dimension-less) 
self-energy-like contribution
$ \frac{1}{2}\, V_{P\alpha\sigma}^{(1)}(q)$, 
together with another such contribution arising from the 
mirror vertex (not shown),
after trivial manipulations
gives rise
to the dimensionful quantity 
\be
\Pi_{P\alpha\beta}^{(1)}(q) =  
V_{P\alpha\sigma}^{(1)}(q)t_{\beta}^{\sigma}(q) \, .
\label{PP1}
\ee 
$\Pi_{P\alpha\beta}^{(1)}(q)$
will be added to the
conventional one-loop two-point function $\Pi_{\alpha\beta}^{(1)}(q)$,
to give rise to the 
the PT one-loop gluon self-energy 
$\widehat{\Pi}_{\alpha\beta}^{(1)}(q)$:
\be
\widehat{\Pi}_{\alpha\beta}^{(1)}(q) =  \Pi_{\alpha\beta}^{(1)}(q)
+ \Pi_{P\alpha\beta}^{(1)}(q) \, .
\label{PTprop1}
\ee
In particular,
suppressing color indices throughout, we have that 
$\Pi_{\alpha\beta}^{(1)}(q)$
is given by the graphs of Fig 2a and Fig 2b, namely
\be
\Pi_{\alpha\beta}^{(1)}(q) = \frac{1}{2} \int [dk] J_1(q,k)
L_{\alpha\beta}(q,k)\, ,
\label{conprop}
\ee
where
\be
L_{\alpha\beta}(q,k) \equiv  
\Gamma_{\alpha}^{(0)\sigma\rho}(q,k,-k-q)
\Gamma_{\beta\sigma\rho}^{(0)}(q,k,-k-q) - 2 k_{\alpha}(k+q)_{\beta}\, ,
\label{Tense1}
\ee
and thus the PT one-loop gluon self-energy 
$\widehat{\Pi}_{\alpha\beta}^{(1)}(q)$ (Fig.2)
assumes the closed form \cite{PT2}
\be
\widehat{\Pi}_{\alpha\beta}^{(1)}(q) = \frac{1}{2} \int [dk] J_1(q,k)
\widehat{L}_{\alpha\beta}(q,k) \, ,
\label{PTprop2}
\ee
with
\be
\widehat{L}_{\alpha\beta}(q,k) \equiv 
\Gamma_{F\alpha}^{(0)\sigma\rho}(q,k,-k-q)
\Gamma_{F\beta\sigma\rho}^{(0)}(q,k,-k-q) - 
2 (2k+q)_{\alpha}(2k+q)_{\beta} \, .
\label{Tense2}
\ee
Notice that in general both 
$\Pi_{\alpha\beta}^{(1)}(q)$ and $\Pi_{P\alpha\beta}^{(1)}(q)$ on the
RHS of Eq.(\ref{PTprop1})
depend explicitly on the GFP 
in such a way as to give an GFP-independent sum. 

In addition, gauge-invariance is encoded in the WI 
\be 
q_{\alpha} {\Pi}_{\alpha\beta}^{(1)}(q) = 
q_{\alpha} \widehat{\Pi}_{\alpha\beta}^{(1)}(q)=0 \, .
\ee

The following important
points have been discussed in detail in the literature
(i) 
$\widehat{\Pi}_{\alpha\beta}(q)$ is independent of the gauge-fixing parameter
in any gauge-fixing scheme. 
(ii) 
As happens in QED for the photon self-energy
\cite{ER},      
the  gluon  self-energy     $\widehat{\Pi}_{\alpha\beta}^{(1)}(q)$
captures the   leading    logarithms of  the  theory   at   that order
\cite{PTCPT,PT} ; therefore  the  coefficient in  front of the  single
logarithm  coming   from      $\widehat{\Pi}_{\alpha\beta}^{(1)}(q)$,
coincides  with the first    coefficient of the  QCD $\beta$  function
\cite{HDP}.  (iii) $\widehat{\Pi}_{\alpha\beta}^{(1)}(q)$ can be Dyson
resumed,  following the diagrammatic  algorithm presented in
\cite{PP1}. (iv) The combination $\alpha_{eff} (q,\mu) \sim g^{2}(\mu)
\widehat\Delta   (q/\mu)$,  where   $\widehat\Delta    (q/\mu)  =  [1-
\widehat\Pi^{(1)} (q/\mu)]^{-1}$ is a  renormalization-group-invariant
quantity,  and  constitutes  the  non-abelian analogue  of   the QED
concept  of   an effective  charge   \cite{PTCPT,PT,NJW2}.  Additional
properties for the gluon two-point function have been presented in the
literature  for the case   of  non-abelian gauge  theories with  Higgs
mechanism     \cite{pap1,papsm,PP1,PP2,PRW,PP3},        and        the
(non-renormalizable) Kunimasa-Goto-Slavnov-Cornwall \cite{KSC} massive
Yang-Mills model \cite{FPP}.

The PT quark-gluon vertex 
$\widehat{\Gamma}_{\alpha}^{(1)}(Q,Q')$ is the sum
of the non-abelian and abelian one-loop graphs, shown in Fig.3a and  
Fig.3b,
which we will denote by
$\widehat{\Gamma}_{\alpha}^{(1),nab}(Q,Q')$ and
$\widehat{\Gamma}_{\alpha}^{(1),ab}(Q,Q')$, respectively.
In addition to being GFP-independent, by virtue of Eq.\ (\ref{WI2B})
$\widehat{\Gamma}_{\alpha}^{(1)}(Q,Q')$ satisfies the following QED-like WI
\be
q_{\alpha}\widehat{\Gamma}_{\alpha}^{(1)}(Q,Q')=
\widehat{\Sigma}^{(1)}(Q)-\widehat{\Sigma}^{(1)}(Q'),
\label{WI1}
\ee
where $\widehat{\Sigma}^{(1)}$ is the PT one-loop
quark self-energy, which coincides with the conventional 
one computed in the RFG. \cite{pap2}.

Finally, the PT one-loop $n$-point functions coincide with those 
computed in the BFMFG (``tilded'' quantities)
\cite{BFMPT} , i.e.
\bea
\widehat{\Pi}_{\alpha\beta}^{(1)}(q) &=&
\widetilde{\Pi}_{\alpha\beta}^{(1)}(q,\xi_Q=1) 
\label{Pi1}\\
\widehat{\Gamma}_{\alpha}^{(1)}(Q,Q') &=&
\widetilde{\Gamma}_{\alpha}^{(1)}(Q,Q',\xi_Q=1)
\label{G1}\\
\widehat{\Sigma}^{(1)}(Q) &=& 
\widetilde{\Sigma}^{(1)}(Q,\xi_Q=1) = \Sigma^{(1)}(Q,\xi=1)
\label{S1}
\eea

In addition, exactly analogous properties have been established for 
the one-loop gluon
three-point function \cite{PT2} and  
four-point function  
(\cite{4g} and second paper in \cite{BFMPT}).

\setcounter{equation}{0}
\section{ The full two-loop construction}

Here we present the full two-loop construction. 
The basic observation is the following: 
if one carries out the
decomposition for the bare three-gluon vertex
described in Eq.\ (\ref{decomp}) to all {\it external} vertices
(to be defined in sub-section IIIB)
appearing in the Feynman diagrams contributing 
to the two-loop $S$-matrix element for the process
$q\bar{q} \to q\bar{q}$, then two-loop sub-amplitudes will
emerge, with precisely the same properties as the one-loop
PT effective Green's functions.

Throughout this section we have used the following formulas, valid
in dimensional regularization: 
\bea
&&\int  \frac{[dk]}{k^2} = 0 \, ,\nonumber\\
&&\int  [dk] \frac{k_{\alpha}k_{\beta}}{k^4}
=\bigg(\frac{1}{4-2\epsilon}\bigg)\int \frac{[dk]}{k^2} = 0 
\, ,\nonumber\\
&&\int   [dk] (2k+q)_{\alpha} J_1(q,k) = 0 \, ,\nonumber\\
&&\int  [dk] \frac{\ln^N (k^2)} {k^2} = 0 \, ~~~N=0,1,2,... 
\label{dimreg}
\eea
The last relation guarantees the absence of
tadpole and seagull contributions 
order by order in perturbation theory.
In the two-loop calculation presented in this paper only 
the case $N=1$ is relevant.
Notice however that nowhere have we used 
the slightly subtler dimensional regularization
result 
\be
\int \frac{[dk]}{k^4} = 0 ~,
\ee
which is often employed in the literature. 
We also use the group theoretical identities
\bea 
[{\bf \tau}^{a}, {\bf \tau}^{b}]  &=& i f^{abc} {\bf \tau}^{c} \, ,\nonumber\\
f^{aex}f^{bex} &=& C_A \delta^{ab} \, ,\nonumber\\
f^{axm}f^{bmn}f^{cnx}  &=& \frac{1}{2}\, C_A f^{abc}\, ,
\eea
where ${\bf \tau}^{a}$ are the gauge group generators in the fundamental
representation; in the case of QCD ${\bf \tau}^{a} = \lambda^{a}/2$,
where $\lambda^{a}$ are the Gell-Mann  matrices.

The identity
\bea
\Gamma_{\alpha\mu\nu}^{(0)}\Gamma^{(0)\beta\mu\nu} &=& 
\Gamma_{F\alpha\mu\nu}^{(0)}\Gamma_{F}^{(0)\beta\mu\nu}
+  
\Gamma_{P\alpha\mu\nu}^{(0)}\Gamma_{F}^{(0)\beta\mu\nu}
+ \Gamma_{F\alpha\mu\nu}^{(0)}\Gamma_{P}^{(0)\beta\mu\nu} 
+ \Gamma_{P\alpha\mu\nu}^{(0)}\Gamma_{P}^{(0)\beta\mu\nu}
\nonumber\\
&=& 
\Gamma_{F\alpha\mu\nu}^{(0)}\Gamma_{F}^{(0)\beta\mu\nu}
+  
\Gamma_{P\alpha\mu\nu}^{(0)}\Gamma^{(0)\beta\mu\nu}
+ \Gamma_{\alpha\mu\nu}^{(0)}\Gamma_{P}^{(0)\beta\mu\nu} 
-\Gamma_{P\alpha\mu\nu}^{(0)}\Gamma_{P}^{(0)\beta\mu\nu}
\label{iden}
\eea
may also be found useful at intermediate steps.

The two-loop integration symbol  
\be
(\mu^{2\epsilon})^2
\int\int \frac{d^d k}{(2\pi)^d} \frac{d^d \ell}{(2\pi)^d} 
\ee
will be suppressed throughout. We define the following
quantities
\bea
iI_{1} &=& g^4 C_A^2 
[\ell^2 (\ell-q)^2 k^2 (k+\ell)^2 (k+\ell-q)^2]^{-1}\, ,\nonumber\\
iI_{2}  &=& g^4 C_A^2 
[\ell^2 (\ell-q)^2 k^2 (k+q)^2]^{-1} \, ,\nonumber\\
iI_{3} &=& g^4 C_A^2 
[\ell^2 (\ell-q)^2 k^2 (k+\ell)^2]^{-1} \, ,\nonumber\\
iI_{4} &=& g^4 C_A^2 
[\ell^2 \ell^2 (\ell-q)^2 k^2 (k+\ell)^2]^{-1} \, ,\nonumber\\
iI_{5} &=& g^4 C_A^2 [\ell^2 k^2 (k+q)^2]^{-1}\, ,
\label{2LInt}
\eea
which will be used extensively in what follows.

\subsection{ The one-particle reducible graphs}

As has been explained in detail in \cite{PP1} the resummability of
the one-loop PT self-energy requires the conversion of 
one-particle reducible (1PR) strings of 
conventional self-energies $\Pi^{(1)}$
into strings containing PT self-energies  $\widehat\Pi^{(1)}$.
The process of converting conventional strings into PT strings gives rise
to left-overs, which are effectively one-particle irreducible (1PI), 
and must be alloted
to the genuine 1PI two-loop structures. 
Various self-consistency arguments supporting the validity
of this method
have been presented in the literature \cite{phsir}; 
as we will see  at the end of the third sub-section,
the extension of the PT to two loops provides the ultimate test
for the self-consistency of this procedure.
It is straightforward to establish that the set of 1PI graphs 
(Fig.4a - Fig.4d)  
may be converted into the equivalent set of 1PI
PT graphs (Fig.4e - Fig.4h), up to some missing pieces:

\bea
(4a) &=& (4e) - R^{(2)}_{P\,\alpha\beta}(q)
\label{1prse} \\
(4b)+(4c)+(4d) &=&(4f)+(4g)+(4h) - F^{(2)}_{P\,\alpha}(Q,Q')
\label{RandF}
\eea
with
\bea
iR^{(2)}_{P\,\alpha\beta}(q) &=& 
\Pi^{(1)}_{\alpha\rho}(q) V_{P\beta}^{(1)\rho}(q) + 
\frac{3}{4} \,\Pi_{P\,\alpha\rho}^{(1)}(q)V_{P\,\beta}^{(1)\rho}(q)\, ,
\label{RP} \\
F^{(2)}_{P\,\alpha}(Q,Q') &=&
\Pi_{P\alpha}^{(1)\beta}(q) d(q)
\widehat{\Gamma}_{\beta}^{(1)}(Q,Q') +  
 Y_{P\alpha}^{(2)}(Q,Q') \, ,
\label{1prmixed}
\eea
with
\be
 Y_{P\alpha}^{(2)}(Q,Q') \equiv
 X_{1\alpha}^{(1)}(Q,Q')\Sigma^{(1)}(Q)+
 X_{2\alpha}^{(1)}(Q,Q')\Sigma^{(1)}(Q') \, .
\label{FP}
\ee
The above terms originate from carrying out the vertex
decomposition of Eq.\ (\ref{decomp})
at all conventional 1PI diagrams.
For example,
the term  $Y_{P\alpha}^{(2)}(Q,Q')$ (Fig 5b)
originates after imposing  Eq.\ (\ref{decomp})
on the diagram of Fig.5a ; in addition, one obtains the 
PT counterpart of Fig.5a, namely Fig.5$a^F$, and the graph of
Fig.5c, which is part of Fig.4f.
The terms $R^{(2)}_{P\,\alpha\beta}(q)$ 
has been derived in detail in \cite{PP1}.
Notice that 
\be
 R^{(2)}_{P\,\alpha\beta}(q)  = 
I_2 \Bigg[ L_{\alpha\beta}(q,k)+ 3t_{\alpha\beta}(q)\Bigg] \, .
\label{RP2}
\ee

\subsection{ The two-loop vertex}

In this sub-section we will demonstrate the construction
of the two-loop PT quark-gluon vertex 
$\widehat{\Gamma}_{\alpha}^{(2)}(Q,Q')$, which turns out
to have the exact same properties as its one-loop counterpart 
$\widehat{\Gamma}_{\alpha}^{(1)}(Q,Q')$. At the same time
we will determine the two-loop propagator-like
contributions 
$V_{P\alpha\sigma}^{(2)} \gamma_{\sigma}$,  which will be subsequently
converted into 
$\Pi_{P\alpha\sigma}^{(2)}$, i.e. the
two-loop version of 
$\Pi_{P\alpha\sigma}^{(1)}$
of Eq.(\ref{PP1}). In addition, out of this procedure
the terms  $Y_{P\alpha}^{(2)}(Q,Q')$
of Eq.(\ref{FP}) will emerge.

The construction proceeds as follows:
The Feynman graphs contributing to $\Gamma_{\alpha}^{(2)}(Q,Q')$
can be classified into two sets.
(a) those containing an ``external'' three-gluon vertex 
i.e. a three-gluon vertex where the momentum $q$ is incoming,
as shown Fig.6 and Fig.7,
(b) those which do not have an ``external'' three-gluon vertex.
This latter set contains either graphs with no three gluon vertices
(abelian-like), or graphs with
three-gluon vertices whose all three
legs are irrigated by virtual momenta, i.e. $q$ never enters  
alone into any of the legs; such would be for example the
abelian graph of Fig.3b, if one was to insert a 
one-loop self-energy correction to the internal gluon line  
\cite{comment}). Of course, all  three-gluon vertices appearing
in the computation of the one-loop $S$-matrix are external, and
so are those appearing in the 1PR part of the two-loop $S$-matrix
(see previous section).
Then one
carries out the decomposition of Eq. (\ref{decomp}) 
to the external three-gluon vertex   
of all graphs belonging to set (a), leaving {\it all} their other
vertices unchanged, and identifies the propagator-like pieces
generated at the end of this procedure.

The calculation is straightforward, but lengthy; it is more economical
to identify the sub-structure of the one-loop three-gluon vertex 
$\Gamma_{\alpha\mu\nu}^{(1)}(q,p_1,p_2)$ (Fig K)
 nested inside the
two-loop graph, Fig(N$a1$), 
and use the 
results presented in
section VI, Eq. (\ref{decomp2}).
To that end we must set 
 $p_1 \to -\ell$ and  
$p_2 \to \ell - q $, and $J_3(q,-p_1,k)\to J_3(q,\ell,k)$, 
and rewrite the $\Gamma_{P\,\alpha\mu\nu}^{(1)}$ 
of Eq.(\ref{decomp2})
as follows
\bea
\Gamma_{P\,\alpha\mu\nu}^{(1)}(q,-\ell,\ell-q) &=&
\,\, \Bigg[
J_3 \,[\Gamma_{\nu\rho\alpha}^{(0)}
(\ell-q,k,-k-\ell+q)+k_{\nu}g_{\alpha\rho}]
\ell^{\rho} - \frac{i}{2}\, V_{P\alpha\nu}^{(1)}(q)\Bigg]
\ell_{\mu} \nonumber\\
&+&
\Bigg[
J_3 \,
[\Gamma_{\mu\alpha\rho}^{(0)}(-\ell,k+\ell,-k)+k_{\mu}g_{\alpha\rho}]
(\ell -q)^{\rho} - \frac{i}{2}\, V_{P\alpha\mu}^{(1)}(q)\Bigg]
(\ell-q)_{\nu}\nonumber\\
&-& \Bigg[
J_3 \, 
[\Gamma_{\nu\mu\alpha}^{(0)}(\ell-q,k,-k-\ell+q)+k_{\nu}g_{\alpha\mu}]
+ \frac{1}{2}\, d(q) q_{\alpha}V_{P\mu\nu}^{(1)}(q)\Bigg]\ell^2
\nonumber\\
&-& \Bigg[
J_3 \, 
 [\Gamma_{\mu\alpha\nu}^{(0)}(-\ell,k+\ell,-k)+k_{\mu}g_{\alpha\nu}]
- \frac{1}{2}\, d(q) q_{\alpha}V_{P\mu\nu}^{(1)}(q)\Bigg](\ell-q)^2
\nonumber\\
&&{}
\eea

Thus we have (we omit the external spinors, and use $Q'=Q+q$)
\bea
(6a_1) &=& (6a_1^F)
+ \frac{1}{2}\Pi_{P\alpha}^{(1)\beta}(q) d(q)
\Gamma^{(1),nab}_{\beta}(Q,Q')
\nonumber\\  
&& 
+ \frac{1}{2} \Bigg[2 I_{2}g_{\alpha\sigma} 
-I_1 [\Gamma_{\rho\sigma\alpha}^{(0)}(-k,-\ell,k+\ell)
+k_{\sigma}g_{\alpha\rho} ] (\ell-q)^{\rho}\Bigg]
\gamma^{\sigma} \nonumber\\
&&
+ D_1 + D_2 + D_3 + D_4 
\label{Vert2a} \\
(6b_1) &=& (6b_1^F) 
+ \frac{1}{4} I_{3}g_{\alpha\sigma} \gamma^{\sigma}
- D_1 + D_5 + D_6 
\label{Vert2b} \\
(6b_1)' &=& (6b_1^F)' 
+ \frac{1}{4} I_{3}g_{\alpha\sigma} \gamma^{\sigma}
- D_2 + D_7 + D_8  
\label{Vert2c} \\
(6c_1) &=& (6c_1^F) + 
\frac{1}{2} I_4 L_{\alpha\sigma}(\ell,k) \gamma^{\sigma}  
\label{Vert2d} \\
(7r_i) &=& (7 r_i^F) 
+ \frac{1}{2}\Pi_{P\alpha}^{(1)\beta}(q) d(q)
\Gamma^{(1),ab}_{\beta}(Q,Q')
+\frac{1}{2}Y_{P\alpha}^{(2)}(Q,Q')
\nonumber\\
&&- (D_3 + D_4 + D_5 + D_6 + D_7 + D_8)
\label{Vert2e}
\eea
where
\bea
D_1 &=& 
-I_{3}
[\Gamma_{\alpha\mu\nu}^{(0)}(\ell,k,-\ell-k)+k_{\nu}g_{\alpha\mu}]
\{ \gamma^{\nu}S(Q+\ell+k)\gamma^{\mu}\} \, ,
\nonumber\\
D_2  &=& I_3  
[\Gamma_{\alpha\nu\mu}^{(0)}(\ell,-\ell- k,k)+k_{\mu}g_{\alpha\nu}]
\{ \gamma^{\nu}S(Q'-\ell-k)\gamma^{\mu} \} \, ,
\nonumber\\
D_3  &=& -\frac{1}{2}I_{5} q_{\alpha}
\{ \gamma_{\mu}S(Q+\ell)\gamma^{\mu}\} \, ,
\nonumber\\
D_4 &=& \frac{1}{2}I_{5} q_{\alpha}
\{ \gamma_{\mu}S(Q'+\ell)\gamma^{\mu} \} \, ,
\nonumber\\
D_5 &=& - I_{5} \{ \gamma_{\alpha}S(Q-k)
\gamma_{\mu}S(Q-k-\ell)\gamma^{\mu}\} \, ,
\nonumber\\
D_6 &=& I_{5} \{ \gamma_{\alpha}S(Q-k)
\gamma_{\mu}S(Q+\ell)\gamma^{\mu}\} \, ,
\nonumber\\
D_7 &=& - I_{5} \{ \gamma_{\mu}S(Q-k-\ell)
\gamma^{\mu}S(Q-k)\gamma_{\alpha}\} \, ,
\nonumber\\
D_8 &=& I_{5} \{ \gamma_{\mu}S(Q+\ell)
\gamma^{\mu}S(Q-k)\gamma_{\alpha}\} \, ,
\label{Ds} 
\eea
Before we proceed the following comments are 
warranted:

(i) In the above formulas appropriate shiftings and relabellings of the
integration momenta have been carried out, in order for
the answer to be expressed in terms of the five basic 
denominators $I_i$ defined in Eq.(\ref{2LInt}).

(ii) The topologies of  
$D_1$, $D_2$, $D_3$, $D_4$, $D_5$, and $D_6$ are shown in
Fig.6$a_4$, Fig.6$a_5$,  Fig.6$a_6$, Fig.6$a_7$, 
Fig.6$b_4$, and Fig.6$b_5$, respectively.
$(6b_1)'$ corresponds to the figure obtained from Fig.6$b_1$ by drawing the 
internal three-gluon vertex on the other leg of the
external three-gluon vertex, i.e. the leg which hooks onto 
the external spinor carrying momentum $Q'$ (not shown);
$D_7$ (shown in Fig.7$h$)
and $D_8$ (not shown)
are the analogues of $D_5$ and $D_6$ for 
the $(6b_1)'$ topology.
Furthermore, the second and third term on the RHS of
Eq.(\ref{Vert2a}) are depicted in Fig.6$a_2$ and
Fig.6$a_3$ respectively,
the second term on the RHS of Eq.(\ref{Vert2b}) 
is shown
in
Fig.6$b_2$, the second term on the RHS
of Eq.(\ref{Vert2d}) is shown
in
Fig.6$c_2$, and the second term 
on the RHS
of Eq.(\ref{Vert2e}) in Fig.7$c_a$. 

(iii) The graph in  Fig.7$r_2$ is accompanied by 
the graph with the abelian vertex correction on the 
other side (not shown). The graph containing the 
bare four-gluon vertex is not shown either.

(iv) Notice that the propagator-like part in  Fig.7$a$
is connected to the rest of the graph by a factor
$g_{\alpha\beta}$; since this term will become part
of the 1PR graph of Fig. 4c it must be supplemented
a longitudinal component
$q_{\alpha}q_{\beta}$. 
This contribution is proportional to  
the terms shown in Fig.7$a_6$ and Fig.7$a_7$
(remember that the latter are proportional 
to $q_{\alpha}$), but notice that the color factors are different;
the terms  shown in Fig.6$a_6$ and Fig.6$a_7$
are proportional to $C_A^{2}$, whereas the term
missing in Fig.(S$a$) will have both $C_A^{2}$ and 
$C_A C_f$, where  $C_f$ is the Casimir for the 
$\tau_{\alpha}$ representation of the (external) quarks.
It is easy to show that 
these
terms vanish 
 when the external fermions
are on shell, which is our case.  In order to prove that
one does not have to assume that $q_{\alpha}$ hits a
conserved current on the other side of the graph; 
instead one notices
that the terms in question are proportional to  
$[\Sigma^{(1)}(\not\! Q,m)-\Sigma^{(1)}(\not\! Q' ,m)]$,
which vanishes when $\not\! Q=\not\! Q'=m$.

(v) Notice the appearance
of propagator-like terms 
(Fig.6$a_3$, Fig.6$b_2$, and Fig.6$c_2$)

Adding Eq.(\ref{Vert2a}) -- Eq.(\ref{Vert2e}),
by parts, and using   Eq.(\ref{FP}), 
we find 
\be
\Gamma_{\alpha}^{(2)}(Q,Q')
=
\frac{1}{2} F^{(2)}_{P\,\alpha}(Q,Q')
+\frac{1}{2}V_{P\alpha\sigma}^{(2)} \gamma^{\sigma}
+\widehat{\Gamma}_{\alpha}^{(2)}(Q,Q')
\label{VPT2}
\ee
with
\be
V_{P\alpha\sigma}^{(2)}(q) = I_4 L_{\alpha\sigma}(\ell,k)
+ (2 I_{2} + I_{3}) g_{\alpha\sigma}
-I_1 \Bigg[k_{\sigma}g_{\alpha\rho}+  
\Gamma_{\rho\sigma\alpha}^{(0)}(-k,-\ell,k+\ell)\Bigg] (\ell-q)^{\rho}
\label{BRes1}
\ee

The interpretation of the three terms appearing on the RHS of
Eq.(\ref{VPT2}) is as follows:

(i)  The term 
$\frac{1}{2} F^{(2)}_{P\,\alpha}(Q,Q')$ 
is half of the 
vertex-like part necessary to cancel the
corresponding term appearing in Eq.(\ref{RandF}), during
the conversion of 
conventional 1PR graphs into their PT counterparts. The other
half will come from the mirror vertex (not shown).

(ii)  $\frac{1}{2}V_{P\alpha\sigma}^{(2)} \gamma^{\sigma} $
is  
the total propagator-like term originating from the 
two-loop quark-gluon vertex; together with the
equal contribution  
from
the mirror set of two-loop vertex graphs (not shown) will
give rise to the self-energy term
\be
\Pi_{P\alpha\beta}^{(2)}(q) = V_{P\alpha\sigma}^{(2)}(q)t^{\sigma}_{\beta}(q) , 
\label{2LPinch}
\ee
which 
will be part of the 
effective two-loop PT gluon self-energy,
to be constructed in the next sub-section

(iii)
$\widehat{\Gamma}_{\alpha}^{(2)}(Q,Q')$ is the PT two-loop
quark-gluon vertex; it {\it coincides} with the 
corresponding two-loop quark-gluon vertex computed
in the BFMFG, i.e.
\be
\widehat{\Gamma}_{\alpha}^{(2)}(Q,Q') = 
\widetilde{\Gamma}_{\alpha}^{(2)}(Q,Q',\xi_Q =1)
\ee
as happens in the one-loop case , Eq.(\ref{G1}). 
Either by virtue of the above equality and 
the formal properties of the BFM, 
or by means of
an explicit diagrammatic calculation 
where one acts with $q_{\alpha}$ on individual diagrams,
one 
can establish that 
$\widehat{\Gamma}_{\alpha}^{(2)}(Q,Q')$ satisfies
the following QED-like WI
\be
q_{\alpha}\widehat{\Gamma}_{\alpha}^{(2)}(Q,Q')=
\widehat{\Sigma}^{(2)}(Q)-\widehat{\Sigma}^{(2)}(Q') \, ,
\label{WI2}
\ee
which is the exact two-loop 
analogue of Eq.(\ref{WI1}). 
$\widehat{\Sigma}^{(2)}(Q)$ is the two-loop PT fermion self-energy
which satisfies 
\be
\widehat{\Sigma}^{(2)}(Q) = 
\widetilde{\Sigma}^{(2)}(Q,\xi_Q =1) \, 
 =\Sigma^{(2)}(Q,\xi=1) 
\ee
Again, this is the precise generalization of the one-loop result
of Eq.(\ref{S1}). 
At this point this result comes as no surprise, since 
all three gluon vertices
appearing in the Feynman graphs contributing to
$\Sigma^{(2)}(Q,\xi)$ are internal; therefore, at $\xi=1$ there
will be no pinching \cite{NJW3}.

We emphasize that the above result is non-trivial; indeed,
even if one accepts that the appearance of the
first and third term in Eq.(\ref{VPT2}), for example, 
could be forced, there is no a-priory
reason why the remainder should turn out to be purely
self-energy-like.
Notice also that in deriving the above results 
no integrations (or sub-integrations) over virtual momenta
have been carried out.

\subsection{The two-loop self-energy}
The construction of 
${\widehat\Pi}^{(2)}_{\alpha\beta}(q)$ proceeds as follows:
To the conventional two-loop gluon self-energy 
$\Pi^{(2)}_{\alpha\beta}(q)$ we add two additional terms;
(i) the propagator-like term  
$\Pi_{P\alpha\beta}^{(2)}(q)$ derived in the previous sub-section,
 Eq.(\ref{2LPinch}), and (ii)
the propagator-like part $- R^{(2)}_{P\,\alpha\beta}(q)$ 
given in Eq.(\ref{RP}),
stemming from the conversion of the conventional string
into a PT string; this term must be removed from the  
1PR reducible set and be alloted to
${\widehat\Pi}^{(2)}_{\alpha\beta}(q)$, as described in \cite{PP1}. 
Thus, ${\widehat\Pi}^{(2)}_{\alpha\beta}(q)$ reads
\be
{\widehat\Pi}^{(2)}_{\alpha\beta}(q) =
\Pi^{(2)}_{\alpha\beta}(q) + \Pi_{P\alpha\beta}^{(2)}(q) -
R^{(2)}_{P\,\alpha\beta}(q) \, .
\label{FinalA}
\ee
 Up to minor notational modifications,
this last equation is in fact identical to Eq.(3.5) in the second paper
of  \cite{PP1}, except that now we know
the exact closed expression for the term $\Pi^{(2)}_{P\alpha\beta}(q)$.
 
It is a lengthy but 
relatively straightforward exercise to establish that 
in fact
\be
{\widehat\Pi}^{(2)}_{\alpha\beta}(q) = 
{\widetilde\Pi}^{(2)}_{\alpha\beta}(q,\xi_Q =1)
\label{FinalB}
\ee
To see that in detail, we simply start out with 
the diagrams contributing to $\Pi^{(2)}_{\alpha\beta}(q)$
and convert them into the corresponding diagrams contributing
to ${\widetilde\Pi}^{(2)}_{\alpha\beta}(q,\xi_Q =1)$; in doing
so we only need to carry out algebraic manipulations
in the numerators of individual Feynman diagrams,
and the judicious use of the identity of 
Eq.(\ref{iden}). The individual diagrams yield:

\bea
(8a) &=& (9a)+(9m)+(9n)+(9o)+(9p)+(9q)+(9r)+(9s) \nonumber\\
     && + I_1 \Bigg[ \ell_{\beta}\ell^{\rho}k^{\sigma}
          \Gamma_{\alpha\sigma\rho}^{(0)}(q,k+\ell-q,-k-\ell) 
+(\ell\cdot q-\ell^2) (k+\ell-q)_{\alpha}(\frac{3}{2}\ell-q)_{\beta}\Bigg] 
\nonumber\\
&& + I_1 [k_{\sigma}g_{\alpha\rho}+  
\Gamma_{\rho\sigma\alpha}^{(0)}(-k,-\ell,k+\ell)] (l-q)^{\rho}
t^{\sigma}_{\beta}(q) -(2 I_{2} + I_{3}) t_{\alpha\beta}(q)\nonumber\\
&& + I_2 \Bigg [ \frac{3}{8} t_{\alpha\beta}(q)
+\frac{9}{16} q_{\alpha}q_{\beta}\Bigg]
-\frac{9}{2} I_{3}\Bigg [\ell_{\alpha}(\ell-q)_{\beta}+
(\ell\cdot q-\ell^2) g_{\alpha\beta}\Bigg] 
\nonumber\\
&& + I_2 [L_{\alpha\beta}(q,k) + 2 k_{\alpha}(k+q)_{\beta} ]
-I_3 [L_{\alpha\beta}(\ell,k) + 2 k_{\alpha}(k+\ell)_{\beta}]
\nonumber\\
&& - I_3 \Bigg[3 \ell_{\alpha}\ell_{\beta} - 
\frac{21}{8} q_{\alpha}\ell_{\beta} + 
q_{\alpha}q_{\beta}\Bigg]
\nonumber\\
(8b)+(8c) &=& (9b)+(9c)
+ \frac{9}{2} I_{3}
\Bigg[\ell_{\alpha}(\ell-q)_{\beta}+(\ell\cdot q-\ell^2)g_{\alpha\beta}\Bigg]
\nonumber\\
(8d) &=& (9d)+ \frac{21}{8}I_2 t_{\alpha\beta}(q)
\nonumber\\
(8e)+(8f) &=& (9e)+(9f) 
- I_{1}\ell_{\beta}\ell^{\rho}k^{\sigma}
\Gamma_{\alpha\sigma\rho}^{(0)}(q,k+\ell-q,-k-\ell) \nonumber\\
&& -I_{2}\Bigg [2 k_{\alpha}(k+q)_{\beta}+\frac{1}{2}q_{\alpha}q_{\beta}\Bigg]
+ I_{3}\Bigg [2 k_{\alpha}(k+\ell)_{\beta}+\frac{1}{4}q_{\alpha}\ell_{\beta}
\Bigg]
\nonumber\\
(8g)+(8h) &=& (9g)+(9h) 
+I_3 L_{\alpha\beta}(\ell,k)
-I_4 L_{\alpha\sigma}(\ell,k)t^{\beta}_{\sigma}(q)\nonumber\\
(8i) &=& (9i)
- I_1~(\ell\cdot q-\ell^2)(k+\ell-q)_{\alpha}
(\frac{3}{2}\ell-q)_{\beta}
 - \frac{1}{16}I_{2}q_{\alpha}q_{\beta} + 
\frac{1}{8}I_{3} q_{\alpha}\ell_{\beta}
\nonumber\\
(8j)+(8k) &=& (9j)+(9k)+ I_3 [3 \ell_{\alpha}\ell_{\beta}
- 3 \ell_{\alpha}q_{\beta} + q_{\alpha}q_{\beta}]
\nonumber\\
(8\ell) &=& (9\ell)
\eea

Adding the above equations by parts we find
\bea
\Pi^{(2)}_{\alpha\beta}(q) &=& 
\widetilde\Pi^{(2)}_{\alpha\beta}(q,\xi_{Q}=1) 
+ I_2 
\Bigg[ L_{\alpha\beta}(q,k)+ 3t_{\alpha\beta}(q)\Bigg]
-I_4 L_{\alpha\sigma}(\ell,k)t^{\beta}_{\sigma}(q)
\nonumber\\
&& + I_1 \Bigg[k_{\sigma}g_{\alpha\rho}+  
\Gamma_{\rho\sigma\alpha}^{(0)}(-k,-\ell,k+\ell)\Bigg] (l-q)^{\rho}
t^{\beta}_{\sigma}(q) - (2 I_{2} + I_{3}) t_{\alpha\beta}(q)
\nonumber\\
&&{}
\label{FinalC}
\eea
Using Eq.(\ref{BRes1}) and Eq.(\ref{RP2}),
Eq.(\ref{FinalC})
 becomes
\be
\Pi^{(2)}_{\alpha\beta}(q)= 
{\widetilde\Pi}^{(2)}_{\alpha\beta}(q,\xi_Q =1)
-\Pi^{(2)}_{P\,\alpha\beta}(q)
+R^{(2)}_{P\,\alpha\beta}(q) . 
\label{Final}
\ee
From Eq.(\ref{Final}) and Eq.(\ref{FinalA}) we arrive immediately
to Eq.(\ref{FinalB}). 
Again, no integrations over virtual momenta need be carried out,
except for identifying 
vanishing contributions by means of the formulas listed
in Eq.(\ref{dimreg}).

Since we have used dimensional regularization throughout
and no integrations have been performed 
the results of this section do not depend on the value $d$ of the
space-time; 
in particular they are valid for $d=3$, 
which is of additional field-theoretical interest
\cite{3d}. Clearly, when $d\to 4$ the renormalization program
needs be carried out; this will be the subject of the next section. 

\setcounter{equation}{0}
\section{Renormalization}

In this section we will carry out the renormalization for the
two-loop PT Green's functions constructed in the previous section. 
There is of course no doubt that if one supplies
the correct counterterms within the conventional 
formulation, the total $S$-matrix will continue being renormalized,
even after the PT rearrangement of the (unrenormalized)
two-loop Feynman graphs.
The point of this section is to show a stronger 
version of renormalizability, i.e. that the new Green's function
constructed through the PT rearrangement are {\it individually}
renormalizable.
The general methodology is as follows: We start out with the
counterterms which are necessary to renormalize
individually the conventional Green's functions contributing to the
two-loop $S$-matrix. Then we will show that by simply rearranging 
them, following the PT rules, we will arrive at 
renormalized two-loop PT Green's functions.

We will use the following notation:
$Z_1$ is the vertex renormalization constant for the
conventional quark-gluon vertex $\Gamma_{\alpha}$, 
$Z_2$ is the wave-function renormalization for the (external) quarks,
$Z_A$ the gluon wave-function renormalization
corresponding to the conventional gluon self-energy $\Pi$,
$\widehat{Z}_A$ the gluon wave-function renormalization
corresponding to the PT gluon self-energy $\widehat\Pi$,
$Z_3$ is the vertex renormalization constant for the
conventional one-loop three-gluon vertex $\Gamma_{\alpha\mu\nu}^{(1)}$, and
$Z_{3F}$ the vertex renormalization constant for 
the $\Gamma_{F\,\alpha\mu\nu}^{(1)}$,  
${\bar Z}_2$ is the ghost wave-function renormalization,
and ${\bar Z}_1$ the ghost-gluon vertex renormalization 
constant.
Equivalently, one can carry out the renormalization program
using appropriately defined counter-terms.
The corresponding counterterms, which, when added to the
above $n$-loop quantities 
render them UV finite, are, respectively
$K_1^{(n)}$, $K_2^{(n)}$, $K_A^{(n)}$,
$\widehat K_A^{(n)}$,
$K_3^{(n)}$, $K_{3F}^{(n)}$
$\bar{K}_2^{(n)}$, and $\bar{K}_1^{(n)}$. 
In addition, mass counterterms $\delta m$  must be supplied if 
the quarks are considered to be massive. In a moment we will
also introduce the counterterm
$K_P^{(n)}$, which renders $V_P^{(n)}$ ultra-violet finite.
Notice that because of the QED-like WI it satisfies
$\widehat{\Gamma}_{\mu}$ becomes
ultraviolet finite when the counterterm $K_2$ is added to it.
The $Z$s and the $K$s are related as follows :
\bea
Z_i &=& 1+ \sum_{j=1} K_i^{(j)} ~~~~i=1,2,3 
\nonumber\\
Z_A &=& 1+ \sum_{j=1} K_A^{(j)}
\nonumber\\
\widehat{Z}_A &=& 1+ \sum_{j=1} \widehat{K}_A^{(j)}
\nonumber\\
\bar{Z}_i  &=& 1+ \sum_{j=1} \bar{K}_i^{(j)}~~~~i=1,2
\nonumber\\
Z_g &=& 1+ \sum_{j=1} K_g^{(j)}
\nonumber\\ 
\delta m  &=& \sum_{j=1} \delta m^{(j)}
\label{R4}
\eea

We first begin with the 1PR part of the $S$-matrix.
It is more convenient to work with dimension-less quantities;
to that end we define the dimension-less gluon self-energy
$\Pi^{(1)}$ simply through 
$\Pi^{(1)}_{\alpha\beta}=\Pi^{(1)}t_{\alpha\beta}$. 
In order to renormalize the extra pieces 
which must be alloted to
the conventional 1PR graphs in order to convert them into their PT form,
i.e. the $iR_P+F_P$ terms in 
Eq.\ (\ref{RP}) -- Eq.\ (\ref{FP}), we must have
(we use the supescript ``R'' to denote renormalizaed quantities)
\bea
iR_{P}^{(2)R}+F_{P}^{(2)R} &=& 
\Pi^{(1)R} V_{P}^{(1)R} + 
\frac{3}{4}V_{P}^{(1)R}V_{P}^{(1)R}
+ V_{P}^{(1)R}\widehat{\Gamma}^{(1)R} + Y_{P}^{(1)R}
\nonumber\\
&=&
\frac{3}{4}(K_{P}^{(1)}+V_{P}^{(1)})^2 + 
 (\Pi^{(1)}-K_A^{(1)})(V_P^{(1)}+K_P^{(1)}) 
\nonumber\\
&&+ (V_P^{(1)}+K_P^{(1)})(\widehat{\Gamma}^{(1)}+ K_2^{(1)}) 
+ (X_1^{(1)}+X_2^{(1)})\delta m^{(1)}
\nonumber\\
&=& iR_P^{(2)}+F_{P}^{(2)} 
+ U_1^{(2)} + U_2^{(2)} 
\label{R1}
\eea
with
\bea
U_1^{(2)} &=& V_P^{(1)} \Bigg [K_2^{(1)}-K_A^{(1)}\Bigg ]
+ K_{P}^{(1)}\Bigg[\Pi^{(1)}+ \Gamma^{(1)}
+V_P^{(1)}\Bigg] 
+(X_1^{(1)} +X_2^{(1)})\delta m^{(1)}
\nonumber\\ 
U_2^{(2)} &=& K_P^{(1)} 
\Bigg[\frac{3}{4} K_P^{(1)}+K_2^{(1)}-K_A^{(1)}\Bigg] \, .
\label{R2}
\eea
Thus, the terms $U_1^{(2)}$ and $U_2^{(2)}$ need be supplied. 
Notice that the terms contained in $U_2^{(2)}$ are
momentum independent (quadratic in the counterterms),
whereas those contained in  $U_1^{(2)}$
or momentum-dependent (linear in the counterterms).
The latter will cancel against parts of the one-loop
counterterms appearing inside the two-loop expressions for
the conventional self-energy (Fig.10)
and vertex (Fig.12),
cancelling their sub-divergences.
As for the former, they will become part of the final
renormalization counterterm of the two-loop PT self-energy, 
i.e. the counterterm necessary for cancelling the 
remaining divergence, after the sub-divergences have been 
taken care of. Another way of saying this is that, since the extra terms
$R_P$ and $F_P$ 
will be alloted to the PT gluon-self-energy and vertex, so should
the counter-terms necessary to renormalize them.

We next show how 
the terms in $U_2^{(2)}$ are to be accounted for, and, at the 
same time, derive some useful relations among the various counterterms.
The QED-like WI of equations Eq.\ (\ref{WI1}) and Eq.\ (\ref{WI2}) 
relating $\widehat{\Gamma}_{\alpha}^{(1)}$ and 
$\widehat{\Sigma}^{(1)}$,
and $\widehat{\Gamma}_{\alpha}^{(2)}$  and
$\widehat{\Sigma}^{(2)}$, respectively,
imposes 
the following QED-like relation between
the renormalization constants 
$\widehat{Z}_1$ and $\widehat{Z}_2$,  up to order $g^{4}$: 
\be
\widehat{Z}_1 = \widehat{Z}_2 \, .
\label{Z1Z2}
\ee
In addition,
from the WI of  Eq.\ (\ref{WI2B2}), we have to order $g^{2}$ (at least)
\be
{Z}_{3F} = {Z}_A \, .
\label{Rex}
\ee
Notice also that  Eq.\ (\ref{WI2B2}) dictates
that the ultraviolet-divergent part of  
$\Gamma_{F\,\alpha\mu\nu}^{(1)}$ is proportional to 
$\Gamma_{F\,\alpha\mu\nu}^{(0)}$ rather than
$\Gamma_{F\,\alpha\mu\nu}^{(0)}$; had it been the other way around 
there would be no longitudinal ultraviolet-divergent 
pieces on the RHS of Eq.\ (\ref{WI2B2}). As we will see, this ``mismatch''
will generate the pieces which in the BFM language give rise to the
gauge-fixing renormalization of the vertices (Fig.10a and Fig.10b)

Furthermore, 
the renormalization constants 
before and after the PT rearrangements 
are related to the 
gauge coupling renormalization as follows:  
\bea
Z^2_g &=& Z_1^2 Z_{2}^{-2} Z_A^{-1} 
\nonumber\\
&=&   \widehat{Z}_1^{2} \widehat{Z}_2^{-2}\widehat{Z}_A^{-1}
\nonumber\\
&=& \widehat{Z}_A^{-1} \, ,
\label{R3}
\eea
where Eq.\ (\ref{Z1Z2}) has been used.
After substituting the expressions given in Eq.\ (\ref{R4})
into Eq.\ (\ref{R3}) and equating powers
of the coupling constant $g$ we arrive at 
the following relations for the corresponding counterterms
\bea
\widehat{K}_A^{(1)} &=& K_A^{(1)} - 2 ( K_1^{(1)}-K_2^{(1)})\, ,
\label{R4b}\\
\widehat{K}_A^{(2)} &=& K_A^{(2)} - 2 ( K_1^{(2)}-K_2^{(2)})
+ K_1^{(1)}[3 K_1^{(1)} -4 K_2^{(1)} -2 K_A^{(1)}]
+ K_2^{(1)}[2K_A^{(1)}+K_2^{(1)}] \, .
\label{R5}
\eea

Substituting the relations 
\be
K_1^{(j)} =  \frac{1}{2} K_P^{(j)} + K_2^{(j)} , ~~~~~~~j=1,2
\label{R6}
\ee
into the above equations we obtain
\bea
\widehat{K}_A^{(1)} &=& K_A^{(1)} - K_P^{(1)}\, ,
\label{R7}\\
\widehat{K}_A^{(2)} &=& K_A^{(2)} - K_P^{(2)} + U_2^{(2)}\, ,
\label{R8}\\
\widehat{K}_A^{(1)} &=& -2 K_g^{(1)} \, .
\label{R9}
\eea

Notice that if $Z_1 = Z_2$, or equivalently, $K_P^{(j)}=0$, then
$Z_A = \widehat{Z}_A$, which is simply the QED case.

It is now clear what the r\^ole 
of the $U_2^{(2)}$ terms is; they must
be added to the conventional two-loop self-energy counterterm
 $K_A^{(2)}$ and the part of the vertex counterterm corresponding
to $V_P^{(2)}$; these two counterterms are already present, whereas  
the term $U_2^{(2)}$ must be borrowed from some other part of the
$S$-matrix, in order for the 
initial equation Eq.\ (\ref{R3}) to be enforced. 

We next turn to the $U_1^{(2)}$ terms. We will show that 
they will cancel precisely against terms originating
from the rearrangement of the graphs shown in Fig.10 and Fig.12,
in order to convert them into the graphs shown in Fig.11 and Fig.13.
The former set contains the counterterms necessary for cancelling
the one-loop sub-divergences inside the conventional two-loop
gluon self-energy and quark-gluon vertex, the latter the
counterterms needed for the two-loop PT 
gluon self-energy and quark-gluon vertex.

We begin with two-loop gluon self-energy shown in Fig.10.
From the Slavnov-Taylor identity \cite{ST}
we have that \cite{WEC,DRTJ}
\be
\frac{Z_3}{Z_A} = \frac{\bar{Z}_1}{\bar{Z}_2}\, ,
\label{STI}
\ee
from which we obtain \cite{expl}
\be
K_3^{(1)}-K_A^{(1)} = \bar{K}_{1}^{(1)}
-\bar{K}_{2}^{(1)} = \frac{1}{2} K_{P}^{(1)}\, .
\label{STI2}
\ee
This relation is important for what follows. 

Then we have for the graphs of Fig.10:
\bea
(10a)+(10b) &=&  K_3^{(1)} \int [dk] J_1(q,k) 
       \Gamma_{\alpha\mu\nu}^{(0)}
       \Gamma_{\beta\mu\nu}^{(0)}
\nonumber\\
(10c) &=& (10c_1) + (10c_2) \nonumber\\
(10d)+(10e) &=& 
\bar{K}_{1}^{(1)} \int [dk] J_1(q,k) k_{\alpha}(k+q)_{\beta}
\nonumber\\
(10f)+(10g) &=& - \bar{K}_{2}^{(1)} 
\int [dk] J_1(q,k) k_{\alpha}(k+q)_{\beta}
\eea
with
\bea
(10c_1) &=&   
-K_A^{(1)} \int [dk] J_1(q,k) \Gamma_{\alpha\mu\nu}^{(0)}
       \Gamma_{\beta\mu\nu}^{(0)}\nonumber\\
(10c_2) &=&
       K_A^{(1)} \int [dk] J_2(q,k) k_{\mu}k_{\sigma}
       \Gamma_{\alpha\mu\nu}^{(0)}
       \Gamma_{\beta\sigma\nu}^{(0)} \, ;
\label{b1b2}
\eea
$J_2$ has been defined in Eq.\ (\ref{somedef1}).
Thus, using Eq. (\ref{STI2}) and  Eq. (\ref{conprop}) we obtain
\bea
[(10a)+(10b) + (10c_1)] +[(10d)+(10f)] + [(10e)+(10g)] =
K_{P}^{(1)} \Pi^{(1)}_{\alpha\beta}(q)
\eea
Next, 
we write $(10c_2)$ as follows
\be
(10c_2) = (11a)+(11b)+(11c)
 - 2t_{\alpha\sigma}(q) \int [dk] J_2(q,k) 
k_{\beta}k^{\sigma}
\label{Ac2}
\ee
where 
\bea
(11a)+(11b) &=& K_A^{(1)}\int [dk] J_1(q,k)
\Gamma_{P\alpha\mu\nu}^{(0)}\Gamma_{F\beta}^{(0)\mu\nu}
\nonumber\\
(11c) &=& K_A^{(1)}\int [dk] J_2(q,k) k^{\mu}k_{\sigma}
\Gamma_{F\alpha\mu\nu}^{(0)}\Gamma_{F\beta}^{(0)\sigma\nu}
\label{R11}
\eea

We next convert the renormalization constants for the
conventional two-loop quark-gluon vertex
 $\Gamma_{\alpha}^{(2)}$ (Fig.12)
into those necessary for
the PT 2-loop quark-gluon vertex 
$\widehat{\Gamma}_{\alpha}^{(2)}$ 
finite (Fig.13). We have:
\bea
(12a) &=& 
K_3^{(1)} \Gamma^{(1)} = (K_{3F}^{(1)}+\frac{1}{2} K_{P}^{(1)})\Gamma^{(1)} =
(13a) + \frac{1}{2} K_{P}^{(1)} \Gamma^{(1)}
\nonumber\\
(12c)+(12d) &=& K_{1}^{(1)} V^{(1)}_{P} + 
(13c)+(13d)
\nonumber\\
(12e) &=& - \frac{1}{2} K_{2}^{(1)} V^{(1)}_{P} + 
\frac{1}{2}(X_1^{(1)} +X_2^{(1)})\delta m^{(1)} +(13e)
\nonumber\\
(12b) &=& - K_A^{(1)} 
\int [dk] J_2(q,k) \Gamma_{\alpha\sigma\nu}^{(0)} t_{\mu}^{\sigma}(k)
\rho^{\mu\nu}(Q+k)
\nonumber\\
&=& - K_A^{(1)} 
\int [dk] J_2(q,k) (\Gamma_{F\alpha\sigma\nu}^{(0)} 
+\Gamma_{P\alpha\sigma\nu}^{(0)})
t_{\mu}^{\sigma}(k)\rho^{\mu\nu}(Q+k)
\nonumber\\
 &=& (13b) - K_A^{(1)}\int [dk] J_2(q,k) t_{\alpha\mu}(k) \gamma^{\mu}
\nonumber\\
&=& (13b) - \frac{1}{2} K_A^{(1)} V^{(1)}_{P} +
K_A^{(1)} 
t_{\alpha\sigma}(q) \int [dk] J_2(q,k) k_{\beta}k^{\sigma} \gamma^{\beta}
\nonumber\\
&& {}
\label{R12}
\eea

Notice that
the last term on the RHS of  Eq. (\ref{R12}), together with an
equal term from the mirror vertex graphs (not shown)
will cancel against the last term in Eq.(\ref{Ac2}).
Using that 
$K_{1}^{(1)}-\frac{1}{2} K_{2}^{(1)} = 
 \frac{1}{2} (K_{P}^{(1)} + K_{2}^{(1)})$
we finally arrive at
\be
K_{(Fig.10)} + 2 K_{(Fig.12)}  = K_{(Fig.11)} +2 K_{(Fig.13)} + U_1^{(1)} 
\label{R13}
\ee
where
the factors of 2 multiplying  $K_{(Fig.12)}$ and $K_{(Fig.13)}$
account for the mirror vertex contributions.
Evidently, the PT rearrangement gives rise to a term $U_1^{(1)}$,
as announced.

The  counterterms resulting from  the  above rearrangement are exactly
those    required    to     cancel      all   sub-divergences   inside
$\widehat{\Gamma}_{\alpha}^{(2)}(Q,Q')$; the latter coincide of course
with                the           sub-divergences               inside
${\widetilde\Gamma}_{\alpha}^{(2)}(Q,Q',\xi_Q  =1)$.   To  demonstrate
that the counterterms shown  in Fig.13 are  in fact identical to those
obtained  when  carrying out   the   BFM renormalization  program as
explained  in  \cite{Abb}, i.e.   renormalizing  only  the  background
gluons,  the external  quarks,    the   coupling constant $g$, and the
GFP ($\xi_Q$), one may proceed  as follows: (i) start
with  graph (13b), and  separate the $t_{\sigma\mu}(k)$  into the part
proportional  to   $g_{\sigma\mu}$   and   the part  proportional   to
$k_{\sigma}k_{\mu}$;  the  second  part  is  simply  the  gauge-fixing
renormalization  to the self-energy, as  explained in \cite{Abb}. (ii)
Half  of the piece proportional   to $g_{\sigma\mu}$ must be added  to
(13a); the latter  is  proportional  to $\Gamma^{(0)}$.   Using   that
$K_{3F}^{(1)}=K_A^{(1)}$  (from   Eq.\   (\ref{Rex})),    the    total
contribution           is               proportional                to
$K_A^{(1)}(\Gamma^{(0)}-\Gamma^{(0)}_{F})=K_A^{(1)}\Gamma^{(0)}_{P}$,
which is the contribution from the gauge-fixing renormalization of the
elementary   BFM  three-gluon  vertex.    (iii)   The  remaining  half
proportional to $g_{\sigma\mu}$ from   step (i) must be  split equally
between $(13c)+\frac{1}{2}(13e)$ and $(13d)+\frac{1}{2}(13e)$. Each of
these two combinations will  then give a contribution proportional  to
$K_1^{(1)}-\frac{1}{2}K_2^{(1)}-\frac{1}{2}K_A^{(1)}=     -\frac{1}{2}
\widehat{K}_A^{(1)}+\frac{1}{2}K_2^{(1)}                    =K_g^{(1)}
+\frac{1}{2}K_2^{(1)}$,   where  Eq.  (\ref{R6}), Eq. (\ref{R7}),  and
Eq. (\ref{R9})  have been used  in the  last steps.  The wave-function
renormalization  for   the   external    fermions  will  then   cancel
$\frac{1}{2}K_2^{(1)}$, and  $K_g^{(1)}$ will  be re-absorbed into the
gauge-coupling renormalization.

Notice that  the correspondence between  the PT and the  BFMFG Green's
functions established  in   the    previous section persists     after
renormalization; the resulting expressions  are the BFMFG renormalized
quantities,  as  derived in \cite{Abb}.   An  immediate consequence of
this  \cite{Abb} is that  the coefficient multiplying the logarithm of
the PT two-loop self-energy is  equal the the two-loop coefficient  of
the QCD $\beta$ function \cite{WEC,DRTJ}, i.e. $(34/3)\, C_A^{2}$.  It
is also  interesting to see how  the  PT rearrangement leads  into the
interpretation of  counterterms generated from   the QCD Lagrangian in
the renormalized Feynman-gauge  \cite{DRTJ}, as GFP
renormalizations of the BFM Lagrangian \cite{Abb} .

\setcounter{equation}{0}
\section{The absorptive construction: General formalism}

In the next two sections we will show in detail how one may construct 
the two-loop PT effective Green's functions using unitarity and
analyticity arguments. This derivation generalizes
the method first presented in \cite{PP2} and \cite{PRW} for the one-loop case,
and constitutes a non-trivial self-consistency check for the
entire approach. In this section we will set up the formalism,
and discuss in detail the one-loop case, which will serve as
the general paradigm for the two-loop generalization, to be presented
in the following section.
Apart from some minor modifications,
in the next two sections we will adopt 
the notation used in \cite{PP2}.

The optical theorem for the case of forward scattering assumes the form 
\be
\Im m 
\langle a | T | a \rangle = \frac{1}{2}
\sum_{i} (2\pi)^{4} \delta^{(4)}(p_{a}-p_{i})
{\langle i | T | a \rangle}^{*}
\langle i | T | a \rangle \, ,
\label{ab1}
\ee
where the sum $\sum_{i}$ should be understood to be over the entire phase space
of all allowed on-shell intermediate states $i$.
After expanding the $T$ matrix in powers of $g$, i.e.
$ T = \sum_{n=2} T^{[n]} $, we have that
\be
\Im m 
\langle a | T^{[n]} | a \rangle = \frac{1}{2}
\sum_{i} (2\pi)^{4} \delta^{(4)}(p_{a}-p_{i})
\sum_{k} 
{\langle i | T^{[k]} | a \rangle}^{*}
\langle i | T^{[n-k]} | a \rangle \, .
\label{ab2}
\ee
In particular, if in the initial states we have a $q$-$\bar{q}$ pair, 
i.e. $ |a\rangle = |q\bar{q}\rangle$
we have for the first non-trivial orders, $n=4$ and $n=6$ 
\be
\Im m 
\langle q\bar{q}| T^{[4]} | q\bar{q} \rangle =
\frac{1}{2} \Bigg(\frac{1}{2!}\Bigg) \int (dPS)_{2g}
{\langle 2g | T^{[2]} |q\bar{q} \rangle}^{*}
\langle 2g | T^{[2]} |q\bar{q} \rangle \, ,
\label{ab3}
\ee
and
\bea
\Im m 
\langle q\bar{q}| T^{[6]} | q\bar{q} \rangle &=&
\frac{1}{2} \Bigg(\frac{1}{3!}\Bigg) \int (dPS)_{3g}
{\langle 3g | T^{[2]} |q\bar{q} \rangle}^{*}
\langle 3g | T^{[2]} |q\bar{q} \rangle
\nonumber\\
&+& \frac{1}{2} \Bigg(\frac{1}{2!}\Bigg)\int (dPS)_{2g}
2 \Re e \Bigg( {\langle 2g | T^{[4]} |q\bar{q} \rangle}^{*}
\langle 2g | T^{[2]} |q\bar{q} \rangle \Bigg) \, ,
\label{ab4}
\eea
respectively  \cite{intquarks}.
$(dPS)_{2g}$ and $(dPS)_{3g}$ stand for the two- and three-body
phase space for mass-less gluons, respectively.
Next we introduce the short-hand notation 
\bea
{\cal A}^{[n]} &\equiv &
\Im m \langle q\bar{q}| T^{[n]} | q\bar{q}\rangle \, ,
\nonumber\\
{\cal T}_{m}^{[k]} &\equiv & 
\langle m g | T^{[k]} |q\bar{q} \rangle 
~~~~~~~~~~m,k = 2,3, ... 
\label{ab5}
\eea
Then, Eq. (\ref{ab3}) and Eq. (\ref{ab4})
become respectively
\be
{\cal A}^{[4]} = \frac{1}{2} \Bigg(\frac{1}{2!}\Bigg)
\int (dPS)_{2g} {{\cal T}_{2}^{[2]}}^{*}{\cal T}_{2}^{[2]} \, ,
\label{ab6}
\ee
\bea
{\cal A}^{[6]} &=& \frac{1}{2} \Bigg(\frac{1}{3!}\Bigg) 
\int (dPS)_{3g} {{\cal T}_{3}^{[3]}}^{*}{\cal T}_{3}^{[3]}
+ \frac{1}{2} \Bigg(\frac{1}{2!}\Bigg)\int (dPS)_{2g}
2 \Re e \Bigg ({{\cal T}_{2}^{[4]}}^{*}{\cal T}_{2}^{[2]}\Bigg) 
\nonumber\\
&=&
{\cal A}^{[6]}_{3} + {\cal A}^{[6]}_{2} \, .
\label{ab7}
\eea
The quantities defined above have the explicit form 
\bea
{\cal A}^{[4]} &=&\frac{1}{2}\bigg(\frac{1}{2!}\bigg) 
 [{\cal T}_{2s}^{[2]}+{\cal T}_{2t}^{[2]}]_{\mu\nu}^{ab}
P^{\mu\mu'}(p_1)P^{\nu\nu'}(p_2)
[{\cal T}_{2s}^{[2]}+{\cal T}_{2t}^{[2]}]_{\mu'\nu'}^{ab*} \, ,
\label{ab8}\\
{\cal A}^{[6]}_{3} &=&
\frac{1}{2} \Bigg(\frac{1}{3!}\Bigg)
[{\cal T}_{3s}^{[3]}+{\cal T}_{3t}^{[3]}]_{\mu\nu\rho}^{abc}
P^{\mu\mu'}(p_1)P^{\nu\nu'}(p_2)P^{\rho\rho'}(p_3)
[{\cal T}_{3s}^{[3]}+{\cal T}_{3t}^{[3]}]_{\mu'\nu'\rho'}^{abc*} \, ,
\label{ab9}\\
{\cal A}^{[6]}_{2} &=& \bigg(\frac{1}{2!}\bigg) \Re e \Bigg (
[{\cal T}_{2s}^{[4]}+{\cal T}_{2t}^{[4]}]_{\mu\nu}^{ab}
P^{\mu\mu'}(p_1)P^{\nu\nu'}(p_2)
[{\cal T}_{2s}^{[4]}+{\cal T}_{2t}^{[4]}]_{\mu'\nu'}^{ab*}\Bigg ) \, ,
\label{ab10}
\eea
where we have suppressed the phase space integrations.
$P_{\mu\nu}$ is the
polarization tensor for mass-less gluons,
\be
P_{\mu\nu}(p,n,\eta )\ =\ -g_{\mu\nu}+ \frac{n_{\mu}p_{\nu}
+n_{\nu}p_{\mu} }{n p} -
\eta \frac{p_{\mu}p_{\nu}}{{(np)}^2}\, ,
\label{PhotPol}
\ee
with $n_{\mu}$ is 
an arbitrary four-vector, and $\eta$ a gauge parameter.

We will now 
study the process
$q(k_1)\bar{q}(k_2)\to g(p_1)g(p_2)$ at tree-level, using the
equations derived above. This study will further elucidate the  r\^ole 
of the Eq.\ (\ref{decomp}) in enforcing
perturbative unitarity at the level of individual Green's functions,
and will set up the stage for the two-loop generalization.

The BRS symmetry \cite{BRS} of the original Lagrangian leads to the
following identities \cite{ChengLi}:
\bea
p_{1}^{\mu} {\cal T}_{2\mu\nu}^{ab} &=& (S_{2}^{\{12\}})^{ab}  p_{2\nu}
\, ,\nonumber\\
p_{2}^{\nu} {\cal T}_{2\mu\nu}^{ab} &=& (S_{2}^{\{21\}})^{ab}  p_{1\mu}
\, ,\nonumber\\
p_{2}^{\nu}p_{1}^{\mu} {\cal T}_{2\mu\nu}^{ab} &=& 0 \, .
\label{ab11}
\eea
If we split the amplitude into an $s$-channel 
and $t$-channel contribution ($s=q^2= (k_1+k_2)^2=(p_1+p_2)^2$, 
and $t=(k_1-p_1)^2=(k_2-p_2)^2$),
the first of the identities in Eq.\ (\ref{ab11})
becomes
\be
p_{1}^{\mu} \bigg({\cal T}_{2s}+{\cal T}_{2t}\bigg)_{\mu\nu}^{ab}
 = \bigg(S_{2s}^{\{12\}} + S_{2t}^{\{12\}}\bigg)^{ab} p_{2\nu}\, .
\label{ab12}
\ee
The remaining two identities
are exactly analogous and will be suppressed
throughout.

The PT rearrangement of the amplitude amounts to a special choice
for ${\cal T}_{2s}$ and ${\cal T}_{2t}$, which
will be denoted by ${\cal T}_{2s}^{F}$ and ${\cal T}_{2t}^{F}$,
respectively.
After defining these ``Feynman'' amplitudes, 
Eq.\ (\ref{ab12}) reads
\be
p_{1}^{\mu}\bigg({\cal T}_{2s}^{F}+{\cal T}_{2t}^{F}\bigg)_{\mu\nu}^{ab}
 = \bigg( S_{2s}^{F\{12\}} + S_{2t}^{F\{12\}}\bigg)^{ab}p_{2\nu} \, ,
\label{ab13}
\ee
and 
to order $k$,
\be
 p_{1}^{\mu}\bigg({\cal T}_{2s}^{[k]F}+{\cal T}_{2t}^{[k]F}\bigg)_{\mu\nu}^{ab}
 = \bigg( S_{2s}^{[k]F\{12\}} + S_{2t}^{[k]F\{12\}} \bigg)^{ab}p_{2\nu} \, .
\label{ab14}
\ee
We will next study the case $k=2$ (tree-level, Fig.14).

For ``on-shell'' gluons, i.e. $p^2=0$, 
$P_{\mu\nu}(p,n,\eta)$ 
of  Eq.\ (\ref{PhotPol})   
satisfies the 
transversality condition $p^{\mu} P_{\mu\nu}= 0$.
Thus 
one may immediately 
eliminate the  
$\Gamma_{P\alpha\mu\nu}^{(0)}(q,p_1,p_2)$ 
part of $\Gamma_{\alpha\mu\nu}^{(0)}(q,p_1,p_2)$,
which vanishes when contracted with the term
$P_{\mu\mu'}(p_1) P_{\nu\nu'}(p_2)$, and effectively
replace $\Gamma_{\alpha\mu\nu}^{(0)}(q,p_1,p_2)$ 
by $\Gamma_{F\alpha\mu\nu}^{(0)}(q,p_1,p_2)$, as in
Fig.14a. One then proceeds by recognizing 
that the longitudinal parts of the  
$P_{\mu\mu'}(p_1)$ and $P_{\nu\nu'}(p_2)$ trigger
a fundamental cancellation \cite{PP2,PRW}
involving the $s$- and $t$- channel graphs 
(Fig.14a and Fig.14d),
which is a consequence of the underlying BRS symmetry. 
In particular, the action of $p_{1}^{\mu}$ 
or $p_{2}^{\nu}$ on  
$\Gamma_{F\alpha\mu\nu}^{(0)}$ gives 
 \bea
p_{1}^{\mu} \Gamma_{F\alpha\mu\nu}^{(0)}(q,p_1,p_2) &=&
(p_2-p_1)_{\alpha}p_{2\nu}
+ (p_1^2-p_2^2)g_{\alpha\nu} + t_{\alpha\nu}(q)
\nonumber\\
p_{2}^{\nu} \Gamma_{F\alpha\mu\nu}^{(0)}(q,p_1,p_2) &=&
(p_2-p_1)_{\alpha}p_{1\mu} 
+ (p_1^2-p_2^2)g_{\alpha\mu} - t_{\alpha\mu}(q)  
\label{WI1L}
\eea
The first term on the RHS of either equation
cancels against an analogous contribution
from the $t$-channel graph, 
whereas the second terms vanish 
for on-shell gluons. Finally, the  
terms proportional to $p_{2\nu}$ and $p_{1\mu}$ (Fig.14b) 
are such that 
all dependence on the unphysical four-vector $n_{\mu}$ 
and the parameter $\eta$
vanishes, as it should.
In addition, 
a residual ($s$-dependent)
contribution emerges from these latter terms, 
which must be added
to the parts stemming from the $g_{\mu\mu'}g_{\nu\nu'}$
part of the calculation.
In particular \cite{PP2}, we have that
\bea
({\cal T}_{2s}^{F})_{\mu\nu}^{ab} &=& 
g\{\gamma_{\alpha}^{m}\}f^{mab}d(q) 
\Gamma_{F\mu\nu}^{(0)\alpha}(q,p_1,p_2)  \, ,\nonumber\\ 
{\cal T}_{2t}^{[2]F} &=& {\cal T}_{2t}^{[2]} \, ,\nonumber\\
S_{2t}^{[2]F\{12\}}&=& S_{2t}^{[2]\{12\}} = 0 \, ,\nonumber\\
p_{1}^{\mu}({\cal T}_{2s}^{[2]F})_{\mu\nu}^{ab} &=& 
(S_{2s}^{[2]F\{12\}})^{ab} p_{2\nu} + 
(\Lambda^{[2]F}_2)_{\nu}^{ab} \, ,\nonumber\\
p_{1}^{\mu} ({\cal T}_{2t}^{[2]F})_{\mu\nu}^{ab}
 &=& - (\Lambda^{[2]F}_2)_{\nu}^{ab} \, ,
\label{ab15}
\eea
with 
\bea
(\Lambda^{[2]F}_2)_{\nu}^{ab} &=& 
g\{\gamma_{\alpha}^{e}\}f^{eab} d(q)t^{\alpha}_{\nu}(q)
\nonumber\\
\bigg(S_{2s}^{[2]F\{12\}}\bigg)^{ab} &=& 
 g\{\gamma_{\alpha}^{e}\}f^{eab} d(q) (p_2-p_1)^{\alpha} \, ,
\label{ab16}
\eea
and 
$\{\gamma_{\alpha}^{a}\} = ig \bar{v}(k_2){\bf \tau}^{a} \gamma_{\alpha}u(k_1)$.

From the above results follows
\be
p_{1}^{\mu}\bigg({\cal T}_{2s}^{[2]F}+{\cal T}_{2s}^{[2]}\bigg)_{\mu\nu}^{ab}
 =     \bigg(S_{2s}^{[2]F\{12\}}\bigg)^{ab} p_{2\nu} \, ,
\label{ab17}
\ee
and thus
\bea
{\cal A}^{[4]} &=& \frac{1}{2} \bigg(\frac{1}{2!}\bigg)
\Bigg[ \bigg({\cal T}_{2s}^{[2]F}+{\cal T}_{2t}^{[2]}\bigg)
\bigg( {\cal T}_{2s}^{[2]F}+{\cal T}_{2t}^{[2]}\bigg)^{*}- 
2 \bigg(S_{2s}^{[2]F\{12\}}\bigg)\bigg(S_{2s}^{[2]F\{12\}}\bigg)^{*}
\Bigg] \nonumber\\
&=&  {\cal A}^{[4]}_{S} +{\cal A}^{[4]}_{V} +{\cal A}^{[4]}_{B} \, ,
\label{ab18}
\eea
with
\bea
{\cal A}^{[4]}_{S} &=&
\frac{1}{2}
\bigg(\frac{1}{2!}\bigg)
\Bigg[ 
{\cal T}_{2s}^{[2]F}{{\cal T}_{2s}^{[2]F}}^{*}
-2\bigg(S_{2s}^{[2]F\{12\}}\bigg)\bigg(S_{2s}^{[2]F\{12\}}\bigg)^{*}
 \Bigg] \, , \nonumber\\ 
{\cal A}^{[4]}_{V} &=&
\frac{1}{2}\bigg(\frac{1}{2!}\bigg) \Bigg[
{\cal T}_{2s}^{[2]F}{{\cal T}_{2t}^{[2]}}^{*}
+{\cal T}_{2t}^{[2]}{{\cal T}_{2s}^{[2]F}}^{*}\Bigg] \, , \nonumber\\
{\cal A}^{[4]}_{B} &=&
\frac{1}{2} \bigg(\frac{1}{2!}\bigg) 
{\cal T}_{2t}^{[2]}{{\cal T}_{2t}^{[2]}}^{*} \, .
\label{ab19}
\eea
In the last step we have defined self-energy (S), vertex (V) and box (B) -like
amplitudes, according to their dependence on the Mandelstam variables
$s$ and $t$ as in the case of a scalar filed theory, or QED; 
${\cal A}^{[4]}_{S}$ depends only on $s$, 
${\cal A}^{[4]}_{V}$ on $s$ and $t$, and ${\cal A}^{[4]}_{B}$ only on $t$.

The next step is to identify these sub-amplitudes as the imaginary 
parts of the effective one-loop self-energy, vertex, and box, 
under construction.
For example, for the effective self-energy 
$\widehat{\Pi}^{\alpha\beta}(q)$
we will proceed as follows:
first write ${\cal A}^{[4]}_{S}$ in the form
\be
{\cal A}^{[4]}_{S} = \{\gamma^{\alpha}\} d(q)
{\cal A}^{[4]}_{S\alpha\beta}(q) d(q)\{\gamma^{\beta}\} \, ;
\label{ab20}
\ee
then identify
\be
\Im m \widehat{\Pi}_{\alpha\beta}^{(1)}(q) =  
{\cal A}^{[4]}_{S\alpha\beta}(q) \, .
\ee
From the last equation follows
\be
\Im m \widehat{\Pi}_{\alpha\beta}(q)=
\frac{1}{2} g^{2} C_A \Bigg[ 
\Gamma_{F\alpha\mu\nu}^{(0)}(q,p_1,p_2)
\Gamma_{F\beta}^{(0)\mu\nu}(q,p_1,p_2) - 2 (p_2-p_1)_{\alpha} 
(p_2-p_1)_{\beta}\Bigg] \, .
\ee

This last equation leads to a well-defined definition of 
$\widehat{\Pi}^{(1)}$, without having to resort
to an intermediate one-loop diagrammatic interpretation: 
after the two-body phase space integrations 
has been carried out using standard results \cite{PP2},
the real parts may be reconstructed
by means of a once-subtracted dispersion relation.
Thus, in the absence of a full (dispersive) one-loop construction,
the $\widehat{\Pi}^{(1)}$ so generated
does not necessarily have to correspond to the imaginary parts of a precisely 
identifiable set of one-loop Feynman diagrams.  
Of course, given that the full one-loop construction 
has been carried out, we know that 
this is actually the case. Thus, when the one-loop (or $n$-loop) construction 
for $\widehat{\Pi}$ exists, 
 Eq.(\ref{ab20})
reduces into a non-trivial self-consistency
check. In particular, one must verify that 
\be 
{\cal A}^{[4]}_{S\alpha\beta}(q) = 
{\cal C}_{2}\Bigg\{\widehat{\Pi}_{\alpha\beta}^{(1)}(q)\Bigg\}
=
{\cal C}_{2}\Bigg\{\widetilde{\Pi}_{\alpha\beta}^{(1)}
(q,\xi_Q =1)\Bigg\} \, ,
\ee
where ${\cal C}_{n}\{...\}$ is the operator which carries out the
$n$-particle Cutkosky cuts 
(for $n=2$ we have two-gluon and two-ghost cuts)
to the quantity appearing inside the
curly bracket (Fig.15). 
This is indeed the case \cite{PP2}, 
as can be directly verified 
from   Eq.(\ref{PTprop2}). 
In fact, the residual contributions
originating from the  terms proportional to  $p_{1}^{\mu}$
and $p_{2}^{\nu}$ 
mentioned above 
correspond precisely 
to the Cutkosky cuts of the one-loop ghost diagrams (Fig.15b).

To fully appreciate the subtlety of the above construction the
following comments are now in order:

(i) Of course, the BRS-driven cancellation
for the process $q\bar{q} \to gg$ 
 takes place regardless 
of the PT rearrangement of the amplitude in general,
and the PT
decomposition of $\Gamma_{\alpha\mu\nu}^{(0)}$ in particular. 
Indeed, if we had 
not eliminated $\Gamma_{P\alpha\mu\nu}^{(0)}(q,p_1,p_2)$ 
but had kept instead
the full vertex $\Gamma_{\alpha\mu\nu}^{(0)}(q,p_1,p_2)$
(as is usually done), the 
WI analogous to Eq. (\ref{WI1L}) would be simply
\bea
p_{1}^{\mu} \Gamma_{\alpha\mu\nu}^{(0)}(q,p_1,p_2) &=&
t_{\alpha\nu}(q) - t_{\alpha\nu}(p_2) \, , \nonumber\\
p_{2}^{\nu} \Gamma_{\alpha\mu\nu}^{(0)}(q,p_1,p_2) &=&
t_{\alpha\mu}(p_1) - t_{\alpha\mu}(q) \, .
\label{ConWI1L}
\eea
The parts which participate in the BRS
cancellation, namely
the parts proportional to 
$t_{\alpha\nu}(q)$ and $t_{\alpha\mu}(q)$ are thus
unaffected. 
What changes after the PT rearrangement
is the resulting 
absorptive part of the effective $n$-point functions under construction. 
So, if one was to define the  
absorptive part of an effective self-energy
keeping the full $\Gamma_{\alpha\mu\nu}^{(0)}(q,p_1,p_2)$,
but still exploiting the BRS cancellation in order to elliminate
the longitudial terms, one
would arrive at the imaginary part of the conventional 
self-energy in the RFG, $\Im m \Pi_{\alpha\beta}(q)$; the latter,
for one thing, does not capture the running of the QCD coupling.

(ii) One could define the absorptive  part of an effective self-energy
before carrying out  the BRS cancellation.  In that case  one would be
led to the absorptive parts of the gluon self-energy in the light-cone
gauges; in particular, the final answer would depend explicitly
on the unphysical quantities $n$ and $\eta$.
A dispersion relation would give rise to a pathological quantity, since 
the  gluon self-energy computed  within the axial
gauges  is   not multiplicatively   renormalizable,   
due to its dependence on  $n_{\alpha}$ \cite{planar}.
Furthermore, spurious  infrared divergences   appear  in the   Feynman
parameter  integrations, which are  artifacts and cancel out only when
full physical quantities are computed \cite{axun}.

Thus, one has to first carry out the PT rearrangement 
at the level of the $S$-matrix element,
then enforce the BRS cancellation at the level
of the cross-section, and, only after these two steps, one 
should define self-energy/vertex/box absorptive parts, 
as one would for a scalar field theory.

Having set up the formalism and discussed
the general methodology, we next proceed with the two-loop
absorptive construction.

\section{The two-loop absorptive construction}
In this section we will show how to extend the methodology established
in the previous section to the two-loop case. This construction involves
two parts: the first part is the study of the one-loop 
amplitude for the process $q(k_1)\bar{q}(k_2)\to g(p_1)g(p_2)$;
the second is the study of the tree-level process
$q(k_1)\bar{q}(k_2)\to g(p_1)g(p_2)g(p_3)$. As we will see in
detail the PT rearrangement (at the level of the $S$-matrix), will
give rise (at the level of the cross-section) to the 
correct Cutkosky cuts.

\subsection{ The one-loop version of $\Gamma_{F}^{(0)}$ and
$\Gamma_{P}^{(0)}$ }

In this sub-section we will first study the 
conventional one-loop three-gluon
vertex, and will see how one can arrive at the one-loop
generalization of the tree-level vertices
$\Gamma_{F\alpha\mu\nu}^{(0)}$ 
and $\Gamma_{P\alpha\mu\nu}^{(0)}$, defined in
Eq.(\ref{GFGP}), 
to be denoted by
$\Gamma_{F\alpha\mu\nu}^{(1)}$ and $\Gamma_{P\alpha\mu\nu}^{(1)}$,
respectively. 
Then we will see in detail why casting
the one-loop
$S$-matrix for the process $q(k_1)\bar{q}(k_2)\to g(p_1)g(p_2)$
into its PT form is crucial for 
enforcing the optical theorem at the level of
 {\it individual} two-loop PT Green's functions. 

We start with the one-loop $S$-matrix element for  
$q(k_1)\bar{q}(k_2)\to g(p_1)g(p_2)$. 

Then Eq.\ (\ref{ab14}) yields ($k=4$) 
\be
p_{1}^{\mu}\bigg({\cal T}_{2s}^{[4]F}+{\cal T}_{2t}^{[4]F}\bigg)_{\mu\nu}^{ab}
 = \bigg( S_{2s}^{[4]F\{12\}} + S_{2t}^{[4]F\{12\}} \bigg)^{ab}p_{2\nu}
\, ,
\label{ab50}
\ee
where ${\cal T}_{2s}^{[4]F}$ and  ${\cal T}_{2t}^{[4]F}$ are the
``Feynman'' versions of 
${\cal T}_{2s}^{[4]}$ and  ${\cal T}_{2t}$; their exact form will 
be specified shortly. The last two equations in (\ref{ab15}) become
\bea
p_{1}^{\mu} ({\cal T}_{2s}^{[4]F})_{\mu\nu}^{ab} &=& 
(S_{2s}^{[4]F\{12\}})^{ab} p_{2\nu} + 
(\Lambda^{[4]F}_2)_{\nu}^{ab} \, ,\nonumber\\
p_{1}^{\mu}({\cal T}_{2t}^{[4]F})_{\mu\nu}^{ab}
 &=& (S_{2t}^{[4]F\{12\}})^{ab} p_{2\nu} 
- (\Lambda^{[4]F}_2)_{\nu}^{ab} \, .
\label{ab21}
\eea
Using Eq.(\ref{ab17}), we have that
\bea
{\cal A}^{[6]}_{2} &=& \bigg(\frac{1}{2!}\bigg)
\Re e \Bigg[ \bigg({\cal T}_{2s}^{[4]F}+{\cal T}_{2t}^{[4]F}\bigg)
\bigg({\cal T}_{2s}^{[2]F}+{\cal T}_{2t}^{[2]}\bigg)^{*}- 
2 \bigg(S^{[4]F\{12\}}_{2s} + S^{[4]F\{12\}}_{2t}\bigg)
 \bigg(S^{[2]F\{21\}}_{2s}\bigg)^{*} \Bigg] \nonumber\\
&=&  {\cal A}^{[6]}_{2S} +{\cal A}^{[6]}_{2V} +{\cal A}^{[6]}_{2B} \, ,
\eea
with
\bea
{\cal A}^{[6]}_{2S} &=&
\bigg(\frac{1}{2!}\bigg)\Re e \Bigg[ 
{\cal T}_{2s}^{[4]F}{{\cal T}_{2s}^{[2]F}}^{*}
-2 \bigg(S^{[4]F\{12\}}_{2s}\bigg)
\bigg({S^{[2]F\{21\}}_{2s}}\bigg)^{*} \Bigg] \, ,\nonumber\\ 
{\cal A}^{[6]}_{2V} &=&
\bigg(\frac{1}{2!}\bigg) \Re e \Bigg[
{\cal T}_{2s}^{[4]F}{{\cal T}_{2t}^{[2]}}^{*}
+{\cal T}_{2s}^{[2]F}{{\cal T}_{2t}^{[4]F}}^{*}
- 2 \bigg(S^{ [2]F\{12\}}_{2s}\bigg)
\bigg(S^{[4]F\{21\}}_{2t}\bigg)^{*}\Bigg] \, ,\nonumber\\
{\cal A}^{[6]}_{2B} &=&
\bigg(\frac{1}{2!}\bigg)\Re e \Bigg[
{\cal T}_{2t}^{[4]F}{{\cal T}_{2t}^{[2]}}^{*}\Bigg] \, .
\eea
From these last quantities one could define, 
for example, the quantity ${\cal A}^{[6]}_{2S\alpha\beta}(q)$
exactly as in Eq.\ (\ref{ab20}), i.e.
\be
{\cal A}^{[6]}_{2S} = \{\gamma^{\alpha}\}\, d(q)
{\cal A}^{[6]}_{2S\alpha\beta}(q)\, 
d(q)\{\gamma^{\beta}\}
\label{again}
\ee
The ${\cal A}^{[6]}_{2S\alpha\beta}(q)$ must then be such that
\be 
{\cal A}^{[6]}_{2S\alpha\beta}(q) = 
{\cal C}_{2}\Bigg\{\widehat{\Pi}_{\alpha\beta}^{(2)}(q)\Bigg\} = 
{\cal C}_{2}\Bigg\{\widetilde{\Pi}_{\alpha\beta}^{(2)}(q,\xi_Q =1)\Bigg\}
\label{agag}
\ee

The question is what is the correct form of  
${\cal T}_{2s}^{[4]}$ and  ${\cal T}_{2t}^{F}$, and the corresponding
$S^{[2]F\{12\}}_{2s}$ and $S^{[2]F\{12\}}_{2s}$ ; the latter
quantities
are automatically determined once the former have been specified.
In particular, if such a rearrangement exist, does it 
correspond to
a structure already known from the one-loop PT analysis ?
The natural candidate for this is clearly the PT rearranged
one-loop matrix element for 
$q(k_1)\bar{q}(k_2)\to g(p_1)g(p_2)$, shown in Fig.16
(individual Feynman graphs are shown in Fig.17, Fig.18, and Fig.19);
if that were the case, one would begin to discern an iterative pattern. 
As we will see in detail, this is indeed what happens.

Throughout this section
we will suppress the one-loop integration symbol  $\int [dk]$,
and will use (see \, Eq.\ (\ref{somedef1}) )
\be
J_3 \equiv J_3 (q,-p_1,k) = \frac{i}{2} g^2 C_A
[k^2 (k-p_1)^2 (k+p_2)^2]^{-1} \,.
\label{J}
\ee
We focus on the part of
the process involving 
the conventional one-loop three-gluon vertex  
$\Gamma_{\alpha\mu\nu}^{(1)}(q,p_1,p_2)$, 
which is diagrammatically shown in Fig.17 .
Let us now carry out the
decomposition of Eq. (\ref{decomp}) to the elementary 
three-gluon vertex $\Gamma_{\alpha\sigma\rho}^{(0)}(q,k+p_2,-k+p_1)$ 
appearing in the diagrams of
Fig.17a and Fig.17d; the 
parts stemming from the $\Gamma_{P\alpha\sigma\rho}^{(0)}$ 
will propagate towards the remaining elementary 
three-and four gluon vertices,
and will trigger further WI of the type shown in 
Eq. (\ref{ConWI1L}).
Reorganizing the terms thusly generated, one can show
that the 
$\Gamma_{\alpha\mu\nu}^{(1)}(q,p_1,p_2)$ of Fig.17 
can be written in the
form
\be
\Gamma_{\alpha\mu\nu}^{(1)}(q,p_1,p_2) = 
\Bigg [\frac{1}{2}\, \Pi^{(1)}_{P\alpha\beta}(q)\Bigg] d(q)
\Gamma_{F\,\mu\nu}^{(0)\beta}(q,p_1,p_2)
+\Gamma_{F\,\alpha\mu\nu}^{(1)}(q,p_1,p_2)
+ \Gamma_{P\,\alpha\mu\nu}^{(1)}(q,p_1,p_2) \, ,
\label{V1}
\ee
with 
\bea
\Gamma_{P\,\alpha\mu\nu}^{(1)}(q,p_1,p_2) &=&
- t_{\mu}^{\rho}(p_1) J_3 \,
 [\Gamma_{\nu\rho\alpha}^{(0)}(p_2,k,-k-p_2)+k_{\nu}g_{\alpha\rho}]
\nonumber\\
&& - 
t_{\nu}^{\rho}(p_2) J_3 \,
 [\Gamma_{\mu\alpha\rho}^{(0)}(p_1,k-p_1,-k)+k_{\mu}g_{\alpha\rho}]
\nonumber\\
&& - \frac{i}{2}\, 
V_{P\alpha\beta}^{(1)}\Gamma_{P\mu\nu}^{(0)\beta}(q,p_1,p_2)
+ \frac{1}{2}\, 
d(q)(p_2^2-p_1^2) \, q_{\alpha} V_{P\mu\nu}^{(1)}(q) \, .
\nonumber\\
&&{}
\label{decomp2}
\eea
The first term is precisely half of the pinch contribution needed
for converting $\Pi_{\alpha\nu}^{(1)}(q)$ into 
$\widehat\Pi_{\alpha\nu}^{(1)}(q)$, as shown in Fig.16a;
the other half will come from the one-loop quark-gluon vertex
shown in Fig.19d, following the usual one-loop PT procedure
presented in section II. 
This part of the construction 
has been first 
carried out in \cite{NJW1}, where the process-independence of
the $\widehat\Pi_{\alpha\nu}^{(1)}(q)$ 
was explicitly demonstrated. 

The second term,
$\Gamma_{P\,\alpha\mu\nu}^{(1)}(q,p_1,p_2)$, is Bose-symmetric 
with respect to $p_1 \leftrightarrow p_2$.
As one can easily verify from  Eq. (\ref{decomp2}),
$\Gamma_{P\,\alpha\mu\nu}^{(1)}(q,p_1,p_2)$
is zero on shell, i.e. when contracting with the 
polarization tensors and using $p_1^2=p_1^2=0$.
Thus $\Gamma_{P\,\alpha\mu\nu}^{(1)}(q,p_1,p_2)$
can be dropped when studying the one-loop process  
$q\bar{q} \to gg$, exactly as  
$\Gamma_{P\,\alpha\mu\nu}^{(0)}(q,p_1,p_2)$ was dropped
in the tree-level case.
As we we have seen in section IIIB,
in the off-shell case, i.e.
when $\Gamma_{\alpha\mu\nu}^{(1)}(q,p_1,p_2)$ is 
inserted into a two-loop quark-gluon vertex, the
parts proportional to $p_{1}^{\mu}$ and $p_{2}^{\nu}$ pinch 
the internal quark propagator and give propagator-like contributions,
whereas the parts proportional to $p_1^2$ and $p_2^2$ cancel
exactly against analogous contributions from the rest of the graphs 
contributing to the two-loop quark-gluon vertex.
Finally,
$\Gamma_{F\,\alpha\mu\nu}^{(1)}(q,p_1,p_2)$,
is exactly the one-loop version of 
$\Gamma_{F\,\alpha\mu\nu}^{(0)}(q,p_1,p_2)$:
It is the one-loop three-point function 
involving
one background ($q$) and two quantum ($p_1$,$p_2$) gluons
as incoming fields, computed using  the BFMFG Feynman rules.
(Fig.18).
Notice that the special ghost structure characteristic 
of the BFM (Fig.18e -- Fig.18h)
emerges {\it automatically}, after following 
the procedure outlined above.

It is straightforward to show that 
$\Gamma_{F\,\alpha\mu\nu}^{(1)}(q,p_1,p_2)$
satisfies the following WI
\be
q^{\alpha} \Gamma_{F\,\alpha\mu\nu}^{(1)}(q,p_1,p_2)
= \Pi_{\mu\nu}^{(1)}(p_1) - \Pi_{\mu\nu}^{(1)}(p_2) , 
\label{WI2B2}
\ee
which is the exact one-loop analogue of 
the tree-level Ward identity
of Eq (\ref{WI2B}); indeed the RHS is the
difference of two conventional one-loop self-energies 
computed in the RFG.

In addition, we have for the action of $p_{1\mu}$ on
$\Gamma_{F\,\alpha\mu\nu}^{(1)}(q,p_1,p_2)$
when $p_1^{2}=p_2^{2}=0$, is given by
\be
p_{1}^{\mu} \Gamma_{F\,\alpha\mu\nu}^{(1)} =
i \widehat\Pi_{\alpha\nu}^{(1)}(q) - i \Pi_{\alpha\nu}^{(1)}(p_2)
+ \lambda^{(1)}_{\nu\sigma}t^{\sigma}_{\alpha}(q)
+ s^{(1)}_{\alpha}p_{2\nu}
\label{2LGH}
\ee
where $\widehat\Pi_{\alpha\nu}^{(1)}(q)$ is given in 
 Eq (\ref{PTprop1}), and
\bea
\lambda^{(1)}_{\nu\sigma} &=& 
J_3 \Bigg [(k-p_1)^{\rho}
\Gamma_{\nu\rho\sigma}^{(0)}(p_2,k,-k-p_2)
-(k+p_2)_{\nu}k_{\sigma}\Bigg] - i \Bigg[2 B(q)+B(p_1)\Bigg] g_{\nu\sigma} 
\label{2LGH1} \\
s^{(1)}_{\alpha} &=& 
J_3 \Bigg [ 
p_{2}^{\sigma} k^{\rho} \Gamma^{(0)}_{F\alpha\sigma\rho}(q,k+p_2,-k+p_1)
- p_2 \cdot (k-p_1) (2k+p_2-p_1)_{\alpha}\Bigg] \nonumber\\ 
&& + \bigg(\frac{1}{8}\bigg)\Bigg[B(p_1)+B(p_2)\Bigg] q_{\alpha}
\label{2LGH2}
\eea
with
\be
B(p)\equiv \int [dk]J_1(p,k) \, .
\ee
Eq. (\ref{2LGH}) is the
one-loop analogue of  Eq. (\ref{WI1L}).
It is important to emphasize that 
$\Gamma_{F\,\alpha\mu\nu}^{(1)}(q,p_1,p_2)$ is {\it not} 
equal to the one-loop PT three-gluon vertex 
$\widehat\Gamma_{\,\alpha\mu\nu}^{(1)}(q,p_1,p_2)$
constructed in \cite{PT2}. 
Notice also that, unlike the tree-level case, now there will be
a $t$-channel ghost (we will not report its closed expression here) 
and that the $t$-channel has been modified
also in order to achieve the PT one-loop rearrangement. 

Thus we have,
\be
{\cal T}_{2s}^{[4]F} = (16a)+(16b)+(16d)+(16e)
\ee
with (we suppress a common factor $g\{\gamma_{\alpha}^{e}\}f^{eab} d(q)$ )
\bea
(16b) &=& 
\Gamma_{F\alpha\mu\nu}^{(1)}(q,p_1,p_2)  \, ,\nonumber\\ 
(16a) &=&  
\widehat\Pi_{\alpha\sigma}^{(1)}(q) d(q)
\Gamma_{F\,\mu\nu}^{(0)\sigma}(q,p_1,p_2) \, ,\nonumber\\ 
(16d) &=& 
\Gamma_{F\alpha\mu\sigma}^{(0)}(q,p_1,p_2) 
 d(p_2) \Pi_{\nu}^{(1)\sigma}(p_2) \, ,\nonumber\\ 
(16e) &=& 
\Gamma_{F\alpha\sigma\nu}^{(0)}(q,p_1,p_2) 
 d(p_1) \Pi_{\mu}^{(1)\sigma}(p_1)
\eea
and so, together with the first equation in (\ref{ab15}) 
\bea
{\cal T}_{2s}^{[4]F} {\cal T}_{2s}^{[2]F*} &=&
\Bigg( (16b) + (16d)+ (16e)\Bigg)\cdot {\cal T}_{2s}^{[2]F*} + 
(16a)\cdot {\cal T}_{2s}^{[2]F*} \nonumber\\
&=& \Bigg((20a) + (20b) + (20d) + (20f) + (20j) + (20g) + (20h)\Bigg) 
+ {\cal C}_{2gluons} \Bigg\{ (4e) \Bigg\}
\eea
i.e. we recover the two-gluon Cutkosky cuts of 
$\widetilde{\Pi}_{\alpha\beta}^{(2)}(q,\xi_Q =1)$ (Fig.20), together
with the corresponding two-gluon cuts of the 1PR graph in Fig.4e, 
as we should.
To account for the remaining  two-ghost Cutkosky cuts in Fig.20 and
Fig.4e
we need, in addition to Eq. (\ref{2LGH}), the following results
\bea
p_{1}\cdot (16a) &=&
-i \widehat\Pi_{\alpha\nu}^{(1)}(q) +
\widehat\Pi_{\alpha\sigma}^{(1)}(q) d(q) (p_2-p_1)^{\sigma}p_{2\nu} \, ,\nonumber\\ 
p_{1}\cdot (16d) &=& t_{\alpha\sigma}(q)\Pi_{\nu}^{(1)\sigma}(p_2) + 
i\Pi_{\alpha\nu}^{(1)}(p_2) \, ,\nonumber\\ 
p_{1}\cdot (16e) &=& 0 \, .
\eea
Then we have
\bea
p_{1}\cdot \Bigg( (16a)+(16b)+(16d)+ (16e)\Bigg)  &=& 
t_{\alpha}^{\sigma}(q) 
\Bigg( \lambda^{(1)}_{\nu\sigma}+ \Pi_{\sigma\nu}^{(1)}(p_2) \Bigg)\nonumber\\
&+& \Bigg( s^{(1)}_{\alpha} + \widehat\Pi_{\alpha\sigma}^{(1)}(q) d(q)
(p_2-p_1)^{\sigma}\Bigg)
     p_{2\nu}
\eea
From the above results one may immediately deduce the closed form of 
$(\Lambda^{[4]F}_2)_{\nu}^{ab}$ and 
$\bigg(S_{2s}^{[4]F\{12\}}\bigg)^{ab}$ 
appearing in Eq. (\ref{ab21}): 
\bea
\bigg(S_{2s}^{[4]F\{12\}}\bigg)^{ab} &=& g\{\gamma_{\alpha}^{e}\}f^{eab} d(q)
 \Bigg( s^{(1)}_{\alpha} + \widehat\Pi_{\alpha\sigma}^{(1)}(q) d(q) 
(p_2-p_1)^{\sigma}\Bigg) 
\nonumber\\
(\Lambda^{[4]F}_2)_{\nu}^{ab} &=& -ig\{\gamma_{\alpha}^{e}\}f^{eab} 
\Bigg( \lambda^{(1)\alpha}_{\nu}+ \Pi_{\nu}^{(1)\alpha}(p_2) \Bigg)
\eea
Then we have
\bea
2S_{2s}^{[4]F\{12\}}S_{2s}^{[4]F\{21\}} &=&  2s^{(1)}_{\alpha} (p_2-p_1)_{\beta}
+ 2\widehat\Pi_{\alpha\sigma}^{(1)}(q) (p_2-p_1)_{\sigma} 
(p_2-p_1)_{\beta} \nonumber\\
 &=& 
\Bigg((20i) + (20e) + (20m) + (20n) + (20q) + (20r) \Bigg) 
+ {\cal C}_{2ghosts} \Bigg\{ (4e) \Bigg\}
\eea
which concludes the proof of Eq (\ref{agag}).

\subsection{ Tree-level $q(k_1)\bar{q}(k_2)\to g(p_1)g(p_2)g(p_3)$, and
the $\Gamma_{F\alpha\mu\nu\rho}^{(0)}(q,p_1,p_2,p_3)$}

Next we consider the tree-level amplitude 
for the process $q(k_1)\bar{q}(k_2)\to g(p_1)g(p_2)g(p_3)$,
shown in Fig.21 . This amplitude must be  
appropriately rearranged, and it will eventually 
furnish the imaginary parts
corresponding to the three-particle Cutcosky cuts of 
$\widehat{\Pi}^{(2)}(q)$.

We start by presenting some general properties of this amplitude.
It is straightforward to verify that
${\cal T}_{\mu\nu\rho}^{abc}$
satisfies the following identities:
\begin{eqnarray}
p^{\mu}_1{\cal T}_{3\mu\nu\rho}^{abc} &=&
(S^{\{12\}}_3)_{\rho}^{abc}p_{2\nu}
+ (S^{\{13\}}_3)_{\nu}^{abc}p_{3\rho} \, ,\nonumber\\
p^{\nu}_2{\cal T}_{3\mu\nu\rho}^{abc} &=&
(S^{\{21\}}_3)_{\rho}^{abc}p_{1\mu}
+(S^{\{23\}}_3)_{\mu}^{abc}p_{3\rho} \, ,\nonumber\\
p^{\rho}_3{\cal T}_{3\mu\nu\rho}^{abc} &=& 
(S^{\{31\}}_3)_{\nu}^{abc}p_{1\mu}
+(S^{\{32\}}_3)_{\mu}^{abc}p_{2\nu} \ .
\label{BRS3g}
\end{eqnarray}
Bose symmetry imposes the following relations among the
${S}_3^{\{ij\}}$ amplitudes \cite{FPP}:
\bea
({S}_3^{\{ij\}})^{a_{i}a_{j}a_{\ell}}_{\sigma}(p_{i},p_{j},p_{\ell}) &=&
({S}_3^{\{ji\}})^{a_{j}a_{i}a_{\ell}}_{\sigma}(p_{j},p_{i},p_{\ell})
\nonumber\\
({S}_3^{\{ij\}})^{a_{i}a_{j}a_{\ell}}_{\sigma}(p_{i},p_{j},p_{\ell}) &=&
({S}_3^{\{i\ell\}})^{a_{i}a_{\ell}a_{j}}_{\sigma}(p_{i},p_{\ell},p_{j}) \, ,
\label{Bose}  
\eea
and 
\be
p_{i}^{\sigma}({\cal S}^{\{j\ell\}})_{\sigma}^{abc}= 
p_{j}^{\sigma}( {{\cal S}}^{\{i\ell\}})_{\sigma}^{abc} 
~, ~~~\ell \neq i\neq j.
\label{Bose2}
\ee

These sets of identities guarantee that all dependence
on the unphysical four-vector $n_{\alpha}$ and the 
gauge parameter $\eta$  
appearing in the polarization tensors will disappear from
the final expression for ${\cal A}^{[6]}_{3}$. 

The amplitude rearrangement is as follows (Fig.21):
\bea
({\cal T}_{3s}^{[3]F})_{\mu\nu\rho}^{abc} &=& 
g^{2}\{\gamma^{e,\alpha}\}d(q) 
\Gamma_{F\alpha\mu\nu\rho}^{(0)eabc}(q,p_1,p_2,p_3) \nonumber\\
({\cal T}_{3t}^{[3]F})_{\mu\nu\rho}^{abc} &=& 
({\cal T}_{3t}^{[3]})_{\mu\nu\rho}^{abc}
\eea
where 
\bea
\Gamma_{F\alpha\mu\nu\rho}^{(0)eabc}(q,p_1,p_2,p_3) &=&
f^{ecx}f^{abx} d(p_1+p_2)
\Gamma_{F\alpha\rho\sigma}^{(0)}(q,p_3,p_1+p_2)
\Gamma_{\mu\nu}^{(0)\sigma}(p_1,p_2,-p_1-p_2)
\nonumber\\
&&+
f^{ebx}f^{acx} d(k_1+k_3)
\Gamma_{F\alpha\nu}^{(0)}(q,p_2,p_1+p_3)
\Gamma_{\mu\rho\sigma}^{(0)\sigma}(p_1,p_3,-p_1-p_3)
\nonumber\\
&&+
f^{eax}f^{bcx} d(k_2+k_3)
\Gamma_{F\alpha\mu\sigma}^{(0)}(q,p_1,p_2+p_3)
\Gamma_{\nu\rho}^{(0)\sigma}(p_2,p_3,-p_2-p_3)
\nonumber\\
&&+\Gamma_{\alpha\mu\nu\rho}^{(0)eabc}
\eea
It is easy to verify now that 
$\Gamma_{F\alpha\mu\nu\rho}^{(0)eabc}(q,p_1,p_2,p_3)$ 
is the analogue of $\Gamma_{F\alpha\mu\nu}^{(0)abc}(q,p_1,p_2)$ appearing in the
first equation of (\ref{ab15}) for the case 
of three on-shell gluons. 
In particular, we have that
\bea
q^{\alpha}\Gamma_{F\alpha\mu\nu\rho}^{(0)eabc}(q,p_1,p_2,p_3) &=&
-f^{ecx}f^{abx} p_3^{2}\, d(p_1+p_2)
\Gamma_{\mu\nu\rho}^{(0)}(p_1,p_2,-p_1-p_2)
\nonumber\\
&&-
f^{ebx}f^{acx} p_2^{2}\, d(k_1+k_3)
\Gamma_{\mu\rho\nu}^{(0)}(p_1,p_3,-p_1-p_3)
\nonumber\\
&&-
f^{eax}f^{bcx} p_1^{2}\, d(k_2+k_3)
\Gamma_{\nu\rho\mu}^{(0)}(p_2,p_3,-p_2-p_3) \, ,
\label{3gWI}
\eea
where we have used the well-known WI relating the bare 
three-and four-gluon vertices \cite{4g}.
Eq.\ (\ref{3gWI}) is the analogue of Eq.\ (\ref{WI2B}).
Clearly, when $p_1^{2} = p_2^{2} = p_3^{2}= 0$,
\be
q^{\alpha}
\Gamma_{F\alpha\mu\nu\rho}^{(0)eabc}(q,p_1,p_2,p_3) = 0 \, ,
\ee
exactly as happens in Eq.\ (\ref{WI2B}).

In addition,
\bea
p^{\mu}_1 ({\cal T}_{3s}^{[3]F})_{\mu\nu\rho}^{abc} &=&
(S^{[3]F\{12\}}_{3s})_{\rho}^{abc}p_{2\nu}
+ (S^{[3]F\{13\}}_{3s})_{\nu}^{abc}p_{3\rho} + 
(\Lambda^{[3]F}_3)_{\nu\rho}^{abc}
\, ,\nonumber\\
p^{\mu}_1 ({\cal T}_{3t}^{[3]})_{\mu\nu\rho}^{abc} &=&
(S^{[3]\{12\}}_{3t})_{\rho}^{abc}p_{2\nu}
+ (S^{[3]\{13\}}_{3t})_{\nu}^{abc}p_{3\rho} - 
(\Lambda^{[3]F}_3)_{\nu\rho}^{abc}
\eea
where
\be
(S^{[3]F\{12\}}_{3s})_{\rho}^{abc} = \{\gamma^{m,\alpha}\}d(q)
\Bigg[(Z_1)_{\alpha\rho}^{mabc} + (Z_2)_{\alpha\rho}^{mabc}
+ (Z_3)_{\alpha\rho}^{mabc}\Bigg]
\ee
with
\bea
(Z_1)_{\alpha\rho}^{mabc} &=& 
g^2 f^{mcx}f^{abx}
d(p_1+p_2) p^{\sigma}_2 
\Gamma_{F\alpha\rho\sigma}^{(0)}(q,p_3,p_1+p_2)\nonumber\\
(Z_2)_{\alpha\rho}^{mabc} &=& g^2
f^{mbx}f^{acx} \Bigg[d(p_1+p_3)(2p_2+q)_{\alpha}(p_1+p_3)_{\rho} 
-g_{\alpha\rho} \Bigg] \nonumber\\
(Z_3)_{\alpha\rho}^{mabc} &=& g^2
f^{max}f^{bcx} d(p_2+p_3)(2p_1+q)_{\alpha}p_{2\rho}
\label{Zs}
\eea
and
\be
(S_{3t}^{[3]\{12\}})^{abc}_{\rho} =
-g^3 \bar{v}(k_2)
\Bigg[
\, \tau^c \gamma_{\rho}\, 
S(k_2+p_3)
\, \tau^e \gamma_{\sigma}
+ \tau^e \gamma_{\sigma}\,
S(k_1+p_3) \, 
\gamma_{\rho} \tau^c \,  \Bigg] u(k_1)
f^{eab}d(p_1+p_2) p_{2}^{\sigma}.
\label{S23}
\ee
Armed with the above relations, it is then straightforward to show that 
\bea
{\cal A}^{[6]} &=& \frac{1}{2} \bigg(\frac{1}{3!}\bigg)
\Bigg[ \bigg({\cal T}_{3s}^{[3]F}+{\cal T}_{3t}^{[3]}\bigg)
\bigg( {\cal T}_{3s}^{[3]F}+{\cal T}_{3t}^{[3]}\bigg)^{*}- 
6 \bigg(S_{3s}^{[3]F\{12\}}+S_{3t}^{[3]\{12\}}\bigg)
\bigg(S_{3s}^{[3]F\{21\}} + S_{3t}^{[3]\{21\}} \bigg)^{*}
\Bigg] \nonumber\\
&=&  {\cal A}^{[6]}_{S} +{\cal A}^{[6]}_{V} +{\cal A}^{[6]}_{B} \, ,
\label{ab30}
\eea
with
\bea
{\cal A}^{[6]}_{S} &=&
\frac{1}{2}
\bigg(\frac{1}{3!}\bigg)
\Bigg[ 
{\cal T}_{3s}^{[3]F}{{\cal T}_{3s}^{[3]F}}^{*}
-6\bigg(S_{3s}^{[3]F\{12\}}\bigg)\bigg(S_{3s}^{[3]F\{21\}}\bigg)^{*}
 \Bigg] \, , \nonumber\\ 
{\cal A}^{[6]}_{V} &=&
\frac{1}{2}\bigg(\frac{1}{3!}\bigg) \Bigg[
\bigg\{ {\cal T}_{3s}^{[3]F}{{\cal T}_{3t}^{[3]}}^{*}
-6 \bigg(S_{3s}^{[3]F\{12\}}\bigg)\bigg(S_{3t}^{[3]\{21\}}\bigg)^{*}\bigg\}
 + \bigg\{ {\cal T}_{3t}^{[3]}{{\cal T}_{3s}^{[3]F}}^{*}
-6 \bigg(S_{3t}^{[3]\{12\}}\bigg)\bigg(S_{3s}^{[3]F\{21\}}\bigg)^{*}\bigg\}
\Bigg] \, , \nonumber\\
{\cal A}^{[6]}_{B} &=&
 \frac{1}{2} \bigg(\frac{1}{3!}\bigg) \Bigg[
{\cal T}_{3t}^{[3]}{{\cal T}_{3t}^{[3]*}}
-6 \bigg(S_{3t}^{[3]F\{12\}}\bigg)\bigg(S_{3t}^{[3]\{21\}}\bigg)^{*}\Bigg]
\, ,
\label{ab31}
\eea
We are now in position to prove that, indeed, 
\be 
{\cal A}^{[6]}_{S\alpha\beta}(q) = 
{\cal C}_{3}\Bigg\{\widehat{\Pi}_{\alpha\beta}^{(2)}(q)\Bigg\}
=
{\cal C}_{3}\Bigg\{\widetilde{\Pi}_{\alpha\beta}^{(2)}(q,\xi_Q =1)\Bigg\}
\, ,
\ee
where 
${\cal A}^{[6]}_{S\alpha\beta}(q)$ is defined from 
${\cal A}^{[6]}_{S}$ exactly as in Eq.\ (\ref{ab20}) and 
Eq.\ (\ref{again}).

To begin with, 
\be
\bigg(\frac{1}{12}\bigg){\cal T}_{3s}^{[3]F}{{\cal T}_{3s}^{[3]F}}^{*}
= (22a)_{c_1} + (22a)_{c_2} + (22b) + (22c) + (22g) + (22\ell)  \, .
\ee
For the cuts involving ghosts we need the closed form of 
$S^{[3]F\{21\}}_{3s}$, which can be obtained from $S^{[3]F\{12\}}_{3s}$
of Eq.\ (\ref{Zs}) by virtue of the relations given in 
Eq.\ (\ref{Bose}). In particular,
\bea
(Z_1')_{\beta\rho}^{nabc} &=& 
g^2 f^{ncr}f^{bar}
d(p_1+p_2)\, p^{\lambda}_1 
\Gamma_{F\beta\rho\lambda}^{(0)}(q,p_3,p_1+p_2)\nonumber\\
(Z_2')_{\beta\rho}^{nabc} &=&
g^2 f^{nar}f^{bcr}\Bigg[d(p_2+p_3)\, (2p_1+q)_{\beta}(p_2+p_3)_{\rho} 
-g_{\beta\rho} \Bigg] \nonumber\\
(Z_3')_{\beta\rho}^{nabc} &=&
g^2 f^{nbr}f^{acr} d(p_1+p_3)\, (2p_2+q)_{\beta}p_{1\rho} \, .
\eea
Then it is straightforward to show that
\bea
\frac{1}{2} Z_1 Z_1' &=& (22h) \, ,
\nonumber\\
\frac{1}{2} Z_1 Z_2' &=& (22e)_{c_2}+ (22k) \, ,
\nonumber\\
\frac{1}{2} Z_1 Z_3' &=& (22e)_{c_1} \, ,
\nonumber\\
\frac{1}{2} Z_2 Z_1' &=& (22f)_{c_1} + (22j) \, ,
\nonumber\\
\frac{1}{2} Z_2 Z_2' &=& (22i)_{c_1} + (22s) + (22n) + (22o) \, ,
\nonumber\\
\frac{1}{2} Z_2 Z_3' &=& (22q) + (22m) \, ,
\nonumber\\
\frac{1}{2} Z_3 Z_1' &=& (22f)_{c_2} \, ,
\nonumber\\
\frac{1}{2} Z_3 Z_2' &=& (22r) + (22p) \, ,
\nonumber\\
\frac{1}{2} Z_3 Z_3' &=& (22i)_{c_2} \, .
\eea
Thus, we have accounted for all three-particle 
Cutkosky cuts appearing in Fig.22 .

\section {Discussion and Conclusions}

In  this  paper we have   presented the  generalization  of the  PT to
two-loops   for   the case of   mass-less   Yang-Mills  theories.  Two
different, but complementary derivations have  been presented.  In the
first derivation we  have followed the diagrammatic approach  employed
in the  one-loop case \cite{PTCPT,PT,PT2},  and have shown  how the PT
properties are  replicated in the next  order.  In the second  we have
pursued the dispersive PT construction established in
\cite{PP2,PRW}  and  have shown  that  the  resulting  structures  are
consistent with those derived with   the first method.  We   emphasize
that  throughout this     entire analysis we   have   maintained  a
diagrammatic   interpretation  of  the    various  contributions.   In
particular, no sub-integrations
had to be carried out.  This additional feature renders the method all
the more  powerful, because unitarity  is manifest, and can  be easily
verified by means  of the Cutkosky  cuts.  The combination of the  two
methods  constrains significantly the  PT construction presented here,
and  restricts severely any possible deviations  from  it.  The reader
should  be able to  recognize, for example,  that any rearrangement of
the (internal) vertices of two-loop box-diagrams  (for example, of the
so-called ``H-diagram'' discussed in the second paper of
\cite{PP1}) cannot    be  reconciled   with  the   arguments   of  the
simultaneous two-  and    three-particle Cutkosky cuts  presented   in
section VI.   Whereas no strict proof has  been given here that the PT
construction  developed in  this  paper is  mathematically  unique, we
consider that  as a very  plausible possibility.  Notice also that the
appearance of characteristic  one-loop structures inside  the two-loop
PT Green's functions suggests the onset of an iterative pattern, which
may provide clues leading to the generalization of the PT algorithm to
all orders in perturbation theory.

The generalization  of  the two-loop  PT  construction to  the case of
Yang-Mills  theories   with  spontaneous symmetry    breaking   (Higgs
mechanism)  in  general, and  the  electroweak sector of  the Standard
Model in particular,    should  proceed precisely  according  to   the
methodology  presented in  this   paper.  Except  for  the  additional
book-keeping  complications stemming from  the presence of gauge-boson
masses (modifications of WI, appearance of  seagull and tadpole terms,
diagrams  with would-be Goldstone  bosons),  no  additional conceptual
obstacles  are expected.  

The results    of this paper  clearly   prove  that the correspondence
between PT and BFMFG \cite{BFMPT}  persists at two-loops.  Notice that
this   proof is based   on  an a-posteriori  comparison with  a result
established through the  systematic diagrammatic  rearrangement of the
physical $S$-matrix computed in the renormalizable gauges, rather than
on an  a-priori formal derivation at  the level  of the BFM generating
functional.    It  would be clearly      important to reach  a  deeper
understanding of   what singles out   the   value $\xi_Q  =  1$.   One
possibility would be to look for special properties  of the BFM action
at $\xi_Q = 1$  \cite{RDP}.  In such a case  one could choose to avoid
the  complications    arising    from  renormalization,     since  the
correspondence is valid also for the super-renormalizable 3-d QCD.

Just as happened in the  one-loop  case, the two-loop PT  self-energy
defined  here  lends itself  as  an  essential ingredient for  the
extension of the notion of the QCD effective charge
\cite{PT,GGr,NJW2} to two-loops (for  a thorough   
discussion of the one-loop case see \cite{NJW2}),  
since  it has  precisely the  same  properties  as  the
corresponding   QED quantity, i.e.    the  vacuum polarization of  the
photon.   First of  all,  the  two-loop  PT self-energy captures   the
leading logarithms of the theory, i.e.  the prefactor of the logarithm
is   the second coefficient of the   QCD $\beta$ function.  Second, by
virtue of the the QED like WI given in Eq.\ (\ref{R3}) the combination
$\alpha_{eff}  (q,\mu) \sim g^{2}(\mu)  \widehat\Delta (q/\mu)$, where
$\widehat\Delta (q/\mu) = [1- \widehat\Pi^{(1)} (q/\mu) -
\widehat\Pi^{(2)}  (q/\mu)]^{-1}$ is a renormalization group invariant
quantity.  Third, it   has  by construction  the  correct unitarity
structure \cite{PRW}.   While the $\alpha_{eff}$ defined above appears
as the obvious  candidate, a  detailed study  needs be  carried out in
order to determine whether or    not there are any   field-theoretical
obstructions \cite{HKS}  which would  prevent  the realization  of the
two-loop construction corresponding to the QCD effective charge.

In    this context notice also   that  the construction presented here
determines uniquely the constant term of the two-loop effective charge
(within a given renormalization  scheme) \cite{PPS}.  To determine its
actual value  one should go  beyond the two-loop calculation presented
in \cite{Abb}, and compute the Feynman diagrams  of Fig.9 keeping also
constant  terms   \cite{GrGa}.   Knowledge of  this  constant  term is
important   when  the QCD  effective charge  is  used in  the field of
renormalon calculus;   in   particular,  it would  be   free   of  the
ambiguities infesting the estimates of the renormalon contributions to
physical observables \cite{NJW2,GaSir}.

In addition, it is well-known \cite{RK} that if one was to compute the
two-loop gluon self-energy in the context of the BFM keeping $\xi_{Q}$
arbitrary, the resulting $\xi_{Q}$-dependent term would be a constant,
i.e.  it would not affect the  coefficient of the logarithm.  While in
such a case  the residual gauge-dependence  could  be thought of  as a
renormalization-scheme ambiguity, i.e.  it could be re-absorbed in the
wave-function  renormalization of the (background)  gluon, this is not
possible when the gauge fields are massive.  In that case unitarity is
even more constraining; as is  known from the studies
on the one-loop electro-weak effective charges \cite{PP2,PRW,PP3}, the
gauge-dependence affects  non-trivially the analytic structure of the
answer, giving rise to unphysical thresholds.

The   calculations   presented in section   VI   constitute  the first
dispersive derivation  of the two-loop  QCD $\beta$ function.  In this
analysis we have made use of the one-to-one correspondence between the
physical $S$-matrix elements and the Cutkosky cuts  of the two-loop PT
self-energy.  This construction involves a very particular combination
of one-loop (section  VIA)   and tree-level graphs (section   VIB); in
addition to furnishing  the  correct $\beta$ function  coefficient,  a
subtle cancellation of   infrared divergences also takes  place: while
both sets  of graphs are  infrared divergent, they  combine to  give a
infrared finite answer.  This can be directly inferred from the simple
observation   that the Cutkosky  cutting  procedure  we have  employed
amounts finally to the determination of the imaginary part of a single
logarithm,  namely  that of the  two-loop   self-energy; the latter is
infrared finite \cite{IR,TM}.   While the Cutkosky formalism furnishes
an intuitive  diagrammatic understanding  and a valuable calculational
short-cut, it would be interesting to reproduce the results of section
IV without resorting to it.  In particular one could study the precise
cancellation mechanism   of the infrared   divergences using  a proper
infrared regularization scheme, and  explicit expressions for the two-
and  three-gluon  phase-space, which  we have  not   needed here.   In
addition, it  would  be interesting   to  attempt a similar   two-loop
derivation using the formalism developed in
\cite{CDE}, and study possible connections.

It has been often advocated that  the non-perturbative QCD effects can
be reliably captured  at  an inclusive level  by means  of an infrared
finite  quantity, which  would  constitute    the  extension of    the
perturbative  QCD  running coupling  to  low energy  scales \cite{DMW}
Early results by Cornwall based on the study  of {\it gauge invariant}
Schwinger-Dyson  equations \cite{PT} involving  this  quantity suggest
that such a description can in fact  be derived from first principles.
According to this  analysis, the self-interaction  of gluons give rise
to a dynamical gluon mass, while preserving at the same time the local
gauge  symmetry  of   the theory.  The  presence   of  the gluon  mass
saturates the  running of the  QCD coupling; so, instead of increasing
indefinitely in the   infrared  as perturbation  theory predicts,   it
``freezes'' at a finite value \cite{PT2,gm2}.  It would be interesting
to  revisit this  issue in  the light of  the  results  derived in the
present paper.   For  example, one could study  the  structure of  the
gauge-invariant Schwinger-Dyson equation for the PT gluon self-energy,
and    in  particular     the    way  the     PT   three-gluon  vertex
$\widehat\Gamma^{(1)}$ enters in  the PT gluon self-energy, using  the
two loop  results  as a  guidance.   In doing  so  one  could  hope to
systematically improve    on  the analysis   of  \cite{PT},  where the
gauge-technique ansatz  for the vertex was  used.  In this context one
may find   it advantageous to  rewrite the   vertex $\Gamma_{F}^{(1)}$
appearing  inside   the two-loop  PT gluon    self-energy  in terms of
$\widehat\Gamma^{(1)}$  ; one   should   then interpret  the  emerging
residual terms as parts of the  two-loop self-energy, even though they
appear to be pinch-like
\cite{FL}, i.e. once  the PT self-energy
has been  fixed it may be recast
into a different form, but no pieces should be re-assigned to vertices or
boxes.

Finally it would be interesting to pursue a
connection with other field- or string-theoretical methods 
\cite{ZB,VMLRM,GG,WLT,EL},
either in order to acquire a more formal understanding of the PT,
or in order to combine various attempts into a coherent framework. 

\vspace{0.7cm}\noindent {\bf  Acknowledgments.}  
I thank  L. Alvarez-Gaume, S. Catani, J.M. Cornwall, 
Yu.L. Dokshitzer, G. Grunberg,
C. Kounnas, L. Magnea, A.H. Mueller, A. Pilaftsis, G. Sterman, and R. Stora,  
for various useful discussions.


\newpage
\centerline{\Large{Figure Captions}}

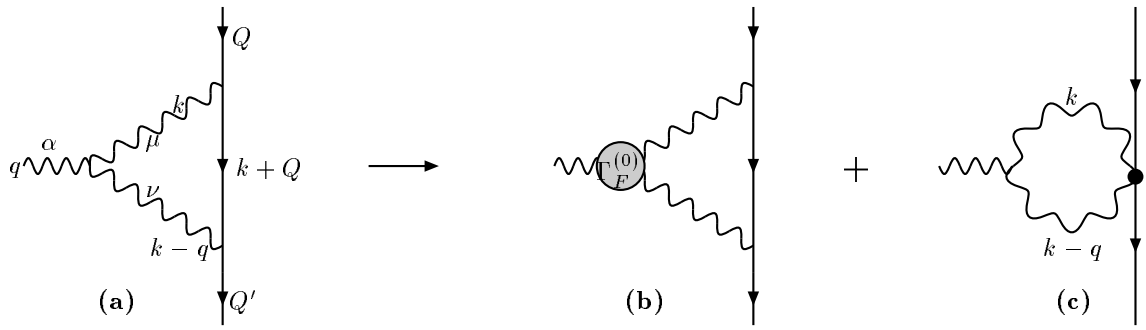
{\bf Fig.1}:
Carrying out the fundamental vertex decomposition inside the 
non-abelian Feynman graph contributing 
to ${\Gamma}_{\alpha}^{(2)}$ ($a$), gives rise to 
a genuine vertex ($b$) and a self-enery-like contribution ($c$). 

\vskip0.5cm

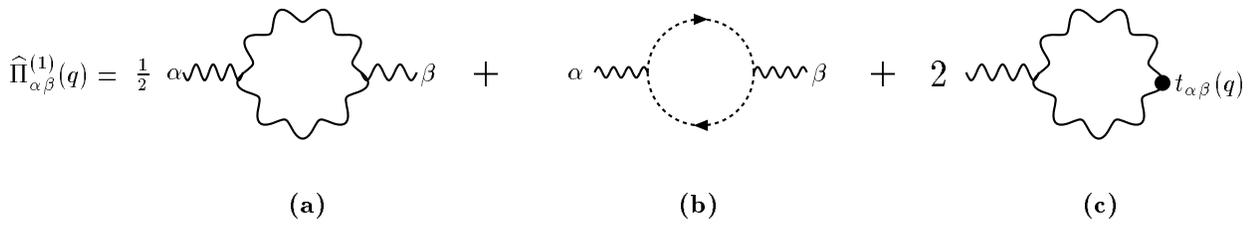
{\bf Fig.2}: 
The diagrammatic representation of
PT one-loop self-energy ${\widehat\Pi}_{\alpha\beta}^{(1)}$,
as the sum of the conventional self-energy $\Pi_{\alpha\beta}^{(1)}$,
graphs ($a$) and ($b$), and the pinch contributions coming from the 
vertices ($c$).

\vskip0.5cm

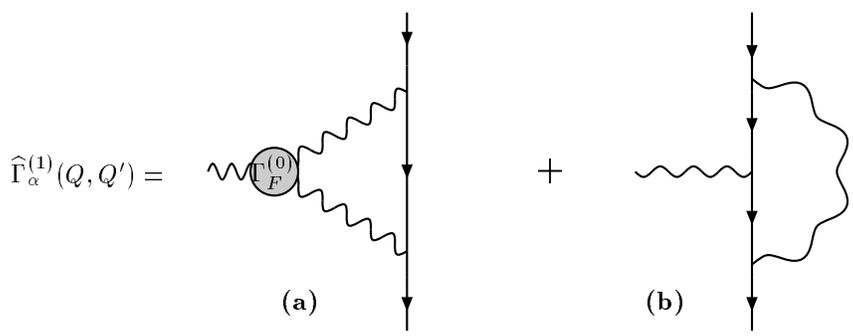
{\bf Fig.3}: 
The PT one-loop vertex ${\widehat\Gamma}_{\alpha}^{(1)}$.

\vskip0.7cm

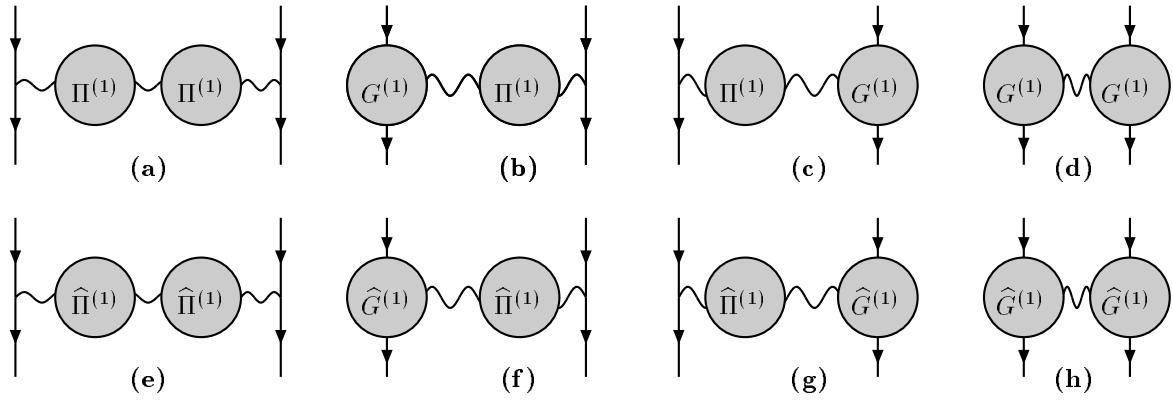
{\bf Fig.4}: 
The one-particle reducible graphs before 
[($a$), ($b$), ($c$), ($d$)]
and after [($e$), ($f$), ($g$), ($h$)] the PT rearrangement.  
$G^{(1)}$ and ${\widehat G}^{(1)}$ denote respectively the conventional
and PT one-loop vertices with one-loop 
self-energy corrections to the external fermions included.
 
\vskip0.5cm

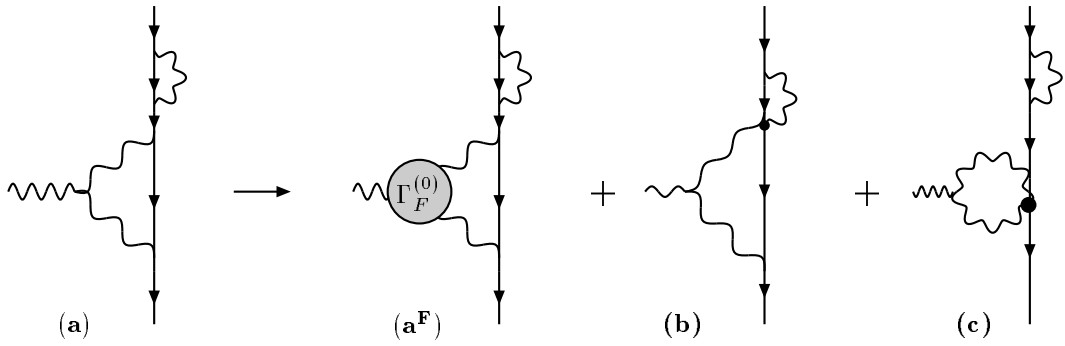
{\bf Fig.5}:
The PT rearrangement of typical one-particle reducible graph ($a$),
giving rise  to its PT counterpart ($a^{P}$), 
and to contributions to the first term of $F_P^{(2)}$ ($c$)
and to $Y_P^{(2)}$ ($b$).

\vskip0.5cm

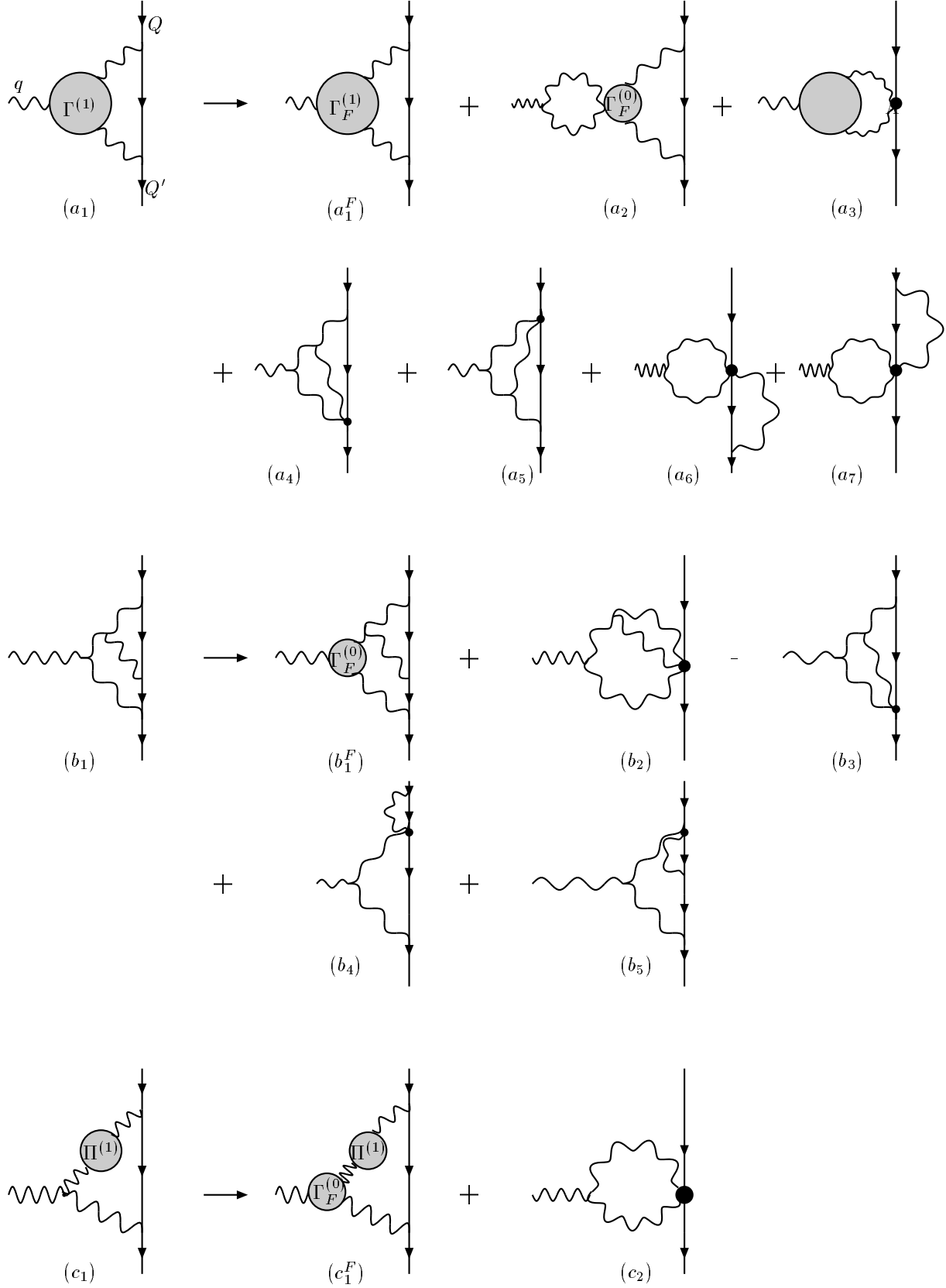
{\bf Fig.6}:
The result of enforcing the PT decomposition on
the external vertices of some of the two-loop Feynman diagrams
contributing the conventional two-loop quark-gluon vertex
${\Gamma}_{\alpha}^{(2)}$.

\vskip0.5cm

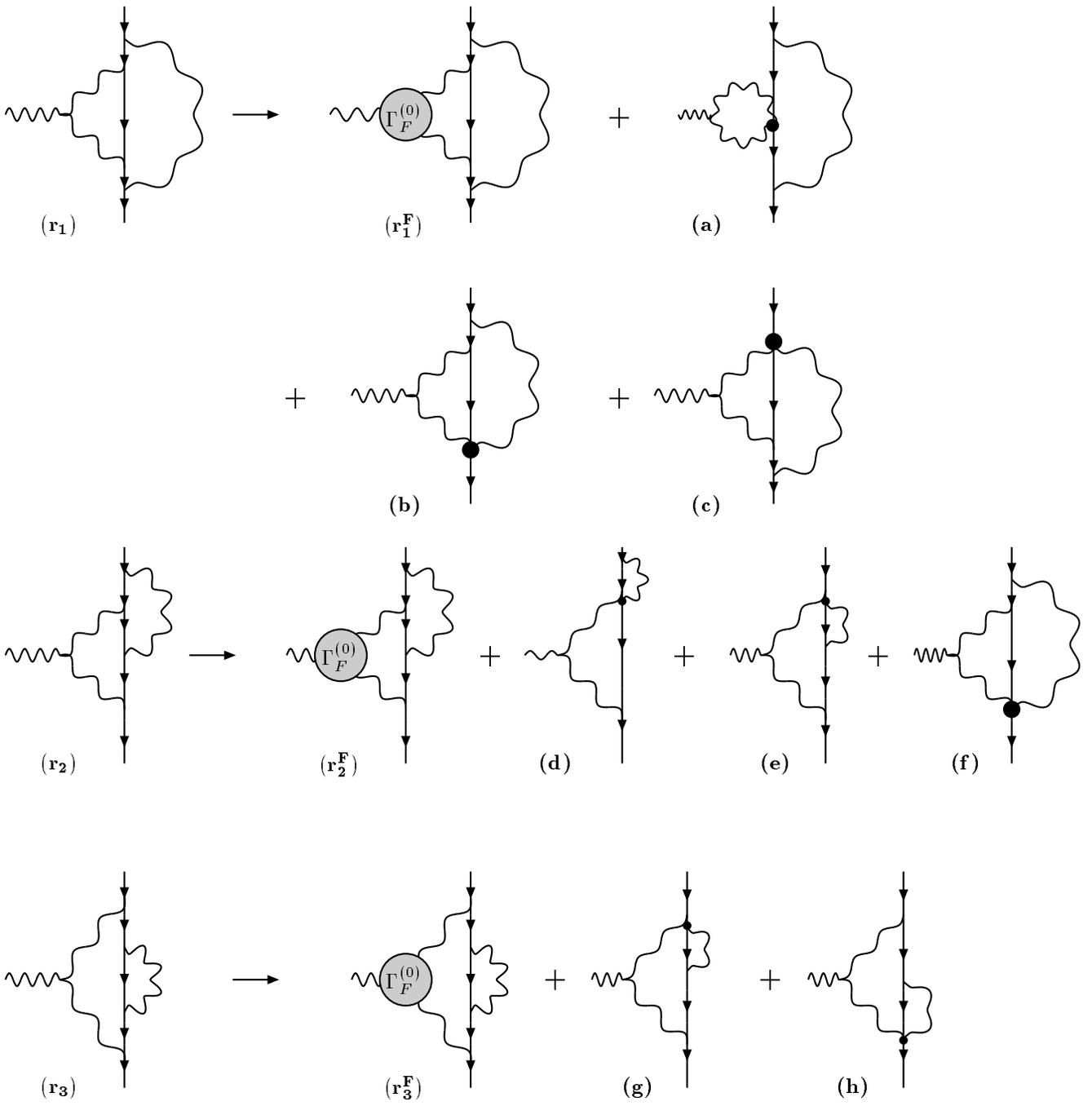
{\bf Fig.7}:
The result of enforcing the PT decomposition on
the external vertices of some of the remaining two-loop vertex graphs.

\vskip0.5cm

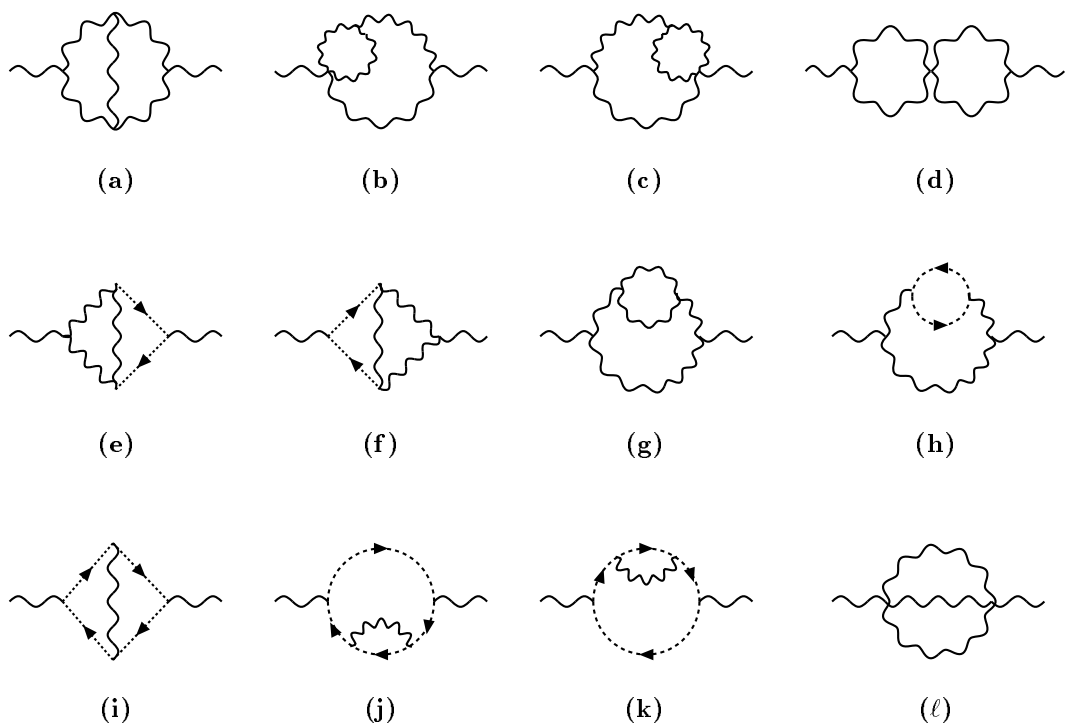
{\bf Fig.8}:
The Feynman diagrams contributing to the conventional
two-loop gluon self-energy $\Pi_{\alpha\beta}^{(2)}$, 
in the $R_{\xi}$ gauges.

\vskip0.5cm

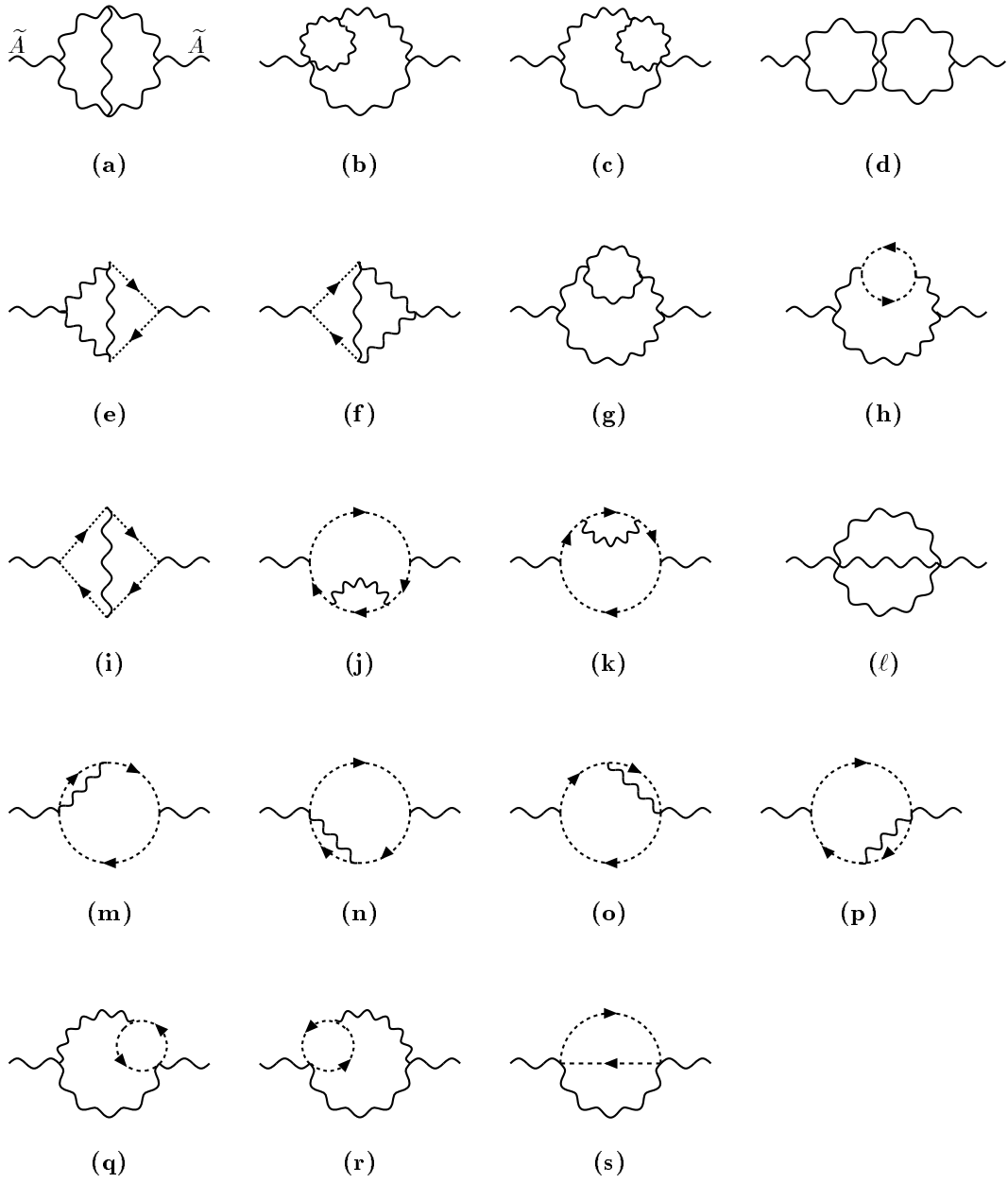
{\bf Fig.9}:
The Feynman diagrams contributing to the  BFM
two-loop gluon self-energy ${\widetilde\Pi}_{\alpha\beta}^{(2)}$.

\vskip0.5cm

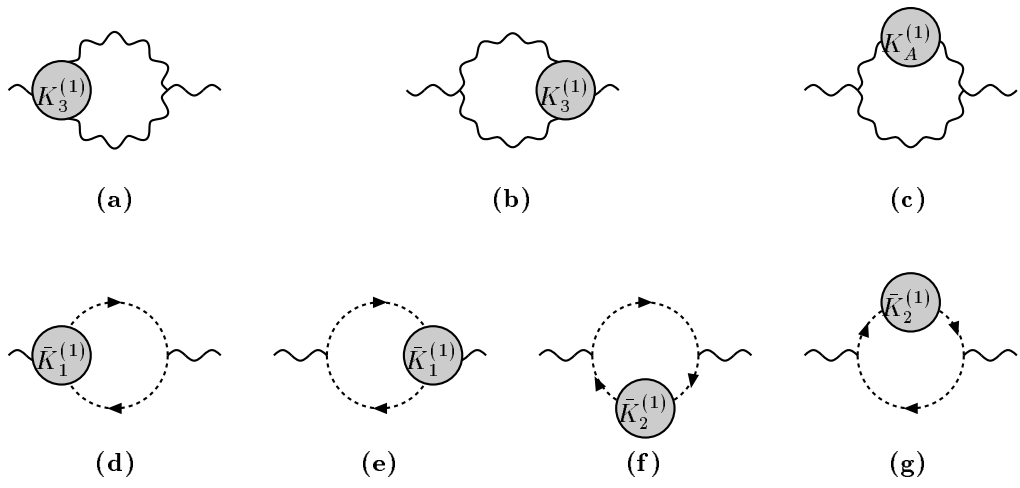
{\bf Fig.10}:
The one-loop counterterms contributing to the 
conventional two-loop  
gluon self-energy $\Pi_{\alpha\beta}^{(2)}. $ 

\vskip0.5cm

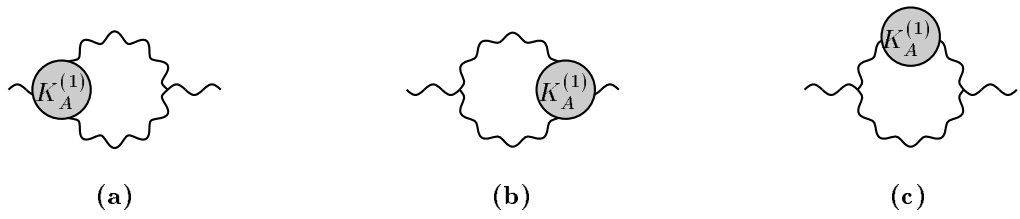
{\bf Fig.11}:
The one-loop counterterms necessary for the
 two-loop gluon self-energy ${\widehat\Pi}_{\alpha\beta}^{(2)} $;
they are identical to those needed for the 
BFM two-loop gluon self-energy ${\widetilde\Pi}_{\alpha\beta}^{(2)} $.

\vskip0.5cm

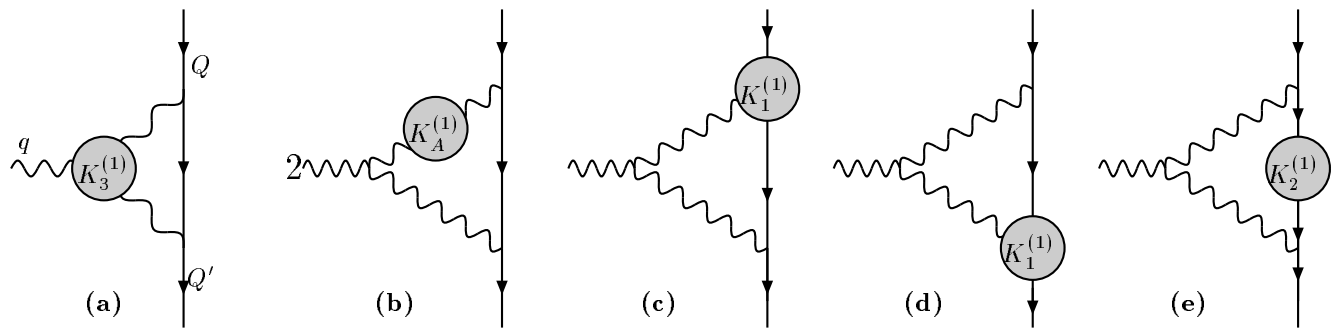
{\bf Fig.12}:
The one-loop counterterms necessary to cancell
the sub-divergences inside the conventional two-loop
quark-gluon vertex ${\Gamma}_{\alpha}^{(2)}$. In the case of massive
fermions the wave-function counterterm $K_2^{(1)}$ should be accompanied by the
appropriate mass counterterm (not shown).

\vskip0.5cm

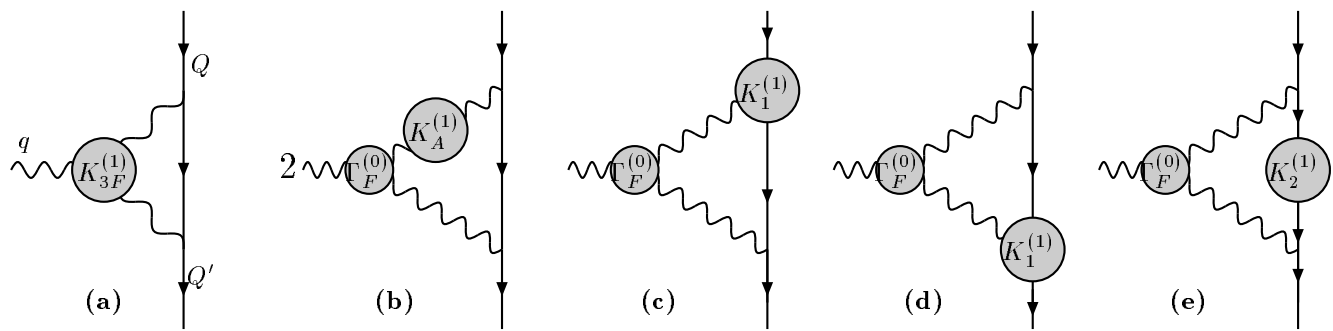
{\bf Fig.13}:
The one-loop counterterms necessary to cancell
the sub-divergences inside the PT two-loop
quark-gluon vertex ${\widehat\Gamma}_{\alpha}^{(2)}$; they
are identical to those neede for the BFM two-loop
quark-gluon vertex  ${\widetilde\Gamma}_{\alpha}^{(2)}$.

\vskip0.5cm

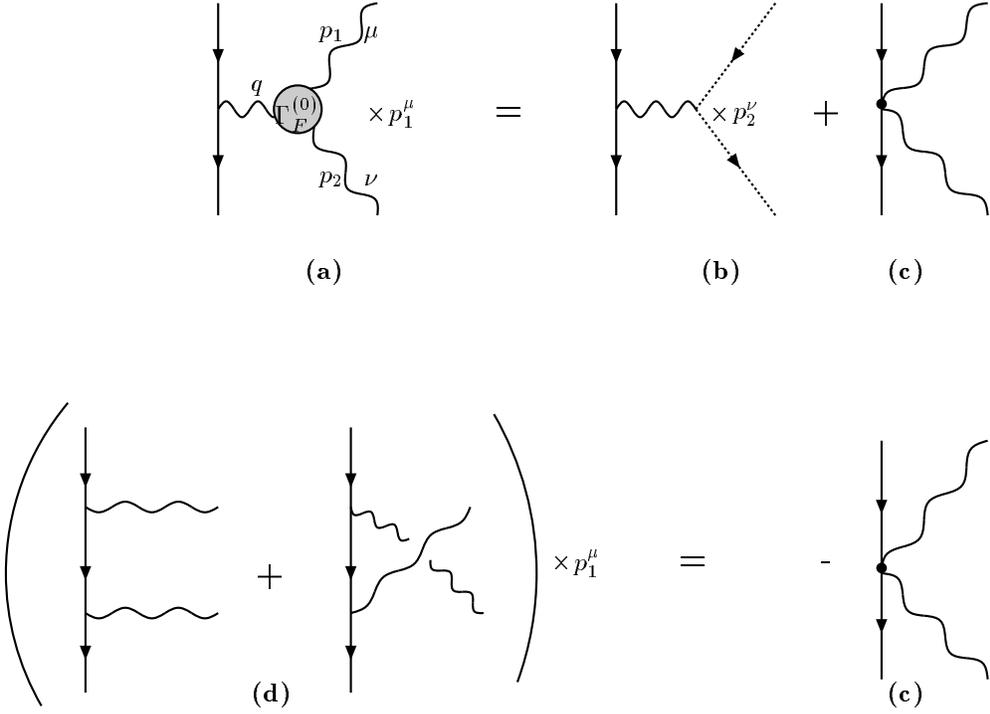
{\bf Fig.14}:
The fundamental BRS-enforced cancellation 
of $s$-channel ($a$)
and $t$-channel [($d_1$) and ($d_2$)]
contributions, instrumental for the
absorptive PT construction. Graph ($b$) gives rise to the
correct ghost-structure.

\vskip0.5cm

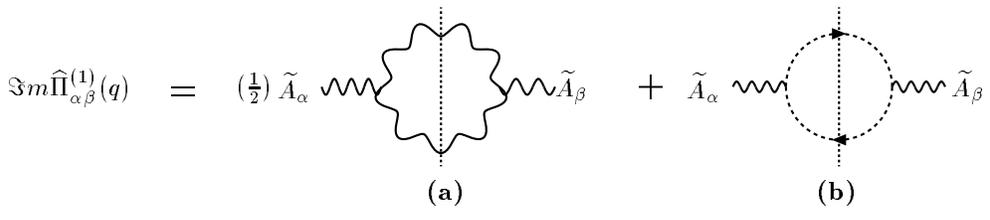
{\bf Fig.15}:
The Cutkosky cuts of the PT (and BFMFG) one-loop gluon self-energy.

\vskip0.5cm

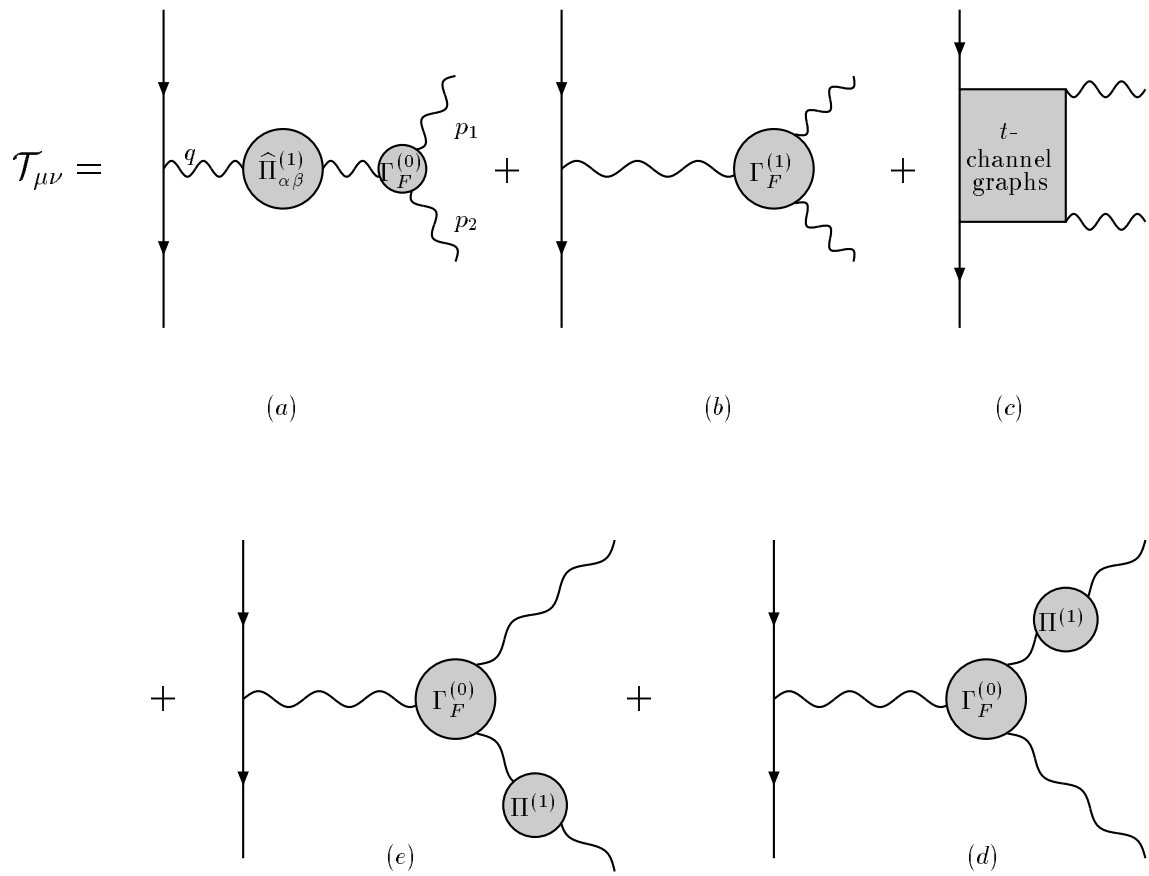
{\bf Fig.16}:
The one-loop amplitude for the process 
$q(k_1) \bar{q}(k_2) \to g(p_1) g(p_2)$, after the PT rearrangement.

\vskip0.5cm

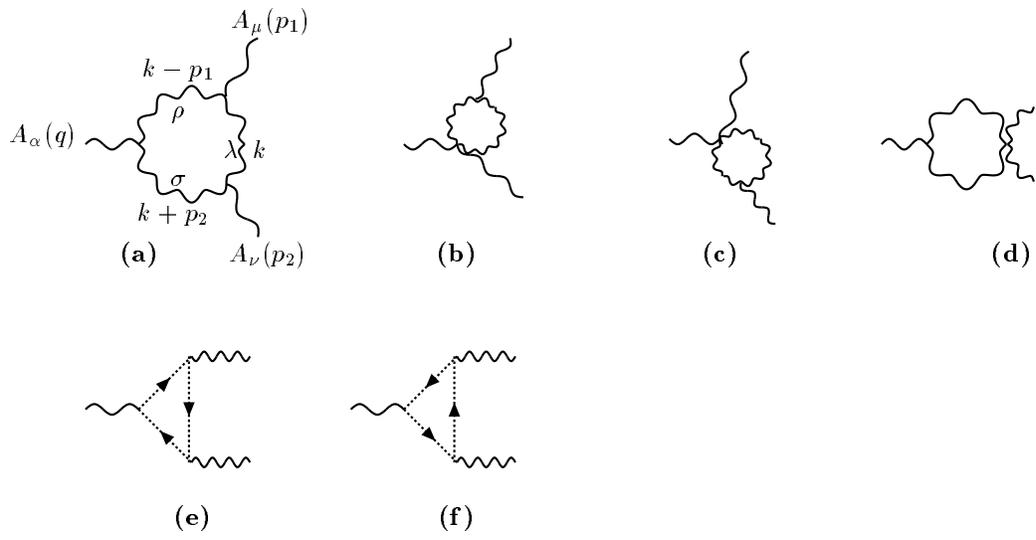
{\bf Fig.17}:
The Feynman diagrans contributing to the conventional one-loop 
three-gluon vertex $\Gamma_{\alpha\mu\nu}^{(1)}$.

\vskip0.5cm

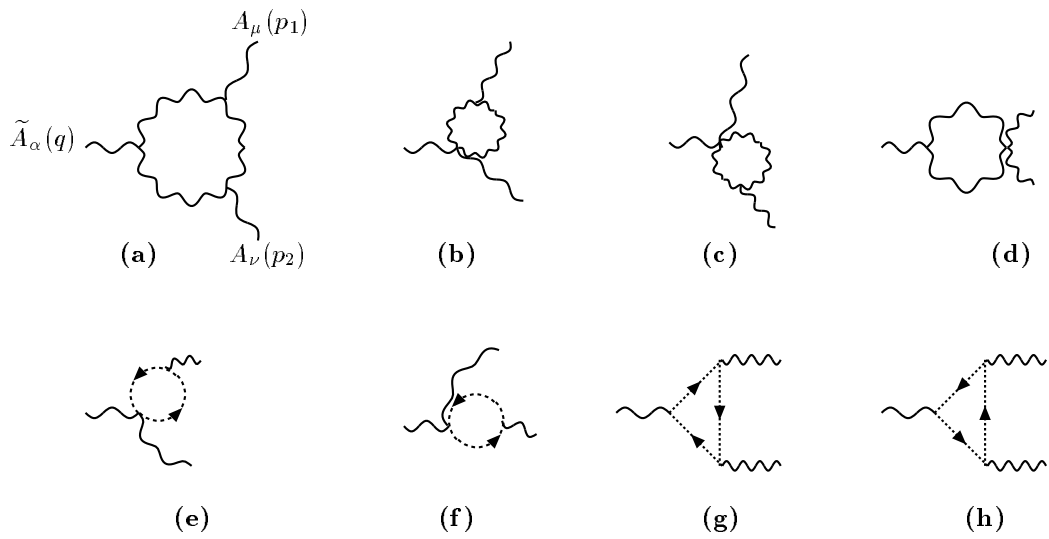
{\bf Fig.18}:
The diagrammatic representation of the one-loop 
three-gluon vertex $\Gamma_{F\alpha\mu\nu}^{(1)}$.

\vskip0.5cm

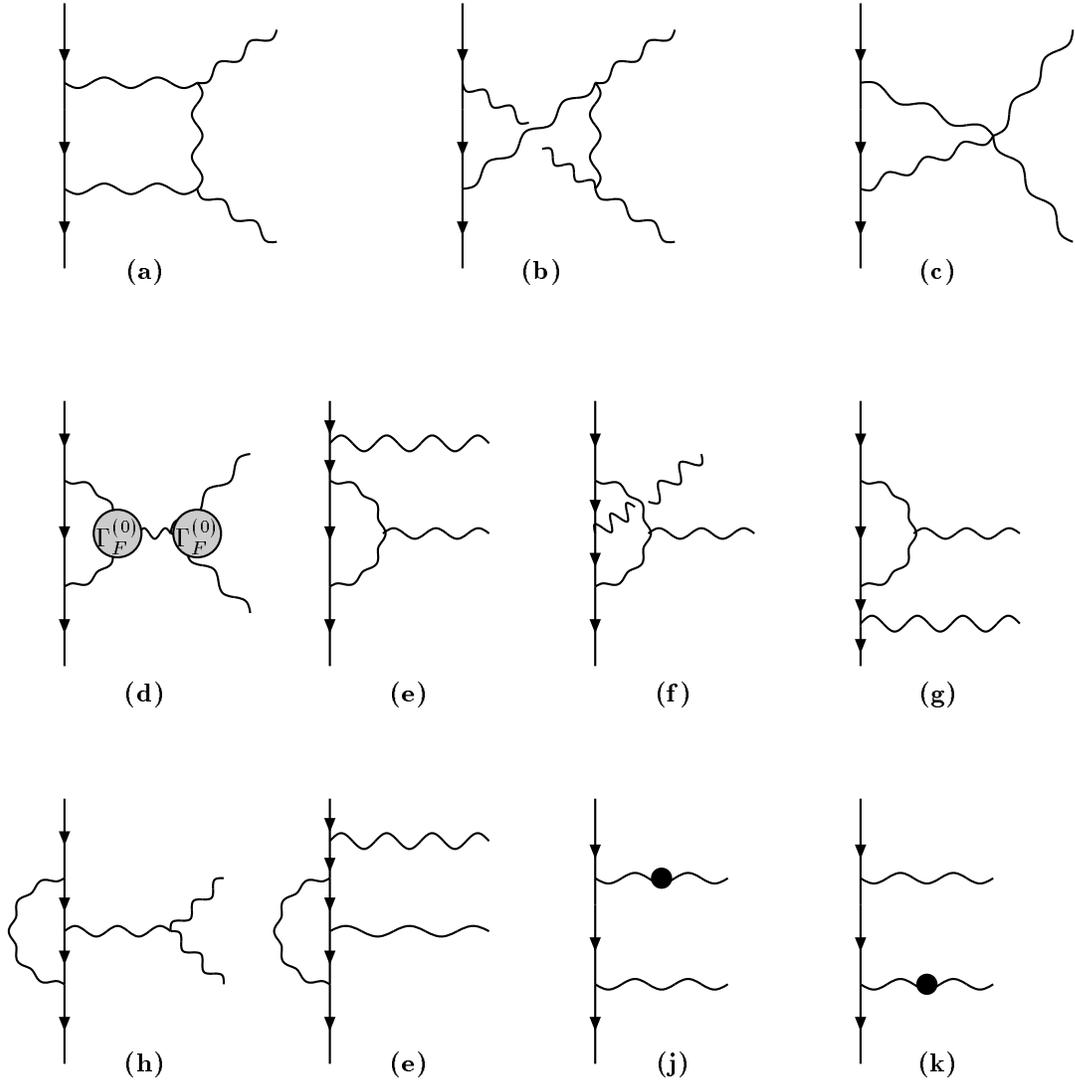
{\bf Fig.19}:
Some of the one-loop $t$-channel graphs contributing to 
$q(k_1) \bar{q}(k_2) \to g(p_1) g(p_2)$. 

\vskip0.5cm

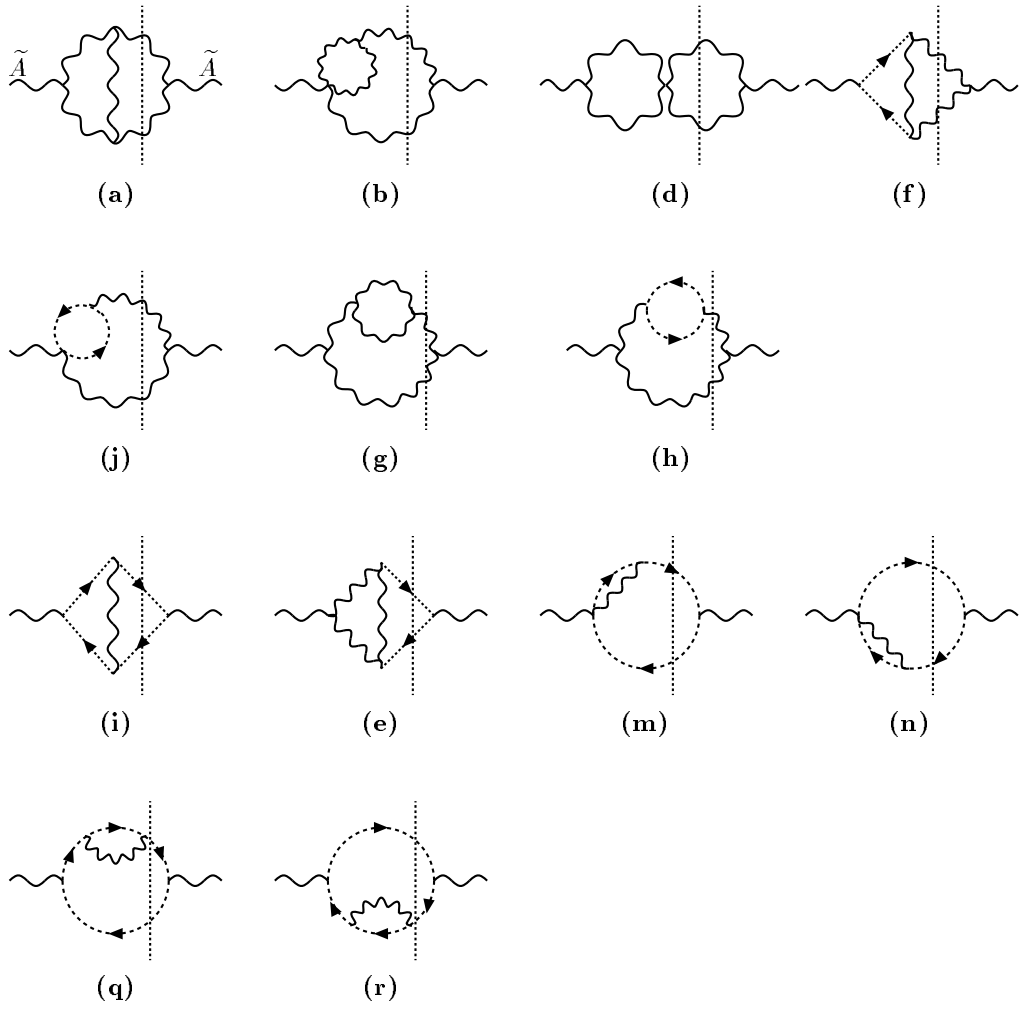
{\bf Fig.20}:
The two-particle Cutkosky cuts of the PT (and BFMFG) two-loop gluon self-energy.
We have used the same labelling of individual diagrams as in 
Fig. 9. The two upper (lower) rows show graphs where two gluon 
(ghost) lines have been cut.

\vskip0.5cm

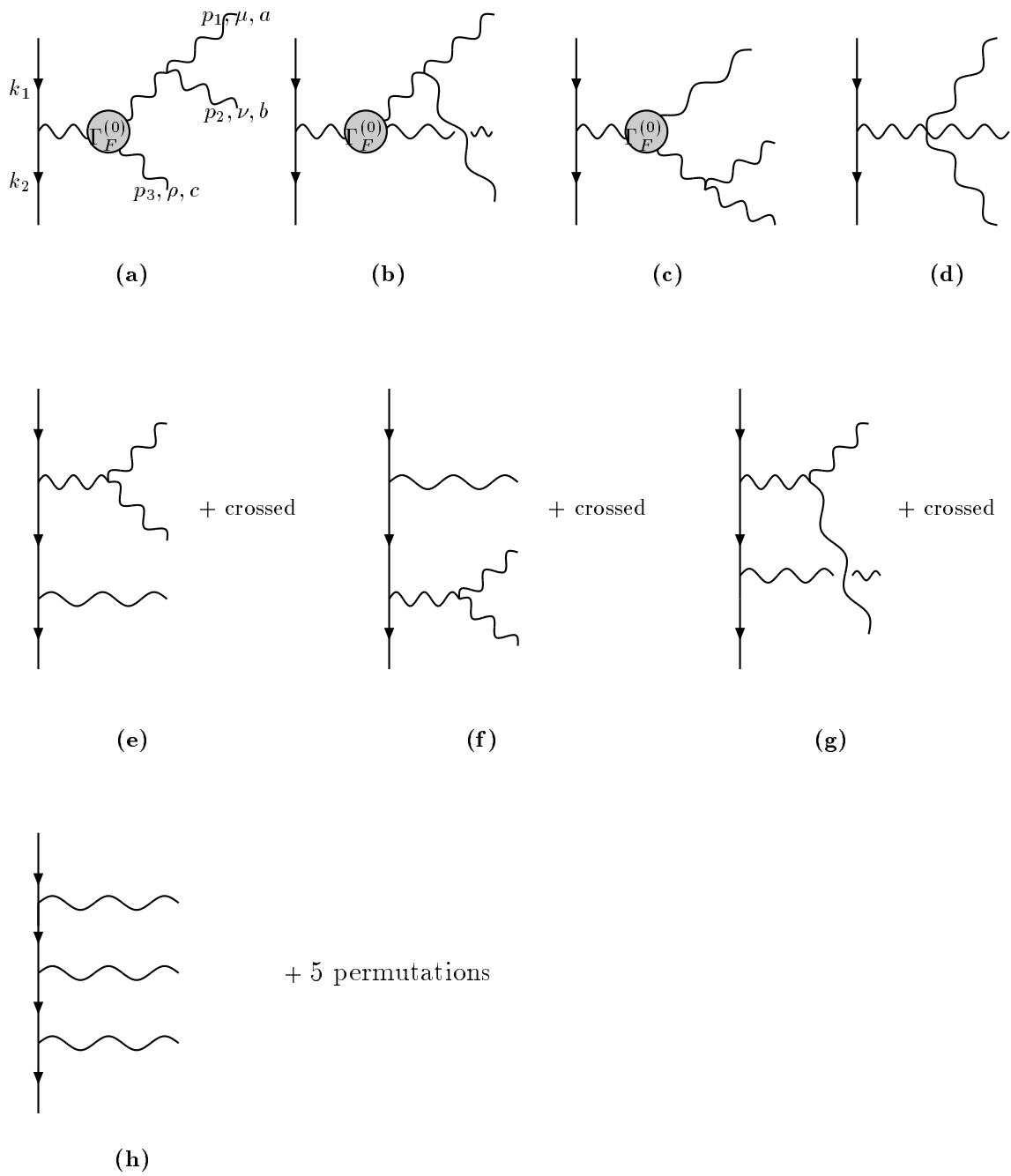
{\bf Fig.21}:
The tree-level graphs contributing to the process
$q(k_1) \bar{q}(k_2) \to g(p_1) g(p_2) g(p_3)$, after the PT rearrangement.

\vskip0.5cm

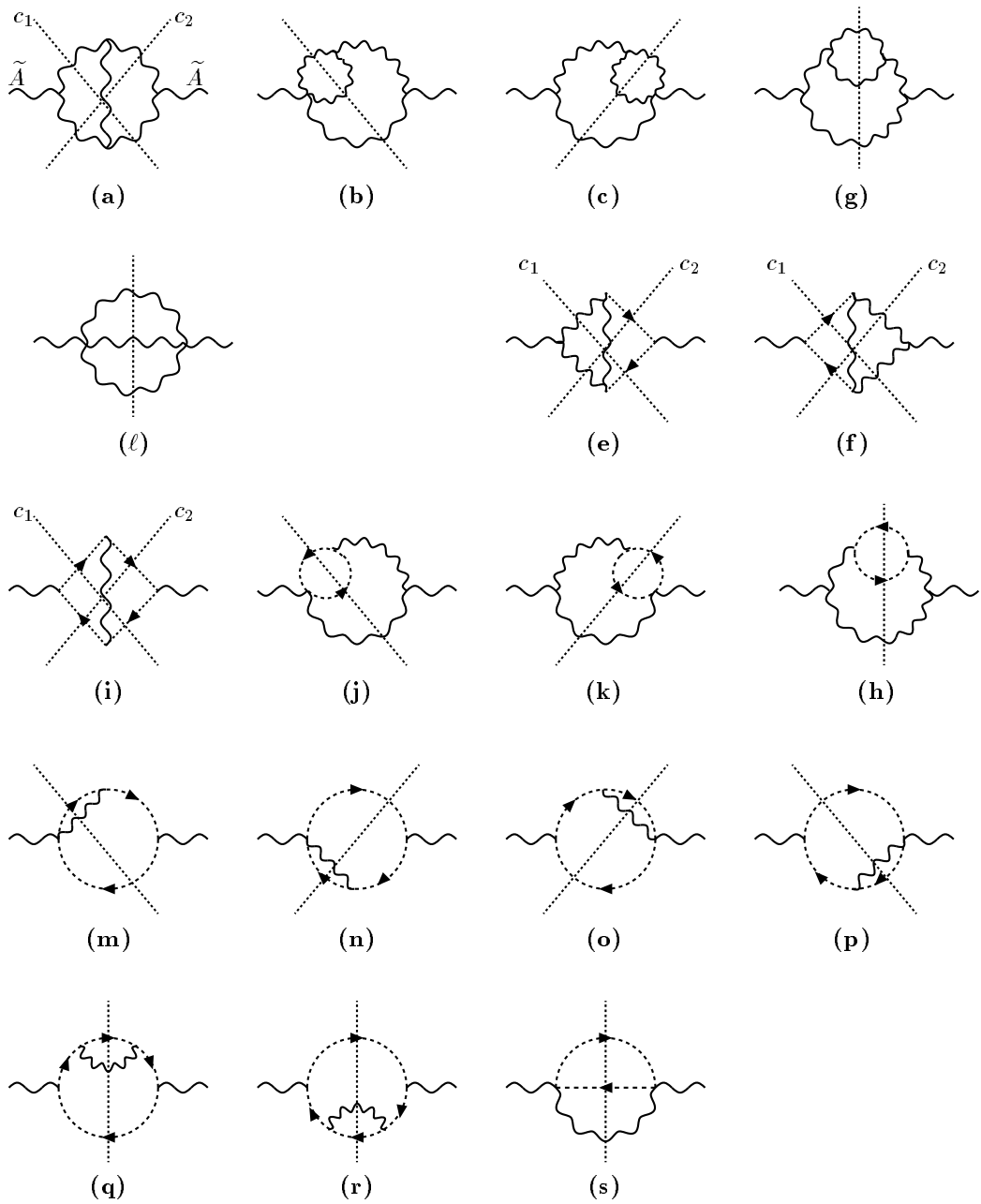
{\bf Fig.22}:
The three-particle Cutkosky cuts of the PT (and BFMFG) two-loop gluon self-energy.
We have used the same labelling of individual diagrams as in 
Fig. 9. The first five graphs have three-gluon cuts, 
the next two have two-gluon-one-ghost cuts,
while the remaining ones have one-gluon-two-ghost cuts.


\newpage

\begin{thebibliography}{99}

\bibitem{PTCPT} J.M. Cornwall, in {\em Proceedings of the French-American
    Seminar    on Theoretical   Aspects  of   Quantum Chromodynamics},
  Marseille,    France, 1981, edited J.W.    Dash  (Centre de Physique
  Th\'eorique, Marseille, 1982).

\bibitem{PT}
J.M. Cornwall, Phys.\   Rev.\ {\bf  D26},   1453 (1982).

\bibitem{PT2}
J.M.\ Cornwall and J.\ Papavassiliou, 
{Phys.\ Rev.}\ {\bf D40}, 3474 (1989).

\bibitem{pap1}
 J.~Papavassiliou, Phys.\ Rev.\ {\bf D41}, 3179 (1990). 

\bibitem{DS}
G.\ Degrassi and A.\ Sirlin, Phys.\ Rev.\ {\bf D46}, 3104 (1992); 

\bibitem{papsm}
J.~Papavassiliou, Phys.\ Rev.\ {\bf D 50}, 5958 (1994).
\bibitem{DKS}
G.~Degrassi, B.~Kniehl, and A.~Sirlin, Phys.\ Rev. \ {\bf D48},
R3963 (1993); 
J.~Papavassiliou and K.~Philippides,
Phys. Rev {\bf D 48}, 4255 (1993);
J.~Papavassiliou and C.~Parrinello,
Phys. Rev {\bf D 50}, 3059 (1994);
K. Hagiwara, S. Matsumoto, D. Haidt, and C.S. Kim 
Z.\ Phys. {\bf C64}, 559,(1994), Erratum-ibid. {\bf C68}, 352 (1995);
J. Papavassiliou, K. Philippides, and K. Sasaki, 
Phys.Rev.{\bf D53}, 3942 (1996). 

\bibitem{BFM1} B.S.\ DeWitt, {\em Phys.\ Rev.}\ {\bf 162}, 1195 (1967); G.\ 't
Hooft, in {\em Acta Universitatis Wratislvensis no.\ 38}, 12th Winter School
of Theoretical Physics in Kapacz, Functional and probabilistic methods in
quantum field theory vol.\ I (1975); B.S.\ DeWitt, {\em A gauge invariant
effective action}, in Quantum gravity II, ed.\ C.\ Isham, R.\ Penrose, and D.\
Sciama (Oxford press, 1981).

\bibitem{Abb} 
L.F. Abbott, {Nucl.\ Phys.}\ {\bf B185}, 189 (1981), and references therein.

\bibitem{AGS} L.F. Abbott, M.T. Grisaru, and R.K. Schaeffer,
Nucl.\ Phys.\ {\bf B229}, 372 (1983).

\bibitem{BFMPT} 
A.  Denner, S. Dittmaier,  and G. Weiglein, 
Phys.\ Lett.\ {\bf B333}, 420 (1994);
S.~Hashimoto, J.~Kodaira, Y.~Yasui, and K.~Sasaki,
Phys.\ Rev.\ {\bf D50}, 7066 (1994); 
E.~de~Rafael and N.~J.~Watson, unpublished.


\bibitem{PP1} J.~Papavassiliou  and A.~Pilaftsis, {  Phys.\ Rev.\
    Lett.}\ {\bf 75}, 3060 (1995);  {Phys.\ Rev.}\ {\bf D53}, 2128
  (1996).

\bibitem{PP2} J.~Papavassiliou  and A.~Pilaftsis,
{Phys.\ Rev.}\ {\bf D54}, 5315 (1996).

\bibitem{PRW} J.  Papavassiliou, E.  de Rafael,  and N.J. Watson, 
Nucl.\ Phys.\ {\bf B503}, 79 (1997).

\bibitem{PP3} J.~Papavassiliou  and A.~Pilaftsis, {  Phys.\ Rev.\
    Lett.}\ {\bf 80}, 2785 (1998);  {Phys.\ Rev.}\ {\bf D58}
  (1998):053002.


\bibitem{Tqcd}
G. Alexanian and V.P.Nair, Phys.\ Lett.\ {\bf B352}, 435 (1995); 
K.Sasaki, Nucl.Phys. {\bf B472} 271 (1996);
{\it ibid} {\bf B490}, 472 (1997).


\bibitem{NJW2}    N.J. Watson, Nucl.\ Phys.\ {\bf B494}, 388
  (1997).

\bibitem{SPEDR} 
S. Peris and E. de Rafael,
Nucl.\ Phys.\ {\bf B500} 325, (1997).


\bibitem{GPT} A. Pilaftsis, Nucl.\ Phys.\ {\bf B487}, 467 (1997).



\bibitem{CP} R.Cruz, B.Grzadkowski, and J.F.Gunion,
 Phys.\  Lett.\ {\bf B289}, 440 (1992); 
D. Atwood, G. Eilam, A. Soni, R.R. Mendel, R. Migneron 
Phys.\ Rev.\ Lett.\ {\bf 70} 1364 (1993).

 
\bibitem{CPPT}
A. Pilaftsis, 
Phys.\ Rev.\ Lett.\ {\bf 77}, 4996 (1996); Nucl.\ Phys.\ 
{\bf B504}, 61 (1997).


\bibitem{RKWJS} R.~Kleiss and W.~J.~Stirling, 
Phys.\ Lett.\ {\bf B182}, 75 (1986);
U.~Baur and E.~W.~N~Glover,
Nucl.\ Phys.\ {\bf B347}, 12 (1990);  
Phys.\ Rev.\ {\bf  D44}, 99 (1991).

\bibitem{Wfus}
K. Philippides and W.J. Stirling,
Eur.\ Phys.\ J.\ {\bf C9}, 181 (1999).

\bibitem{KL} See also 
D.C. Kennedy and B.W. Lynn, Nucl.\ Phys.\ {\bf B322}, 1 (1989).

\bibitem{papCor}
J.~Papavassiliou,
{\it The Pinch Technique Approach to the Physics of Unstable Particles},
hep-ph/9905328, contribution to the 1998 Corfu Summer Institute on Elementary
Particle Physics (JHEP proceedings).

\bibitem{pap2L1}
J.~Papavassiliou, {\it The Pinch Technique at Two Loops}, hep-ph/9912336.


\bibitem{NJWMontp}
N.J. Watson,  Nucl.Phys.Proc.Suppl.74:341-344,1999.

\bibitem{ST}
A. Slavnov, Theor. and Math. Phys. {\bf 10}, 99 (1973);
J.C. Taylor, Nucl.\ Phys.\ {\bf B33}, 436 (1971).

\bibitem{TH} 
G.~'t Hooft, Nucl.\ Phys.\ {\bf B33}, 173 (1971);
J.M. Cornwall and G. Tiktopoulos, Phys.\ Rev.\ {\bf D15}, 2937 (1977).


\bibitem{pap2} 
J.~Papavassiliou, Phys.\ Rev.\ {\bf D51}, 856 (1995). 


\bibitem{masken}
M. Passera and K. Sasaki,  Phys.\ Rev.\ {\bf D54}, 5763 (1996) 


\bibitem{ER} 
E. de Rafael and J.L. Rosner, Ann.\ Phys.\ (NY) {\bf 82}, 369 (1973)


\bibitem{HDP} H.D. Politzer, Phys.\ Rev.\ Lett.\ {\bf 30}, 1346 (1973);
D.\ Gross and F.\ Wilczek, Phys.\ Rev.\ Lett.\ {\bf 30}, 1343 (1973).


\bibitem{KSC} T. Kunimasa and T. Goto, Prog.\ Theor.\ Phys.\ {\bf 37},
452 (1967);
A.A. Slavnov, Theor.\ Math.\ Phys.\ {\bf 10}, 305 (1972);
J.M. Cornwall, Phys.\ Rev.\ {\bf D10}, 500 (1974).

\bibitem{FPP}
J.~R.~Forshaw, J.~Papavassiliou, and C.~Parrinello,
Phys.\ Rev.\ {\bf D59}:074008 (1999).


\bibitem{4g} 
 J.~Papavassiliou, Phys. Rev {\bf D47}, 4728 (1993).



\bibitem{phsir}
K. Philippides and A. Sirlin, Nucl.\ Phys.\ {\bf B477},59 (1996)

\bibitem{comment} 
The notion of ``internal''  and ``external'' three-gluon vertex can be
generalized  also   to include  the   case  where  the gauge-invariant
physical   observable  is not   an $S$-matrix element,   but instead a
gauge-invariant Green's function. For example one  could carry out the
PT   program inside   the  Green's function   $G(x,y)  =  \langle 0  |
T(Tr\{\Phi(x)\Phi(x)^{\dagger}\} Tr   \{\Phi(y)\Phi(y)^{\dagger}\}  |0
\rangle $, where $\Phi(x)$ is a matrix describing a set of scalar test
particles in  an appropriate representation of  the gauge  group; this
latter  quantity  was  employed  by   Cornwall  for the   original  PT
construction \cite{PT}.  In that case the ``external'' momentum is the
momentum transfer  between the two sides  of the scalar loop, i.e., as
explained in  \cite{PT}, one should count  loops as if the $\Phi$ loop
were opened at $x$ and $y$.



\bibitem{NJW3} 
For an attempt to define the 2-loop PT quark self-energy 
by modifying its internal three-gluon vertices see   
N.J. Watson, Nucl.\ Phys.\ {\bf B552}, 461 (1999) .

\bibitem{3d}
W. Buchm\"uller and O. Philipsen, 
Phys. Lett. {\bf B397}, 112 (1997);
R. Jackiw and S.-Y. Pi,  Phys.\ Lett.\ {\bf B 403}, 297 (1997);
J.M. Cornwall, Phys.\ Rev.\ {\bf D57}, 3694 (1998);
Phys.\ Rev.\ {\bf D59} (1999):125015.


\bibitem{WEC} W.E. Caswell, Phys.\ Rev.\ Lett.\ {\bf 33}, 1346 (1973).

\bibitem{DRTJ}
D.R.T. Jones, Nucl.\ Phys.\ {\bf B75}, 531 (1974).

\bibitem{expl} At one loop one can verify
these  relations   explicitly  by computing     the  divergent part  of
$V_P^{(1)}$ in    Eq. (\ref{somedef2}), 
together with  the expressions  for  the various
one-loop   renormalization   constants   listed  in Eq.5 -- Eq.7 of
\cite{DRTJ}   (notice  the   difference in    the  notation, and  that
$d=4-\epsilon$ )


\bibitem{intquarks}
We will not consider the case where one has 
quarks as intermediate states; it can be
straightforwardly deduced from the 
gluonic case. In fact, one should be able to verify 
easily that the absorptive construction in this case
is in full agreement with the direct two-loop PT generalization
for the case where a mass-less quark-loop is present in the
two-loop diagrams, which has been presented in 
S. Bauberger, F.A. Berends, M. B\"ohm, M. Buza and G. Weiglein, 
Nucl.\ Phys.\ Proc.\ Suppl.\ {\bf B37}, 95 (1994). 

\bibitem{BRS} C.\ Becchi, A.\ Rouet, and R.\ Stora, 
Ann.\ Phys.\ (NY) {\bf 98}, 287 (1976).

\bibitem{ChengLi} See, {\em e.g.}, T.-P. Cheng and L.-F. Li, {\em Gauge Theory
of Elementary Particle Physics}, Clarendon Press, Oxford, 1985, p.\ 277.

\bibitem{planar} Yu.L. Dokshitzer, D.I. Dyakonov, 
and S.I. Troyan, Phys.\ Rep.\ {\bf 58}, 269 (1980);  
A. Andra$\breve{\mbox{s}}$i and J.C. Taylor, Nucl.\ Phys.\ {\bf B192},
283 (1981); D.M. Capper and G. Leibbrandt, Phys.\ Rev.\ {\bf D25}, 1002
(1982).
\bibitem{axun} See, {\em e.g.}, 
third paper of \cite{KSC}. 


\bibitem{NJW1}  N.J. Watson, Phys.\  Lett.\ {\bf B349},   155
  (1995).



\bibitem{RDP} R.D.Pisarski, private communication.

\bibitem{GGr}
G. Grunberg, 
Phys.\ Rev.\ {\bf D40}, 680 (1989); Phys.\ Rev.\ {\bf D46}, 2228 (1992). 


\bibitem{HKS}
R. Haussling, E. Kraus, K. Sibold,
Nucl.\ Phys.\ {\bf B539}, 691 (1999).
\bibitem{PPS} 
The corresponding PT constant at one-loop was first presented in
\cite{pap2}. 


\bibitem{GrGa} I thank G. Grunberg and E. Gardi 
for emphasizing this point to me.


\bibitem{GaSir}
P. Gambino and A. Sirlin, Phys.\ Lett.\ {\bf B355}, 295 (1995).



\bibitem{RK} 
R. Kallosh, Nucl.\ Phys.\ {\bf B78}, 293 (1974).


\bibitem{IR}
T.Kinoshita, J.\ Math.\ Phys.\ {\bf 3}, 650 (1962); 
E.C.Poggio and H.R.Quinn, Phys.\ Rev.\ {\bf D14}, 578 (1976);
G.Sterman,  Phys.\ Rev.\ {\bf D14}, 2123 (1976).

\bibitem{TM}
Taizo Muta, {\em Foundations of Quantum Chromodynamics}, 
World Scientific Lecture Notes in Physics --Vol.5, 1987, ch.6.


\bibitem{CDE}
S. Catani and E. D'Emilio, Fortsch.\ Phys.\ {\bf 41}, 261 (1993). 

\bibitem{DMW}
Yu.L. Dokshitzer, G. Marchesini, and B.R. Webber,
Nucl.\ Phys.\ {\bf B469}, 93 (1996)   

\bibitem{gm2} F. Halzen, G. Krein and A.A. Natale, 
Phys.\ Rev.\ {\bf D47}, 295 (1993); M.B. Gay Ducati, 
F. Halzen and A.A. Natale, Phys.\ Rev.\ {\bf D48}, 2324 (1993); 
J.R. Cudell and B.U. Nguyen, Nucl.\ Phys.\ {\bf B420}, 669 (1994); 
A. Donnachie and P.V. Landshoff, Phys.\ Lett.\ {\bf B387}, 637 (1996);
B.Magradze, hep-ph/9911456. 

\bibitem{FL} 
Y.J. Feng, C.S. Lam,  Phys.Rev. {\bf D50}, 7430 (1994).

\bibitem{ZB} 
Z. Bern and D.C. Dunbar, Nucl.\ Phys.\ {\bf B379}, 562 (1992),
and references therein.

\bibitem{VMLRM} P. Di Vecchia, L. Magnea, A. Lerda, R. Russo, and R. Marotta,
Nucl.\ Phys.\ {\bf B469}, 235 (1996). 


\bibitem{GG}
P. Gambino and P.A. Grassi,
{\it The Nielsen Identities of the SM and the Definition of Mass}
hep-ph/9907254. 

\bibitem{WLT}
C. Schubert,  {\em An Introduction to the Worldline Technique
for Quantum Field Theory Calculations},
Lectures given at 36th Cracow School of Theoretical Physics, Zakopane, 
Poland, 1-10 Jun 1996, 
Acta Phys.\ Polon. {\bf B27}, 3965 (1996), and references therein. 

\bibitem{EL}
W. Beenakker, F.A. Berends, A.P. Chapovsky,
{\em An Effective Lagrangian Approach for Unstable Particles} 
hep-ph/9909472.


\end{thebibliography}
\end{document}